\documentclass[floatfix,twocolumn,showpacs,aps,tightenlines,superscriptaddress]{revtex4-1}
\usepackage{graphicx}
\usepackage{color}
\usepackage{psfrag}
\usepackage{dcolumn}
\usepackage{bm}

\usepackage{amssymb}
\usepackage{amsmath}
\usepackage{amsfonts}
\usepackage{longtable}
\usepackage{xspace}
\usepackage{verbatim}

\newcommand{\beq}{\begin{equation}}
\newcommand{\eeq}{\end{equation}}
\newcommand{\KMS}{\rm km\,s^{-1}}

\newcommand{\be}{\begin{equation}}
\newcommand{\ee}{\end{equation}}
\newcommand{\bea}{\begin{eqnarray}}
\newcommand{\eea}{\end{eqnarray}}
\newcommand{\bes}{\begin{subequations}}
\newcommand{\ees}{\end{subequations}}

\newcommand{\ud}{\mathrm{d}}
\newcommand{\MP}{{moving punctures}\xspace}
\newcommand{\MPA}{{moving punctures approach}\xspace}
\newcommand{\hispid}{\textsc{HiSpID}\xspace}
\newcommand{\by}{BY\xspace}

\usepackage{bm}

\begin{document}

\title{Puncture initial data for black-hole binaries with
high spins and high boosts}

\author{Ian Ruchlin} 
\affiliation{Department of Mathematics, West Virginia University,
Morgantown, West Virginia 26506, USA}
\affiliation{Center for Computational Relativity and Gravitation,
  School of Physics and Astronomy,
  Rochester Institute of Technology, 85 Lomb Memorial Drive,
  Rochester,
   New York 14623}
\author{James Healy} 
\author{Carlos O. Lousto}
\author{Yosef Zlochower} 
\affiliation{Center for Computational Relativity and Gravitation,
School of Mathematical Sciences,
Rochester Institute of Technology, 85 Lomb Memorial Drive, Rochester,
 New York 14623}

\date{\today}

\begin{abstract}
We solve the Hamiltonian and momentum constraints of 
general relativity for
two black holes with nearly extremal spins and relativistic boosts
in the puncture formalism.
We use a non-conformally-flat ansatz
with an attenuated superposition of two Lorentz-boosted, conformally Kerr 
or conformally Schwarzschild 3-metrics
and their corresponding extrinsic curvatures.
We compare evolutions
of these data with the standard Bowen-York 
conformally flat ansatz (technically limited to
intrinsic spins $\chi=S/M^2_{\text{ADM}}=0.928$ and boosts $P/M_{\text{ADM}}=0.897$),
finding, typically, an
order of magnitude smaller burst of spurious radiation and agreement with
inspiral and merger. As a first case
study,
we evolve two equal-mass
black holes from rest with an initial separation of $d=12M$ and spins 
$\chi_i=S_i/m_i^2=0.99$, 
compute the waveforms produced by the collision,
the energy and angular momentum radiated, and the recoil of the
final remnant black hole.
We find that the black-hole trajectories curve
at close separations, leading to
the radiation of angular momentum.
We also study orbiting nonspinning and moderate-spin  black-hole
binaries and compare these with standard Bowen-York data.
We find a substantial reduction in 
the nonphysical initial burst of radiation which leads to
cleaner waveforms.
Finally, we study the case of orbiting binary black-hole systems with spin magnitude $\chi_i=0.95$  in an aligned configuration and
 compare waveform and final remnant results with those of the SXS
 Collaboration~\cite{Mroue:2013xna},
finding excellent agreement. This represents the first
\MP evolution of orbiting and spinning black holes 
exceeding the Bowen-York limit.
Finally, we study different choices of the initial lapse and lapse
evolution equation in the \MPA to improve
the accuracy and efficiency of the simulations.
\end{abstract}

 \maketitle

\section{Introduction}\label{sec:intro}

The detection of gravitational waves from merging binary black holes
\cite{Abbott:2016blz,Abbott:2016nmj}, as predicted by numerical
relativity simulations~\cite{Pretorius:2005gq, Campanelli:2005dd, Baker:2005vv},
further highlights that
general relativity is central to the modern understanding of
much of astrophysics, from cosmological evolutions down to the end state of
large stars. Crucial to this is the correctness of the theory
itself and the elucidation of its predictions
\cite{TheLIGOScientific:2016src,TheLIGOScientific:2016pea}.
 While much can be done using analytic techniques,
 one of the
most interesting regimes---the merger phase of compact-object binaries---requires the use of
large-scale numerical relativity simulations
\cite{Lovelace:2016uwp,Abbott:2016apu}. 
In order to evolve these systems, one requires appropriate initial
data that allow for the simulation of binaries with astrophysically
realistic parameters. Perhaps most important and challenging
of all is the inclusion of large spins.

Highly spinning black holes (BHs) are thought to be common.
For example, supermassive BHs with
high intrinsic spins are fundamental to the contemporary understanding of
active galaxies and galactic evolution, in general. 
In units with $c = 1$ and $G = 1$, a BH's spin magnitude $S$
(i.e., intrinsic angular momentum) is bounded by its mass $m$, where
the maximum dimensionless spin is given by $\chi \equiv S / m^2 = 1$.  While it is actually hard to
have an accurate measure of astrophysical  BH spins, in a few cases
the spins have been measured \cite{Reynolds:2013rva} and some were
found to be near the maximal value.  Since galactic mergers are
expected to lead to mergers of highly spinning BHs, it is important to
be able to simulate black-hole binaries (BHBs) with high spins in order
to model the dynamics of these ubiquitous objects.

Spin can greatly affect the dynamics of
merging BHBs. Important spin-based effects
include the {\it hangup} mechanism
\cite{Campanelli:2006uy}, which delays or prompts the merger of the
binary according to
the sign of the spin-orbit coupling; the {\it flip-flop} of spins
\cite{Lousto:2014ida}, which is due to  a spin-spin coupling effect
capable of
completely reversing the sign of individual spins; and
finally, highly spinning binaries may recoil
at thousands of km/s~\cite{Campanelli:2007ew,Campanelli:2007cga}
due to the asymmetrical emission of gravitational radiation induced by the BH
spins~\cite{Lousto:2012su,Lousto:2012gt}.
These effects are
maximized for highly spinning BHs.

As a consequence of spin-orbit and spin-spin interactions,
high spins can have a dramatic effect on the gravitational waveform.
For example, unlike 
low-spin binaries, highly spinning
binaries can radiate more than $11\%$ of their rest mass
\cite{Hemberger:2013hsa, Healy:2014yta}, the majority of which emanates
during the last moments of merger, down to the formation of a final single
spinning BH. Efforts to interpret
 gravitational wave signals from such systems require accurate model gravitational
waveforms~\cite{Aasi:2014tra, Aylott:2009tn, Aylott:2009ya, Ajith:2012az, Hinder:2013oqa}. 

The \emph{\MPA}~\cite{Campanelli:2005dd,Baker:2005vv}
has proven to be very effective in evolving BHBs with similar
masses and relatively small initial separations, as well as small mass ratios \cite{Lousto:2010ut} and large separations~\cite{Lousto:2013oza}. It is also effective
for more general multiple BH systems \cite{Lousto:2007rj}, hybrid BH--neutron-star binaries \cite{Shibata:2006ks},
and gravitational collapse \cite{Baiotti:2006wm}.
However,
numerical simulations of highly spinning BHs have proven to
be very challenging. The most commonly used initial
data to evolve those binaries, which are based on the
Bowen-York (BY) ansatz~\cite{Bowen:1980yu}, 
use a conformally flat 3-metric. This method has a fundamental spin upper limit of $\chi=0.928$~\cite{Cook:1989fb,Dain:2002ee}. Even when
relaxing the BY ansatz (while retaining conformal flatness), the spin is still
bounded above by $\chi=0.932$~\cite{Lousto:2012es}.

In order to exceed this limit and approach maximally spinning
BHs with $\chi=1$, one has to allow for a more general
3-metric that captures the non-conformally-flat aspects of the
Kerr geometry.
Dain showed \cite{Dain:2000hk} that it is possible to
find solutions to the initial value problem representing a pair of
Kerr-like BHs. 
This proposal was implemented for the case of thin-sandwich initial data 
~\cite{York99,Cook:2000vr,Pfeiffer:2002iy}
with excision of the BH interiors, and produces stable evolutions of orbiting
BHBs with $\chi\sim0.97$~\cite{Lovelace:2011nu,Hemberger:2013hsa}, 
and more recently $\chi \sim 0.998$~\cite{Lovelace:2014twa,Scheel:2014ina}. 
Dain's method was also tested for the case of head-on collisions (from rest) of spinning BHs using the moving punctures approach, which does not employ excision. These are compared to the BY data with spins up to $\chi=0.90$~\cite{Hannam:2006zt}.

In this paper we revisit, and numerically implement,
the problem of finding solutions
to the puncture initial value problem representing two
nearly-extremal-spin BHs, and the subsequent evolution using the moving punctures approach---the most widespread method to evolve BHBs, implemented in the
open source {\sc EinsteinToolkit}~\cite{Loffler:2011ay,
einsteintoolkit, cactus_web, Schnetter-etal-03b}.

To solve for these new data, we construct a superposition of two
conformally Kerr 3-metrics with the corresponding superposition
of Kerr extrinsic curvatures. Note that we do not use the Kerr-Schild
slice \cite{Marronetti:2000rw}, 
but rather use Boyer-Lindquist slice, which is amenable for puncture
evolution. To regularize the problem, the
superposition is such that very close to each BH, the metric and
extrinsic curvature are exactly Kerr~\cite{Bonning:2003im}, or exactly
flat (through attenuation). We then simultaneously solve the
 Hamiltonian
and momentum constraints for an overall conformal factor for the
metric and
 nonsingular correction
to the extrinsic curvature
 using a modification of the {\sc TwoPunctures}
\cite{Ansorg:2004ds} spectral initial data solver. We refer to this
extension as HIghly SPinning Initial Data (\hispid, pronounced ``high speed'').

We evolve these data sets for black-hole binaries from rest and 
find that the spurious initial radiation is significantly reduced
compared to BY initial data (for $\chi\leq0.9$). Ideally, one would
want to use data that had incorporated the exact radiation content of
a binary inspiraling from infinity (for post-Newtonian inspired
radiation content into
the initial data ansatz, see for example
\cite{Tichy:2002ec,Kelly:2007uc,Kelly:2009js,Mundim:2010hu}), but
failing this, one at least wants to minimize the nonphysical
radiation. 
Furthermore, it is this spurious radiation (or more precisely, the
part of this radiation that is absorbed by the black holes) that
prevents conformally flat data from modeling black holes with spins
larger than $\sim 0.93$ (see, e.g., Fig.~\ref{fig:initfinal_by} below). 
Thus by using the \hispid data, we get both
more accurate waveforms for moderate spin binaries, and can go beyond
the BY limit to at least
$\chi=0.99$, which has not been possible before for \MP codes.

We also consider Lorentz boosted Schwarzschild BHs in a
quasicircular orbital configuration. Here too, it is important to minimize 
the spurious radiation content of the initial data in order to
achieve a very accurate gravitational waveform computation. 
We again compare with the new data with equivalent BY
binaries.
BY binaries are also limited in the maximum boost the BHs
can  have ($P/M_{\text{ADM}}=0.897$, see Ref.~\cite{Cook90a}), while
the \hispid data can be used to boost BH in excess of
$P/M_{\text{ADM}}=4$~\cite{Healy:2015mla}.

Finally we address the extremely important problem of modeling 
highly spinning black holes in quasicircular orbits. 
For moderate spins ($\chi \lesssim 0.9$), we find that the \hispid
data
reproduces the dynamics seen in BY binaries, but again, with
substantially lower spurious radiation. BY data cannot model
highly spinning binaries. We thus compare our simulation of a
$\chi=0.95$ quasicircular binary with those produced by the SXS
Collaboration~\cite{SXS:catalog, Hemberger:2013hsa, Mroue:2013xna}.
We find excellent agreement between the two methods. The \MPA is used
in the open source EinsteinToolkit and by many groups worldwide. The
\hispid approach thus opens up the  possibility for numerical
evolutions
of highly spinning binaries by numerical relativists worldwide.

This paper is organized as follows. In Sec.\@~\ref{sec:ID} we present the formalism to
solve for the initial data. We choose the standard transverse-traceless version
since it provides the simplest set of equations and allows one
to both achieve a reduction of the spurious initial radiation 
content and overcome the technical limits of the conformally
flat initial data reaching highly spinning BHs and
highly relativistic velocities.
We describe the explicit conformal decomposition and attenuation
functions used to regularize the superposition of boosted conformal
Kerr/Schwarzschild BHs in the puncture approach.

In Sec.\@~\ref{sec:techniques} we describe the numerical techniques
to solve for the initial data as an extension of those used to solve
the Hamiltonian constraint with the \textsc{TwoPunctures} code.
We also provide a summary of the evolution techniques used in
the regime of parameters previously unexplored with the 
\MPA.

In Sec.\@~\ref{sec:SpinningBHs} we show the convergence with spectral collocation points of the spinning initial data to levels of accuracy acceptable for
evolution. We compare waveforms from the new \hispid initial data to
those of the standard spinning BY solution for $\chi=0.90$.
We then evolve highly spinning BHs with $\chi=0.99$
from rest and discuss the results for the radiated energy and momenta
as well as the horizon measures of mass and spin for the individual and final
BHs.

In Sec.\@~\ref{sec:BoostedBHs}
we show the convergence with spectral collocation points of the initial
data for nonspinning Lorentz-boosted
BHs. We compare waveforms for our new initial data with
the standard boosted BY solution in quasicircular orbit
to highlight the benefits of the lower initial spurious radiation
of our data.

In Sec.\@~\ref{sec:spinningorbiting} we pursue the study of more
generic spinning and orbiting black-hole-binary systems, including
binaries with intrinsic
spin magnitude $\chi_i=0.95$.
 We perform two numerical
evolutions at different initial separations and compare them with those
performed by the SXS Collaboration
finding excellent agreement for both the waveforms and the final remnant.

In the discussion in Sec.\@~\ref{sec:discussion} we summarize the results 
and consider the next series of developments
for the initial data and evolution of BHBs.

In the appendixes we study how the initial choice of the lapse
and its subsequent gauge evolution~\cite{Alcubierre02b}
 affects the accuracy of the simulation
at the typical marginal resolutions used to evolve highly spinning
BHs, orbiting BHs, and
in high-energy head-on BH collisions.
We also give the explicit definitions 
used to compute the Arnowitt-Deser-Misner (ADM) mass and momenta of our initial data.

Throughout this paper we use geometric units where $G=c=1$. The
vacuum Einstein equations  are scale invariant in the sense that if we
rescale all masses, times, distances, momenta, etc., by the
appropriate factor, we obtain an equivalent solution. In the case of
a black-hole binary, this amounts to rescaling the total mass while
keeping the mass ratio fixed, keeping the dimensionless spins fixed,
and rescaling the momenta by the same factor as the masses. When
reporting quantities with dimension, we rescale each by an
appropriate power of an arbitrary positive constant $M$ (which has
dimensions of mass).
\section{Initial Data}\label{sec:ID}

In this section, we summarize the initial data formalism used to describe 
BHBs with spin magnitudes up to near maximal.  First, we review 
the conformal decomposition of general relativity's field equations into 
a set of constraint and evolution equations.  Then, we discuss methods
for generating the necessary background metric and extrinsic
curvature and provide detailed expressions for these in simple cases.
Finally, we demonstrate methods for ameliorating effects of singularities
at the punctures that allow us to find solutions to the elliptical
constraint equations using 
 numerical pseudospectral methods.  
\subsection{Constraints}\label{subsec:Constraints}

In the Cauchy problem of general relativity, the four-dimensional
pseudo-Riemannian spacetime manifold is foliated into
three-dimensional spatial hypersurfaces $\Sigma_t$, parametrized as
surfaces where the time function $t$ is constant.  Vacuum solutions to general relativity's field equations on the initial slice $\Sigma_0 = \Sigma_{t = 0}$ must satisfy \cite{Arnowitt62}
\begin{align}
  \mathcal{H} &\equiv R + K^2 - K_{i j} K^{i j} = 0 \label{eq:Ham} \; , \\
  \mathcal{M}^{i} &\equiv D_j (K^{i j} - \gamma^{i j} K) = 0 \label{eq:mom} \; ,
\end{align}
known, respectively, as the Hamiltonian and momentum constraints.
Latin indices represent spatial degrees of freedom.  Here, $\gamma_{i
j}$ is the induced spatial metric tensor on $\Sigma_0$ with the
associated covariant derivative $D_i$, and $R = \gamma^{i j} R_{i j}$ is the trace of the spatial Ricci tensor $R_{i j}$.  The extrinsic curvature tensor of $\Sigma_0$ and its trace (the mean curvature) are denoted by $K_{i j}$ and $K = \gamma^{i j} K_{i j}$, respectively.

In the conformal transverse-traceless (CTT)
formalism~\cite{York99,Cook:2000vr,Pfeiffer:2002iy,AlcubierreBook2008},
the constraints \eqref{eq:Ham} and \eqref{eq:mom} become a set of
elliptic differential equations for a conformal factor, $\psi$, and an
auxiliary vector, $b^i$. 

The conformal factor relates the physical
metric $\gamma_{ij}$ of the initial slice to a conformally related
metric $\tilde \gamma_{ij}$ by
\begin{equation*}
  \label{eq:conformal_metric}
  \gamma_{i j} = \psi^4 \tilde{\gamma}_{i j} \; .
\end{equation*}
 All quantities with a tilde are associated with $\tilde{\gamma}_{i j}$.  The conformal factor $\psi$ is a scalar function that is everywhere positive.  The extrinsic curvature tensor is split into trace and trace-free parts
\begin{equation}
  \label{eq:Kij_decomp}
  K_{i j} = A_{i j} + \frac{1}{3} \gamma_{i j} K \; .
\end{equation}
It is convenient to adopt the conformal rescaling
\begin{equation}
  \label{eq:Aij_conformal}
  A_{i j} = \psi^{-2} \tilde{A}_{i j} \; ,
\end{equation}
while leaving the mean curvature conformally invariant, $K = \tilde{K}$.  CTT splits $\tilde{A}_{i j}$ into a symmetric, trace-free part and a longitudinal part:
\begin{equation}
  \label{eq:general_Aij}
  \tilde{A}_{i j} = \tilde{M}_{i j} + \frac{1}{\tilde{\sigma}} (\tilde{\mathbb{L}} b )_{i j}
\end{equation}
where $\tilde{\sigma}$ is a positive definite scalar and the longitudinal vector derivative acting on $b^{i}$ is defined by
\begin{equation*}
(\tilde{\mathbb{L}} b)_{i j} \equiv \tilde{D}_{i} b_{j} +
\tilde{D}_{j} b_{i} - \frac{2}{3} \tilde{\gamma}_{i j} \tilde{D}_{k} b^{k} \; .
\end{equation*}

In the CTT formalism, the freely specifiable degrees of freedom are contained in $\tilde{\gamma}_{i j}$, $\tilde{M}_{i j}$, $K$, and $\tilde{\sigma}$.
In the case with no boost, the Kerr metric admits spatial hypersurfaces satisfying the maximal slicing condition $K = 0$~\cite{Dain:2000hk}.  We will adopt nontrivial $K$ for the boosted case (see Sec.\@~\ref{subsec:Conformal_Boosted_Schwarzschild}).
For simplicity, we set $\tilde{\sigma} = 1$ everywhere. With these choices, the constraint equations \eqref{eq:Ham} and \eqref{eq:mom} become
\begin{align}
    \tilde{D}^2 \psi - \frac{\psi \tilde{R}}{8} - \frac{\psi^{5} K^{2}}{12} + \frac{\tilde{A}_{i j} \tilde{A}^{i j}}{8 \psi^{7}} &= 0 \label{eq:final_HAM} \; , \\
  \tilde{D}_j \tilde{A}^{i j} - \frac{2}{3} \psi^6 \tilde{\gamma}^{i
  j} \tilde{D}_j K &= 0 \label{eq:MOM_A} \; ,
\end{align}
where $\tilde{D}^2 \equiv \tilde{\gamma}^{i j} \tilde{D}_{i} \tilde{D}_{j}$.

\subsection{Background Metric}

To calculate the spatial metric and
extrinsic curvature associated with a boosted
black hole of mass $m$, linear $3$-momentum $P^i$, and spin $S^i$, we
Lorentz boost the four-dimensional Kerr (Schwarzschild in the case were
$S=0$) line element in
Cartesian coordinates (we discuss specific coordinate systems below).
We then extract from the transformed metric the spatial metric
$\gamma_{i j}^*$, the lapse function $\alpha^*$, and the shift vector
$\beta^{i}_*$ (a super-/subscript $*$ indicates that this is a single
black-hole quantity). We 
then obtain the extrinsic curvature $K_{i j}^*$ on $\Sigma_{0}$ using the evolution equation for the spatial metric
\begin{equation}
  K_{i j}^* = \frac{1}{2 \alpha^*} \left (D_{i}^* \beta_{j}^* +
  D_{j}^* \beta_{i}^* - \partial_{t} \gamma_{i j}^* \right ) \; .
  \label{eq:Kdef}
\end{equation}
CTT separates this into trace and trace-free parts
\begin{equation*}
  K_{i j}^* = \psi^{-2}_* \tilde{A}_{i j}^* + \frac{1}{3} \psi^{4}_*
  \tilde{\gamma}_{i j}^* K^* \; ,
\end{equation*}
where $K^* = \gamma^{i j}_* K_{i j}^*$.
When factoring $\gamma_{ij}$ into $\gamma_{ij} = \psi^4 \tilde
\gamma_{ij}$, any choice of positive function $\psi$ will lead to a
valid elliptical constraint system outside the black holes. In the
puncture approach~\cite{Brandt97b}, one chooses the conformal factor $\psi$ so
that the resulting conformally related metric
is nonsingular. This can be accomplished in several ways.  For example, one can choose to include only
the leading-order contributions to the background conformal factor,
i.e., $\psi^* = 1 + \frac{\sqrt{m^2  - a^2}}{2 r}$, where $r$ is the
quasi-isotropic radius, or more complete expressions, as detailed in
the next section.  

Our black-hole binary initial data is
constructed using a superposition of metric and extrinsic curvature 
terms derived from the above expressions. To distinguish contributions
for the two black holes, we replace the $*$ super-/subscript above with a $+$
or $-$.

The trace-free part of the extrinsic curvature is split
into background terms $\tilde{M}_{i j}$ and a longitudinal
correction
term obtained from a vector $b_i$.
Here
\begin{equation}
  \tilde M_{ij} = \tilde A^{(+)}_{ij} + \tilde A^{(-)}_{ij},
\end{equation}
where $\tilde A^{(+)}_{ij}$ and $\tilde A^{(-)}_{ij}$ are the
trace-free part of the conformal
extrinsic curvature of a single boosted Kerr (Schwarzschild) black holes located at $\vec r =
\vec r_+$ and $\vec r = \vec r_-$.
Note that the trace-free part of the
single boosted black-hole extrinsic curvature
will have a small trace with respect to a metric constructed by 
superimposing  two different 
background metrics. We remove this extra trace term prior to
solving the initial data equations, i.e.,
$\tilde M_{ij} \to \tilde M_{ij} - \frac{1}{3} \tilde \gamma_{ij}
\tilde \gamma^{lm} \tilde
M_{lm}$ (where $\tilde \gamma_{ij}$ is the superimposed background
metric).
The complete trace-free part of the extrinsic curvature for the
superimposed spacetime is given by Eq.~(\ref{eq:general_Aij}).

In practice, we do not derive analytical expressions for $\gamma_{ij}$
and $K_{ij}$ for all possible boosts and spin orientations. Rather,
we evaluate the unboosted Kerr metric and its first and second
derivatives pointwise,  and then apply a boost and rotations to the metric 
(and the corresponding
transformation of the derivatives of the metric)  and then
algebraically solve for $\gamma_{ij}$ and $K_{ij}$ at that point.
 In the subsections
below, we provide explicit formulas for these quantities for the cases
of nonboosted Kerr and boosted Schwarzschild spacetimes.

\subsubsection{Conformal Kerr}\label{subsec:Conformally_Kerr}

In spherical quasi-isotropic coordinates, the Kerr conformal spatial line element is~\cite{Brandt:1996si,Dain:2000hk}
\begin{equation*}
  \ud \tilde{\ell}^2 = \tilde{\gamma}_{i j} \, \ud x^i \, \ud x^j = \ud r^2 + r^2 \, \ud \Omega^2 + a^2 h r^4 \sin^4(\theta) \, \ud \varphi^2 \; ,
\end{equation*}
where $m$ is the puncture mass, $a$ is the angular momentum per unit mass, $r$ is the quasi-isotropic radial coordinate, $\ud \Omega$ is the unit sphere line element, and~\cite{Hannam:2006zt}
\begin{align*}
  \bar{r} &= r + m + \frac{m^2 - a^2}{4 r} \; , \\
  \Sigma &= \bar{r}^2 + a^2 \cos^2(\theta) \; , \\
  \sigma &= \frac{2 m \bar{r}}{\Sigma} \; , \\
  h &= \frac{1 + \sigma}{\Sigma r^2} \; .
\end{align*}
The nonvanishing component of the shift vector is
\begin{equation*}
  \beta^{\varphi} = - \frac{2 a m r}{\left(r^2 + a^2\right)^2 - a^2 \left(r^2 - 2 m r + a^2\right) \sin^2(\theta)} \; .
\end{equation*}

The nonvanishing components of the conformal extrinsic curvature associated with this metric are given by~\cite{Brandt:1996si,Krivan:1998td,Dain:2000hk}
\begin{align*}
  \tilde{A}_{r \varphi} &= \frac{H_E \sin^2(\theta)}{r^2} \; , \\
  \tilde{A}_{\theta \varphi} &= \frac{H_F \sin(\theta)}{r} \; ,
\end{align*}
with the definitions
\begin{align*}
  e^{-2 q} &= \frac{\Sigma}{\bar{r}^2 + a^2 \left[1 + \sigma \sin^2(\theta)\right]} \; ,\\
  H_E &= e^{-q} \frac{a m}{\Sigma^2} \left [\left(\bar{r}^2 - a^2\right)\Sigma + 2 \bar{r}^2 \left(\bar{r}^2 + a^2\right) \right ] \; , \\
  H_F &= e^{-q} \frac{a^3 m \bar{r}}{2 r \Sigma^2}\left (m^2 - a^2 - 4 r^2\right ) \cos(\theta) \sin^2(\theta) \; .
\end{align*}
The Kerr metric in quasi-isotropic coordinates admits spatial hypersurfaces satisfying the maximal slicing condition $K = 0$~\cite{Dain:2000hk}.

The quasi-isotropic Kerr conformal factor is
\begin{equation}\label{eq:ch3_QI_psi}
  \psi_{\text{QI}} = \left (\frac{\Sigma}{r^2} \right )^{1/4} \; .
\end{equation}
Ultimately, only the asymptotic behavior is important, so sometimes just the lowest order terms of $\psi_{\text{QI}}$ are used~\cite{Hannam:2006zt}:
\begin{equation*}
  \psi_{\text{QI}} \approx 1 + \frac{\sqrt{m^2 - a^2}}{2 r} \; .
\end{equation*}

Near the puncture
\begin{equation*}
  \tilde{R} = - \frac{96 a^2}{(m^2 - a^2)^2} + \mathcal{O}(r^2)
\end{equation*}
and
\begin{equation*}
  \tilde{D}^2 \psi_{\text{QI}} = -\frac{6 a^2}{(m^2 - a^2)^{3/2} r} - \frac{12 m a^2}{(m^2 - a^2)^{5/2}} + \mathcal{O}(r) \; .
\end{equation*}
It follows that
\begin{equation}
  \label{eq:Kerr_cancellation}
  \tilde{D}^2 \psi_{\text{QI}} - \frac{1}{8} \psi_{\text{QI}} \tilde{R} = -\frac{288 m^2 a^2 r \sin^2(\theta)}{(m^2 - a^2)^{7/2}} + \mathcal{O}(r^2) \; .
\end{equation}

All fields are transformed to a Cartesian basis, with coordinates related by
\begin{align}
  \label{eq:sphere_to_cart}
  x &= r \sin(\theta) \cos(\varphi) \; , \nonumber \\
  y &= r \sin(\theta) \sin(\varphi) \; , \\
  z &= r \cos(\theta) \; . \nonumber
\end{align}
The conformal spatial metric takes the form
\begin{equation*}
  \tilde{\gamma}_{i j} = \delta_{i j} + a^2 h v_{i j} \; ,
\end{equation*}
where $\delta_{i j}$ is the Kronecker delta and
\begin{equation*}
  v_{i j} = \begin{pmatrix}
    y^2 & - x y & 0 \\
    -x y & x^2 & 0 \\
    0 & 0 & 0
  \end{pmatrix} \; .
\end{equation*}
The nonvanishing Cartesian components of the trace-free conformal extrinsic curvature tensor are
\begin{align*}
  \tilde{A}_{x x} &= - \tilde{A}_{y y} = - \frac{2 H_{1} \sin(\varphi) \cos(\varphi)}{r^3} \; , \\
  \tilde{A}_{x y} &= \frac{H_{1} \cos(2 \theta)}{r^3} \; , \\
  \tilde{A}_{x z} &= \frac{H_{2} \sin(\theta) \sin(\varphi)}{r^3} \; , \\
  \tilde{A}_{y z} &= - \frac{H_{2} \sin(\theta) \cos(\varphi)}{r^3} \; ,
\end{align*}
where
\begin{align*}
  H_{1} &= H_F \cos(\theta) + H_E \sin^2(\theta) \; , \\
  H_{2} &= H_F - H_E \cos(\theta) \; .
\end{align*}
At the puncture, these functions have the series expansion
\begin{align*}
  H_{1} &\sim 3 a m \sin^2(\theta) + \mathcal{O}(r^2) \; , \\
  H_{2} &\sim -3 a m \cos(\theta) + \mathcal{O}(r^2) \; .
\end{align*}
Thus, in Cartesian coordinates $\tilde{A}_{i j} \sim
\mathcal{O}\left(1/r^3\right)$ at the puncture. At this point, the
spin is parallel to the $z$-axis. In our approach, we specify the
direction and magnitude of the angular momentum $\vec S$ of each black hole.
For a given black hole, this means that we choose the Kerr parameter
$a$ such that $a = |\vec S|/m$, and rotate all fields so
that the spin points along $\vec S$. 

\subsubsection{Conformal Lorentz Boosted Schwarzschild}\label{subsec:Conformal_Boosted_Schwarzschild}

To describe a nonspinning BH with and arbitrary linear momentum $P^i$, we begin with the Schwarzschild line element in isotropic Cartesian coordinates $(t_0, x_0, y_0, z_0)$:
\begin{equation*}
  \ud s^2 = - \alpha_0^2 \, \ud t_0^2 + \psi_0^4 \left ( \ud x_0^2 + \ud y_0^2 + \ud z_0^2 \right ) \; ,
\end{equation*}
where
\begin{equation*}
  \alpha_0 = \frac{1 - \frac{m}{2 r_0}}{1 + \frac{m}{2 r_0}}
\end{equation*}
is the lapse,
\begin{equation*}
  \psi_0 = 1 + \frac{m}{2 r_0}
\end{equation*}
is the puncture conformal factor, and $r_0 = \sqrt{x_0^2 + y_0^2 + z_0^2}$.  Next, we perform a Lorentz transformation in the $y_0$-direction, with the associated change of variables
\begin{align*}
  t_0 &= \gamma (t - v y) \; , \\
  x_0 &= x \; , \\
  y_0 &= \gamma (y - v t) \; , \\
  z_0 &= z \; ,
\end{align*}
where $(t, x, y, z)$ are the coordinates of the boosted reference frame, $v$ is the magnitude of the local velocity vector
\begin{equation*}
  v^i = \frac{P^i}{\sqrt{m^2 + P^j P_j}} \; ,
\end{equation*}
and $\gamma = (1 - v^2)^{-1/2}$.  Afterward, all of the fields are rotated such that they are oriented in the desired direction, momentum aligned with $P^i$.

From the boosted spacetime metric, we extract the lapse function, shift vector, and spatial metric.  The only nonvanishing component of the shift is
\begin{equation*}
  \beta^y = - \frac{m v (m^2 + 6 m r + 16 r^2)(m^3 + 6 m^2 r + 8 m r^2 + 16 r^3)}{B^2} \; ,
\end{equation*}
with
\begin{equation*}
  B = \sqrt{(m + 2 r)^6 - 16 (m - 2 r)^2 r^4 v^2} \; .
\end{equation*}
On the $t_0 = 0$ hypersurface, $r_0 \to r = \sqrt{x^2 + y^2 \gamma^2 + z^2}$ and the conformal factor is
\begin{equation*}
  \psi_{\text{B}} = 1 + \frac{m}{2 r}
\end{equation*}
The conformal spatial line element on $\Sigma_{0}$ is
\begin{equation*}
  \ud \tilde{\ell}^2 = \ud x^2 + \gamma^2 \left [1 - \frac{16 (m - 2 r)^2 r^4 v^2}{(m + 2 r)^6} \right ] \, \ud y^2 + \ud z^2 \; .
\end{equation*}

Near to the puncture
\begin{equation}
  \label{eq:Schw_R}
  \tilde{R} = \frac{32 v^2 \left [7 + \cos(2 \theta) + 2 \sin^2(\theta) \cos(2 \varphi) \right ] r^2}{m^4} + \mathcal{O}(r^3)
\end{equation}
and
\begin{equation}
  \label{eq:Schw_lappsi}
  \tilde{D}^2 \psi_{\text{B}} = \frac{8 v^2 \left [\cos^2(\theta) + \sin^2(\theta) \cos^2(\varphi) \right ] r}{m^3} + \mathcal{O}(r^2) \; ,
\end{equation}
with coordinates $\theta$ and $\varphi$ defined by~\eqref{eq:sphere_to_cart}.

The evolution equation for the spatial metric gives us an expression for the extrinsic curvature
\begin{equation*}
  K_{i j} = \frac{1}{2 \alpha} \left (D_{i} \beta_{j} + D_{j} \beta_{i} - \partial_t \gamma_{i j} \right ) \; .
\end{equation*}
The mean curvature is
\begin{equation*}
  K = \frac{32 \gamma m v \left [(m + 2 r)^7 - 32 (m - 2 r)^2 (m - r) r^4 v^2 \right ] r^2 y}{(m + 2 r)^3 B^3} \; .
\end{equation*}
The nonvanishing components of the trace-free, conformal extrinsic curvature tensor are
\begin{align*}
  \tilde{A}_{x x} &= \tilde{A}_{z z} = \frac{\gamma m v (m - 4 r)(m + 2 r)^3 B C y}{3 D r^4} \; , \\
  \tilde{A}_{x y} &= - \frac{\gamma m v (m - 4 r)(m + 2 r)^3 x}{2 B r^4} \; , \\
  \tilde{A}_{y y} &= - \frac{2 \gamma^3 m v (m - 4 r) C y}{3 (m + 2 r)^3 B r^4} \; , \\
  \tilde{A}_{y z} &= - \frac{\gamma m v (m - 4 r)(m + 2 r)^3 z}{2 B r^4} \; ,
\end{align*}
with
\begin{align*}
  C = {} & (m + 2 r)^6 - 8 (m - 2 r)^2 r^4 v^2 \; , \\
  D = {} & (m + 2 r)^{12} - 32 (m - 2 r)^2 r^4 (m + 2 r)^6 v^2 \\
  & + 256 (m - 2 r)^4 r^8 v^4 \; .
\end{align*}
We see in Cartesian coordinates that $\tilde{A}_{i j} \sim \mathcal{O}\left(1/r^2\right)$ and $K \sim \mathcal{O}(r^3)$ at the puncture.

\subsubsection{Kerr with arbitrary spin orientations and boosts}
\label{subsubsec:general_kerr}

In this section we describe how we construct the background metric
($\tilde \gamma_{ij}$) and extrinsic curvature ($K$ and $\tilde
M_{ij}$) for binaries with generic spin orientations and arbitrary
momenta. Due to the complexity of the expressions, we do not calculate
the extrinsic curvature in closed form. Rather, we start by
calculating the Kerr 4-metric in various gauges and then perform
rotations and boosts. Specifically, we start with the Kerr metric in
quasi-isotropic coordinates. In the text below we will refer to
quasi-isotropic coordinates as QI coordinates. QI coordinates have the
unfortunate property that the horizon coordinate size goes to zero as
the spin becomes maximal. We ameliorate this problem by introducing a
radial {\it fisheye}~\cite{Baker:2001sf} transformation that increases
the horizon's radius. A fisheye transformation has the general form
\begin{equation}
  R = r f(r),
\end{equation}
where $R$ is the new radial coordinate, $r$ is the original
radial coordinate, and if $f(r)$ can be expanded as an even power
series in $r$, then the transformation is guaranteed to be $C^\infty$
(this is sufficient, but not necessary). These types of coordinates
have a long history of use for implementing fixed-mesh-refinement
within unigrid codes (see, e.g., \cite{Baker:2001sf,
Zlochower:2005bj, Campanelli:2005dd, Zilhao:2013dta}).
When studying evolutions of single black holes,
Liu, Etienne, and Shapiro ~\cite{Liu:2009al} found a
fisheye-like coordinate system, which we will refer to as LES coordinates
here, that has a finite
horizon coordinate radius for all values of the spin (the metric is,
however, singular for maximal spin).
The LES radius can be obtained from the implicit relationship
\begin{equation}
  R_{\rm LES} \left(1+\frac{r_+}{4 R_{\rm LES}}\right)^2 = r_{\rm QI}
  \left(1+\frac{m+a}{2 r_{\rm QI}}\right)\left(1+\frac{m-a}{2 r_{\rm
  QI}}\right),
\end{equation}
where $R_{\rm LES}$ is the LES radius, $r_{\rm QI}$ is the
quasi-isotropic radius, and $r_+ = m+\sqrt{m^2-a^2}$.

In addition to the
quasi-isotropic and LES coordinates, we also consider another
radial transformation of the quasi-isotropic radius that allows us to
fine-tune the horizon radius. These coordinates, which we will refer
to as FE (for fisheye) coordinates are obtained from
$R_{\rm FE} = r_{\rm QI} [1-A \exp(-r_{\rm QI}^2/s^2)]$, 
where $A$ and $s$ are parameters. These coordinates have the property
that at large $r_{\rm QI}$, $R_{\rm FE} \approx r_{\rm QI}$, and
at small $r_{\rm QI}$, $dR_{\rm FE} = (1-A) dr_{\rm QI}$.
Finally, for brevity, we will refer to the standard
quasi-isotropic coordinates as QI coordinates here.

Our procedure is as follows. We start with the Kerr metric in QI,
LES, or FE coordinates and then transform to Cartesian coordinates
defined by $x= r \sin \theta \cos \phi$, etc., where ($r$, $\theta$,
$\phi$) are the QI, LES, or FE coordinates. If the desired spin
direction is not aligned with the $z$, we perform a rotation about the
center of the BH to align the spin with the desired direction. We then
perform a boost on the resulting metric. In addition to the metric, we
also calculate its first and second derivatives by applying the above
rotation and boost to the known derivatives of the Kerr metric in the
QI, LES, or FE coordinates.

Given the 4-metric and its derivatives in the desired rotated and
boosted coordinates at a given point,
we calculate the 3-metric and extrinsic curvature
via Eq.~(\ref{eq:Kdef}).
There is a complication here because in all cases, the
lapse goes to zero on the horizon. This leads to severe round-off
issues in double precision calculation. To ameliorate this, we
calculate the metric and its derivatives to high precision 
using the MPFR
C++\footnote{http://www.holoborodko.com/pavel/mpfr/} wrapper for the
MPFR\footnote{http://www.mpfr.org/} high-precision library. Due to the expense
of using these libraries, we only use them in a small volume around
the horizon where high precision is needed.

Next we extract a conformal factor $\psi^*$ so that 
$\tilde \gamma_{ij}^*$ has unit determinant. We then calculate the
first and second derivatives of $\psi^*$ and  the metric $\tilde
\gamma_{ij}^*$, as well as the first derivative of the trace-free part
of the conformal extrinsic curvature $\tilde A_{ij}^*$ and the trace
$K$.

We then perform a series of optional modifications to the metric.
First, we can attenuate the metric to a flat metric far from the BH.
For this, we use a simple attenuation function
\begin{eqnarray}
  \tilde \gamma_{ij} - \delta_{ij} \to F(r)  (\tilde \gamma_{ij} -
  \delta_{ij}),\\
  \psi^* -1 \to F(r)  (\psi^* - 1),\\
  K^* \to F(r) K^*,\\
  F(r) = e^{-(r/s)^4},
\end{eqnarray}
where $r$ is the coordinate distance to the BH and the parameter $s$
is of order $20$---100 (note that we do not attenuate $\tilde A_{ij}$).
When applying this attenuation, we ensure that all derivatives
are modified such that they are consistent with the attenuated
functions.

This attenuation ensures that the metric is flat at infinity. If it is
not applied, the coordinate components of the resulting metric
will have an angular dependent monopole term,
i.e., the falloff of the diagonal components 
will be of the form $1+f(\theta,\phi)/r + \cdots$, where $f(\theta,
\phi)$ is a nontrivial function rather than
$M_{\rm ADM}/2$. As we would expect that the leading-order effects of
an extended body on distant geodesics to be spherically symmetric,
this choice of initial data leads to a coordinate system that obscures
this symmetry.
More importantly, the angular dependence of the
background metric at infinity will induce logarithmic terms in the
correction functions, which will reduce the order of convergence of
the solver.

One drawback of this construction is that by avoiding the derivation
of explicit analytical forms for the metric and extrinsic curvature, a
process for removing the singularities in a manner consistent with the
constraint equations is not apparent. In this paper, we address this
issue by modifying the elliptical constraint equations inside the
puncture, as described in the next section.

\subsection{Punctures}
\label{subsec:punctures}
We use an extended version \textsc{TwoPunctures}~\cite{Ansorg:2004ds} thorn to generate 
puncture initial data for black-hole-binary simulations.

If the constraint equations with a zero right-hand side
[Eqs.~(\ref{eq:Ham})
and (\ref{eq:mom})] are solved everywhere, then the resulting spacetime would
have zero ADM mass, linear, and angular momentum. In the puncture
approach, this is circumvented by finding singular solutions to the
constraint equations where the source terms are $\delta$ functions
centered on the two black holes.

In the puncture approach to the CTT
formalism~\cite{York99, Cook:2000vr, Pfeiffer:2002iy,
AlcubierreBook2008}, we decompose the conformal factor into singular
parts plus a finite correction, $u$,
\begin{equation}\label{eq:psi}
  \psi = \psi_{(+)} + \psi_{(-)} - 1 + u \; ,
\end{equation}
where $\psi_{(\pm)}$ represent the conformal factors
associated with the individual, isolated black holes located at 
positions labeled as $(+)$ and $(-)$,
with the spatial metric tensors $\tilde{\gamma}_{i j}^{(\pm)}$.

Similarly, we decompose $\tilde A_{ij}$ into a sum of singular terms (here
denoted by $\tilde M_{ij}$) and a nonsingular correction $(\tilde{\mathbb{L}} b)_{i j}$.

In addition to the required singular longitudinal component of $\tilde
M_{ij}$, there are nonsingular longitudinal components as well. These
extra nonsingular components are removed by the inclusion of
$(\tilde{\mathbb{L}} b)_{i j}$ (see, e.g., \cite{Pfeiffer:2002xz,
AlcubierreBook2008}).

In order to deal with the puncture singularities, 
we introduce modifications (in the form of attenuation functions) to both the  background
metric and the mean curvature, as well as modifications to
the singular source terms
inside the horizons themselves.
The first type of modification  is consistent with the
Einstein constraint equations everywhere and
has the form,
  \begin{align}
    \label{eq:metric_superposition}
    \tilde{\gamma}_{i j} &= \delta_{i j} + f_{(+)} \left(\tilde{\gamma}_{i j}^{(+)} - \delta_{i j}\right) + f_{(-)} \left(\tilde{\gamma}_{i j}^{(-)} - \delta_{i j}\right) \; , \\
    \label{eq:A_superposition}
    \tilde{M}_{i j} &= \tilde{A}_{i j}^{(+)} + \tilde{A}_{i j}^{(-)} \; , \\
    \label{eq:K_superposition}
    K &= f_{(+)}  K_{(+)} + f_{(-)} K_{(-)} \; , \\
    \label{eq:psi_superposition}
    \Psi &= \psi_{(+)} + \psi_{(-)} - 1 \;,
  \end{align}
where
\begin{equation*}
  f_{(\pm)} = 1 - e^{-(r_{(\mp)}/\omega_{(\pm)})^p} \; ,
\end{equation*}
and $r_{(\pm)}$ is the coordinate distance from 
a field point to the location of puncture $(\pm)$. 
The parameters $\omega_{(\pm)}$ control the 
steepness of the attenuation and $p$  controls how many
derivatives of the attenuated function are zero at the origin.
We explain how this can  be effective at removing certain
singularities, and
its limitations, below.

This type of attenuation, by itself is sufficient
for superimposed boosted Schwarzschild BHs or superimposed unboosted
Kerr BHs. However, for the more general case we found that modifying
the equations themselves inside the BHs was needed for satisfactory
convergence of the constraints outside the horizons.
This modification, which also comes in the form of a smooth
attenuation function,
has the effect of introducing
constraint violations inside the horizons. The constraint violations
induced by these modifications are not necessarily small, nor does the
resulting {\it fictitious matter} have to obey any of the standard
energy conditions.
Fortunately,
as shown in~\cite{Brown:2007pg, Brown:2008sb}, constraint violating
modes for standard BSSN and Z4 systems are causal, i.e., they
must stay inside the horizons. Hence the resulting numerical
spacetime will be a valid solution of the vacuum Einstein equations outside
the horizons.

The modified Hamiltonian and momentum constraint equations for
the correction functions $u$ and $b^i$ are 
\begin{subequations}
  \label{eq:constraints}
  \begin{align}
    \tilde{D}^2 u - g \frac{\psi \tilde{R}}{8} - g \frac{\psi^{5} K^2}{12}
    + g \frac{\tilde{A}_{i j} \tilde{A}^{i j}}{8 \psi^{7}} + g
    \tilde{D}^2\left(\psi_{(+)} + \psi_{(-)}\right) &= 0 \; ,
    \label{eq:hamiltonian} \\
    \tilde{\Delta}_{\mathbb{L}} b^i + g \tilde{D}_{j} \tilde{M}^{i
    j} - g \frac{2}{3} \psi^{6} \tilde{\gamma}^{i j} \tilde{D}_{j}
    K &= 0 \; ,\label{eq:momentum}
  \end{align}
\end{subequations}
where $\tilde{\Delta}_{\mathbb{L}} b^i \equiv \tilde{D}_{j}
(\tilde{\mathbb{L}} b)^{i j}$ is the vector Laplacian and $\tilde{R}$
is the scalar curvature associated with $\tilde{\gamma}_{i j}$, and
where the attenuation function $g$ takes the form
\begin{align*}
g &= g_{+}\times g_{-} \; ,\\
  g_{\pm} &= 
          \begin{cases} 
     1 & \mbox{if } r_\pm > r_{\rm max} \\
     0 & \mbox{if } r_\pm < r_{\rm min} \\
            {\cal G}(r_{\pm}) & \mbox{otherwise},
  \end{cases} \; ,\\
  {\cal G}(r_\pm) &= \frac{1}{2}\left[1+ \tanh\left(\tan\left[ \frac{\pi}{2}
  \left(-1 + 2 \frac{r_{\pm}-r_{\rm min}}{r_{\rm max} -
  r_{\rm min}}\right)\right]\right)\right],
\end{align*}
and the parameters $r_{\rm min} < r_{\rm max}$ are chosen
to be within the horizon. In addition, we can optionally attenuate the
background metric itself when calculating the
$\tilde D^2 u$ and $\tilde{\Delta}_{\mathbb{L}} b^i$. To do this
we take
\begin{eqnarray}
  \tilde \gamma_{ij}  \to \delta_{ij} + g(r) (\tilde \gamma_{ij} -
  \delta_{ij}),\\
  \tilde \Gamma^{k}_{\,ij} \to g(r) \tilde \Gamma^{k}_{\,ij}.
\end{eqnarray}
Note that the modified $\tilde \Gamma^{k}_{\,ij}$ is not consistent
with the modified $\tilde \gamma_{ij}$. There is no advantage to
making them consistent because the constraints will be violated in the
attenuation zone regardless. By modifying the metric in this way, we
can ensure that the elliptical system has exactly the form of the flat
space Poisson system in the vicinity of the punctures.

To understand the limitations of using the $f$ attenuation alone,
consider Eqs.~(\ref{eq:hamiltonian}) and (\ref{eq:momentum}) at the location of puncture
$+$ assuming the background contributions of puncture $-$ are
attenuated to zero in the vicinity of puncture $+$. Since the
background fields $\tilde \gamma_{ij} = \tilde \gamma_{ij}^+$, $\psi = \psi_+$, $K=K_+$,
$\tilde M_{ij} = \tilde M_{ij}^+$
obey the constraints exactly, these equations reduce to (note $\psi_-=0$, $M_{ij}^{-} = 0$)
\begin{eqnarray*}
    \tilde{D}^2 u - \frac{u \tilde{R}}{8} - \frac{\psi_{+}^{5}
    K^2}{12} \left(\left(1+ \frac{u}{\psi_+}\right)^5 -
    1\right)\nonumber && \\
  + \frac{\tilde{M}_{i j} \tilde{M}^{i j}}{8
    \psi_+^{7}}\left(\left(\frac{1}{1+u/\psi_+}\right)^7 - 1\right)
    \nonumber && \\
    + \frac{\tilde{A}_{i j} \tilde{A}^{i j} - \tilde{M}_{i j}
  \tilde{M}^{i j}}{8
    \psi_+^{7}}\left(\left(\frac{1}{1+u/\psi_+}\right)^7 \right)
    &=& 0 \; ,
    \\
    \tilde{\Delta}_{\mathbb{L}} b^i  - \frac{2}{3}
    \psi_+^{6}((1+u/\psi_+)^6 - 1) \tilde{\gamma}^{i j} \tilde{D}_{j}
    K &=& 0 \; .
\end{eqnarray*}
Since $\psi_+$ is singular at puncture $+$, these equations are
nonsingular on the puncture only if $\tilde R$ is finite and  $K$ goes to zero
sufficiently rapidly. The latter condition, in particular, cannot
always be guaranteed. In addition, the operators $\tilde D^2$ and
$\tilde{\Delta}_{\mathbb{L}}$  themselves need to be nonsingular on the puncture.
Hence the use of the $g$ attenuation which
explicitly modifies these terms to guarantee smooth behavior.

In addition, the form of the superimposed boosted Kerr background
itself induces logarithmic terms in the $1/r$ expansion
of $u$~\cite{Lovelace:2007zz}. While $u$ may still be formally finite,
these logarithmic terms reduce the order of convergence of spectral
expansions. This can be overcome by attenuating the metric at large
distances so that it is conformally flat. Note that modifying the
background spacetime far away only guarantees that the lowest order
(in $1/r$) singular terms are removed. Higher-order terms can still be
present, in general.

\section{Numerical Techniques}\label{sec:techniques}

\subsection{Initial data solver with Spectral Methods}\label{subsec:Spectral_Methods}

The standard version of the \textsc{TwoPunctures}
thorn~\cite{Ansorg:2004ds} generates conformally flat
($\tilde{\gamma}_{i j} = \delta_{i j}$) initial data via a spectral
expansion of the Hamiltonian constraint on a compactified collocation
point grid. Here we extend this to include both a nonflat background
metric and a vector Poisson equation for the $b^i$ in
Eq.~(\ref{eq:constraints}).

This is achieved by extending \textsc{TwoPunctures} to solve for $u$ and $b^i$ simultaneously at each collocation point.  The solver handles the nonlinearities in the constraint equations by using a linearized Newton-Raphson method. 

The solutions are required to obey the falloff conditions
\begin{equation*}
  \lim_{r \to \infty} u = 0 \quad \text{and} \quad \lim_{r \to \infty} b^i = 0 \; ,
\end{equation*}
which are the only physical boundary conditions for the problem.
These are enforced numerically by noting that the
\textsc{TwoPunctures} compactified coordinate $A$ obeys $A \to 1$ as
$r \to \infty$, thus allowing the asymptotic behavior to be factored
out (e.g.\ write $u = (A - 1) U$ and solve for the auxiliary function
$U$)~\cite{Ansorg:2004ds}. 

\begin{figure}
  \includegraphics[width=0.9\columnwidth]{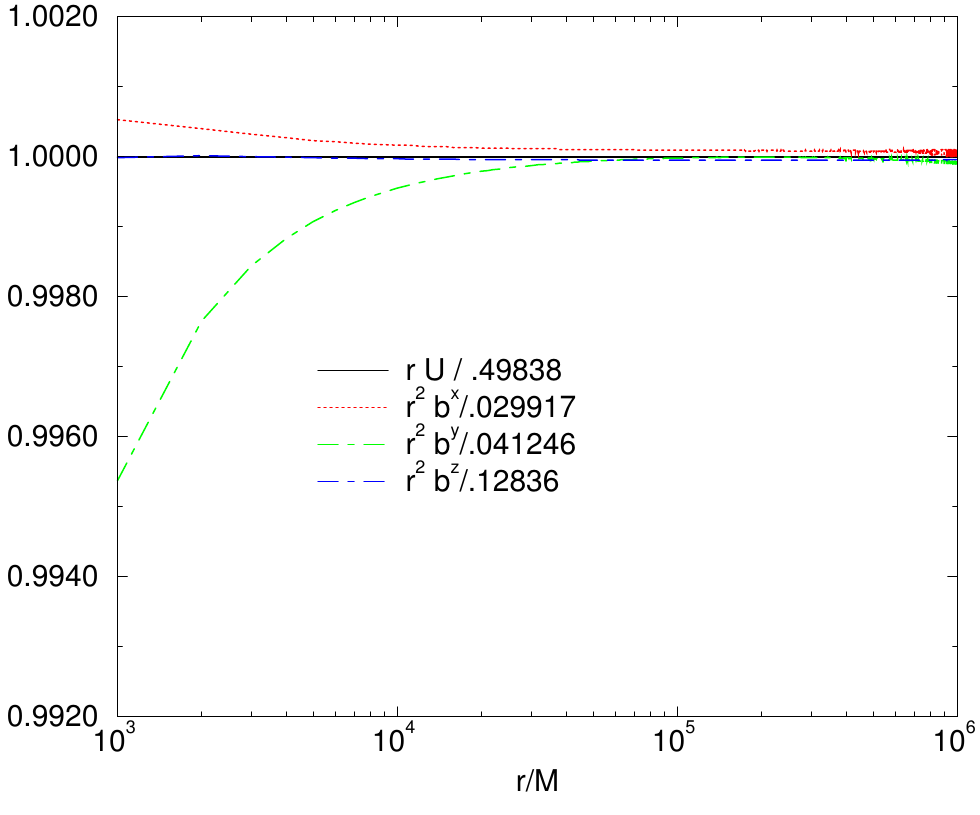}
  \caption{The falloff with radius of the correction fields $u$ and
    $b^i$ for a $\chi = 0.95$ binary along the line $x=y=z$.
    As can be seen, $u$ falls off as
    $1/r$ and all components of $b^i$ fall off as $1/r^2$. Note that
    the boundary conditions for $b^i$ only require that the components go to
  zero at infinity. The quadratic falloff is a consequence of the
initial data not having net linear momentum. In addition to
multiplying each function by the appropriate power of $r$, we rescale
the resulting curves so that all approach $\sim 1$ at large $r$.}\label{fig:falloff}
\end{figure}

 In our implementation, we enforce a
falloff of at least $1/r$ for all fields by requiring that they be
zero at infinity. However, we find that the $b^i$
fields fall off faster, as expected given the form of the
equations, as shown in Fig.~\ref{fig:falloff}.

As mentioned above, there are several  sources of singular behaviors.
Briefly, the differential equations
themselves can become singular at the punctures, and the behavior of
the equations at infinity can induce logarithmic terms in the
solution. 
Regardless, finite-precision effects makes evaluating the Laplacian
operator and source terms inaccurate close to the puncture. We
overcome both  these issues in the general case  by enforcing that the background metric is
identically flat (with $K=0$) at large $r$ and sufficiently close to
the puncture. Furthermore, close to the puncture, we smoothly
attenuate all source terms to zero. Thus in a finite ball surrounding
each puncture, the elliptic equations reduce to a source-free
Laplacian and a vector Laplacian, respectively. The end result is a well
behaved solution to the Einstein constraint system outside the two
horizons.

Note that solutions to Eqs.~(\ref{eq:hamiltonian}) and
(\ref{eq:momentum}) can contain
logarithmic terms of the form $r^{-k} \ln r$, where the integer $k$ is
determined by the falloff rate of the source terms in the two
equations~\cite{Grandclement:2000xv}. If such terms are present in the solution, the convergence
rate of the solver will be algebraic rather than exponential. In
practice, we find that the accuracy of the solver is limited by
small-scale features induced by the attenuation functions inside the
horizons.

In Eq.~(\ref{eq:constraints}), derivatives of the background fields
$\tilde \gamma_{ij}$, $\tilde M_{ij}$, and $\psi_{\pm}$ can be
calculated analytically or numerically.  We have implemented both
approaches. In practice, we use an eighth-order finite difference
operator with a step size of $\leq 10^{-4}$ (always smaller than
the finest grid spacing) for the QI boosted Kerr (fixed spin direction) and boosted
Schwarzschild data, and analytical expressions for the LES and FE
boosted Kerr data derivatives (these latter two are evaluated using the
high-precision libraries mentioned above). 
Note that the accuracy of these derivatives near
the punctures is irrelevant inside the attenuation region as these
terms get multiplied by zero.

\subsection{Evolution and gauges}\label{subsec:Evolution}

We use the extended \textsc{TwoPunctures} thorn to generate puncture initial data~\cite{Brandt97b} for the BHB simulations. These data are
characterized by mass parameters $m_{\text{p}}$ (which
are not the horizon masses), as well as the momentum and spin of
each BH, and their initial coordinate separation.  We evolve these BHB data sets using the {\sc
LazEv}~\cite{Zlochower:2005bj} implementation of the \MPA with the conformal
function $W=\exp(-2\phi)$ suggested by
Ref.\@~\cite{Marronetti:2007wz}.  For the runs presented here, we use
centered, eighth-order finite differencing in
space~\cite{Lousto:2007rj} and a fourth-order Runge-Kutta time
integrator. (Note that we do not upwind the advection terms.) Our code
uses the {\sc Cactus}/{\sc EinsteinToolkit}~\cite{cactus_web,
einsteintoolkit} infrastructure.  We use the {\sc
Carpet} mesh refinement driver to provide a
``moving boxes'' style of mesh refinement.

We locate the apparent horizons using the {\sc AHFinderDirect}
code~\cite{Thornburg2003:AH-finding} and measure the horizon spin
using the isolated horizon (IH) algorithm detailed
in~\cite{Dreyer02a}.

For the computation of the radiated angular momentum components, we
use formulas based on ``flux linkages''~\cite{Winicour_AMGR} and
explicitly written in terms of $\Psi_4$  in
\cite{Campanelli:1998jv, Lousto:2007mh}.

We obtain accurate, convergent waveforms and horizon parameters by
evolving this system in conjunction with a modified 1+log lapse and a
modified Gamma-driver shift condition~\cite{Alcubierre02a,
Campanelli:2005dd,vanMeter:2006vi}. The lapse and shift are evolved with
\begin{subequations}
\begin{align}
\label{eq:gauge}
(\partial_t - \beta^i \partial_i) \alpha &= - \alpha^2 f(\alpha) K \; , \\
 \partial_t \beta^a &= \frac{3}{4} \tilde{\Gamma}^a - \eta \beta^a \; .
\end{align}
\end{subequations}
In the original \MPA we used
$f(\alpha)=2/\alpha$ and an initial lapse $\alpha(t=0) =\psi_{\text{BL}}^{-2}$
\cite{Campanelli:2005dd}  
or $\alpha(t=0)=2/(1+\psi_{\text{BL}}^{4})$ 
\cite{Campanelli:2006uy}, where   $\psi_{\text{BL}}=1+m_{(+)}/(2r_{(+)})+m_{(-)}/(2r_{(-)})$.
In Sec.\@~\ref{sec:BoostedBHs} and Appendix~\ref{sec:gauge} 
we also use $\alpha(t=0)=1/(2\psi_{\text{BL}}-1)$, which seems 
better suited for the highly spinning or highly boosted BH evolutions.
There, we explore other gauge conditions for the lapse in the
form of $f(\alpha)=1/\alpha$ (gauge speed = 1) 
and $f(\alpha)=8/(3\alpha(3-\alpha))$ (shock avoiding) 
\cite{Alcubierre02b} which prove to be more convenient when
dealing with highly boosted moving punctures.

\section{Evolutions of the \hispid data}

In this section we describe results from evolutions of the new \hispid
data for black-hole binaries with spins, boosts, or both.
We start by examining the case of two superimposed Kerr black holes
initially at rest. We then examine superimposed boosted Schwarzschild
black holes in quasicircular binaries.
Next, we compare evolutions of spinning quasicircular binaries with
moderate spins using both \hispid and \by data. Finally, we compare
evolutions of a spinning, quasicircular black-hole binary with
specific spin $\chi=0.95$ to results found by the SXS~\cite{SXS:catalog,
Hemberger:2013hsa, Mroue:2013xna} Collaboration for
similar systems.

\subsection{Headon collisions of Spinning Black
Holes}\label{sec:SpinningBHs}

We begin our analysis with the case of two spinning black holes
initially at rest. This type of system was first analyzed
in~\cite{Hannam:2006zt} using a very similar construction (i.e.,
superimposed Kerr black holes in a puncture gauge).

\begin{figure}
  \includegraphics[angle=270,width=\columnwidth]{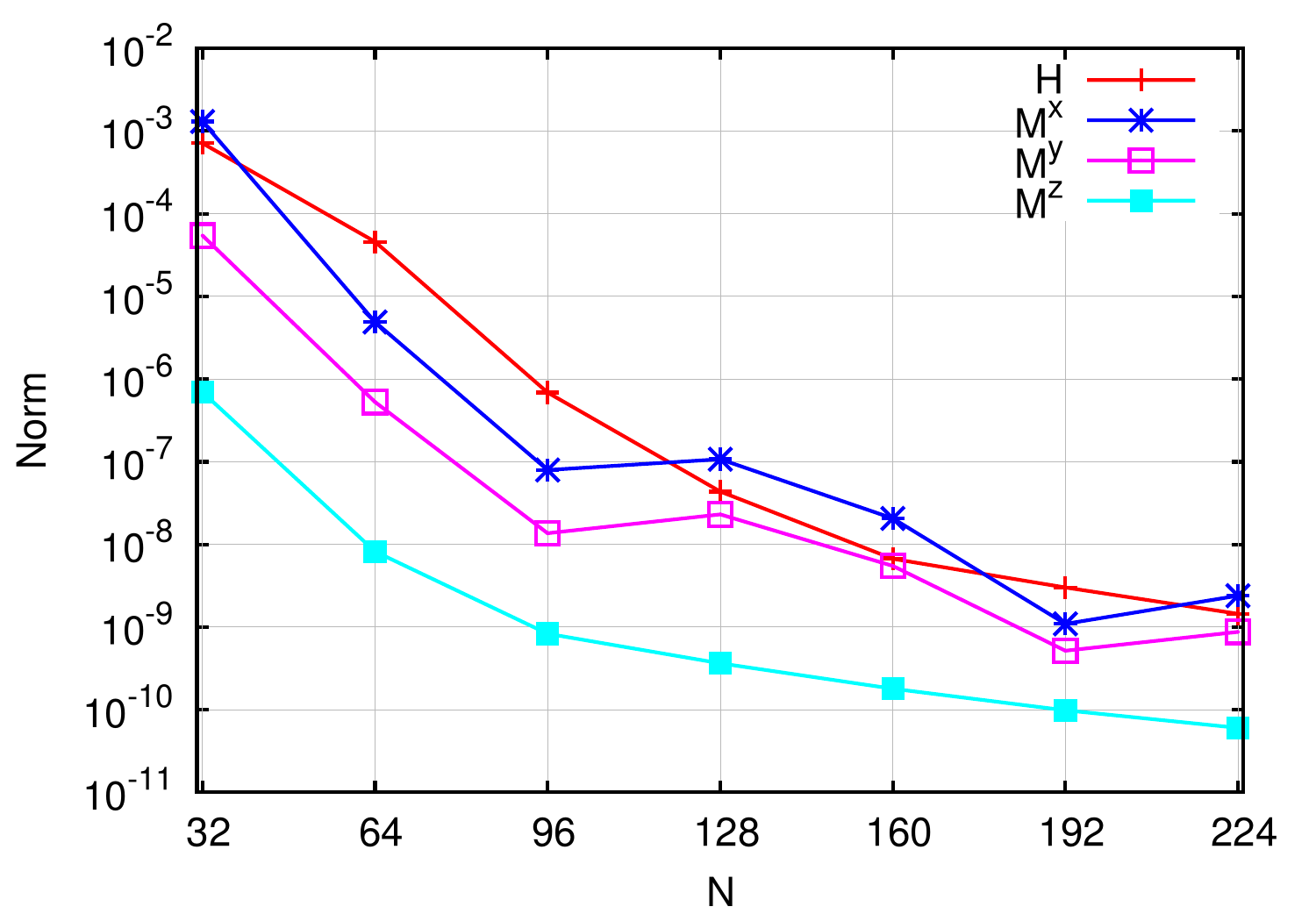}
  \includegraphics[angle=270,width=\columnwidth]{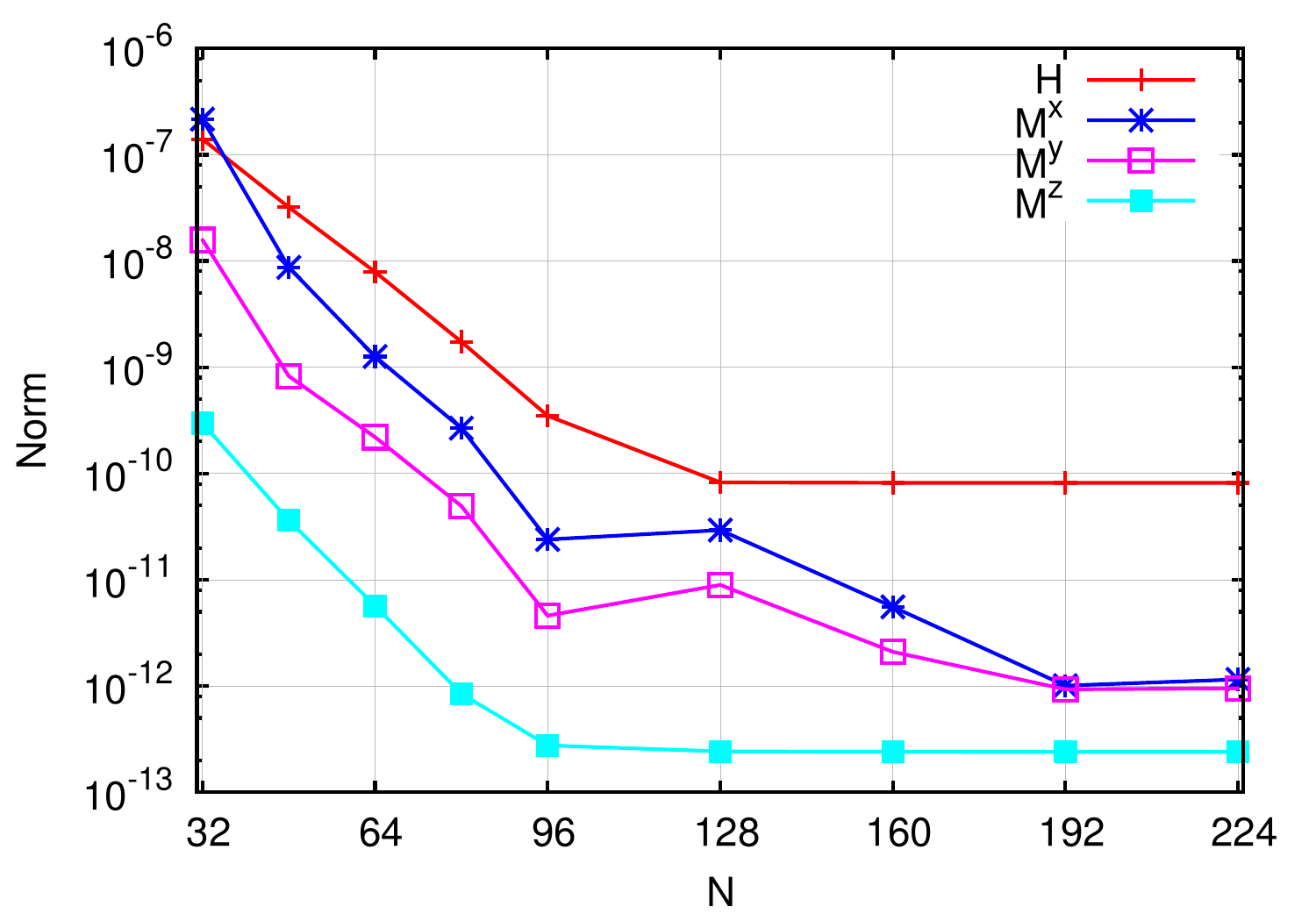}
  \caption{Convergence of the residuals of the Hamiltonian and momentum constraints 
versus the number of collocation points $N$ for 
BHBs with $\chi=0.99$ in the UU configuration with exponential attenuation parameters $\omega_{(\pm)}=1.0$ and $p=4$.
(Top panel) For a small grid along $x$-axis 
($4.25 \leq x \leq 4.75$, $-0.25 \leq y \leq 0.25$, and $-0.25 \leq z \leq 0.25$).
(Bottom panel) For the full numerical evolution grid.
}
  \label{fig:conv99}
\end{figure}

We begin by demonstrating the convergence of the initial data. As
shown in Fig.~\ref{fig:conv99}, even at very high spins (here
$\chi=0.99$) the constraints converge to round-off levels. In order to
reach acceptable levels of constraint satisfaction, we had to use
relatively large numbers of collocation points. Here we used up to
224 collocation points in the two directions orthogonal to the
symmetry axis. Because a simple $L^2$ norm of the constraints may {\it
hide} issues near the black holes, we construct $L^2$ norms in both a
small box near each black hole, and in the bulk of the simulation
domain (out to a distance of $30M$ from the origin). Close to the black holes, the constraints continue to
reduce with the number of collocation points, falling down to ${\cal
O}(10^{-9})$. In the bulk, the much smaller $L^2$ norms fall to
round-off levels ${\cal O}(10^{-10}) - {\cal O}(10^{-12})$ and then
remain constant.

\begin{figure}
  \includegraphics[angle=270,width=0.49\columnwidth]{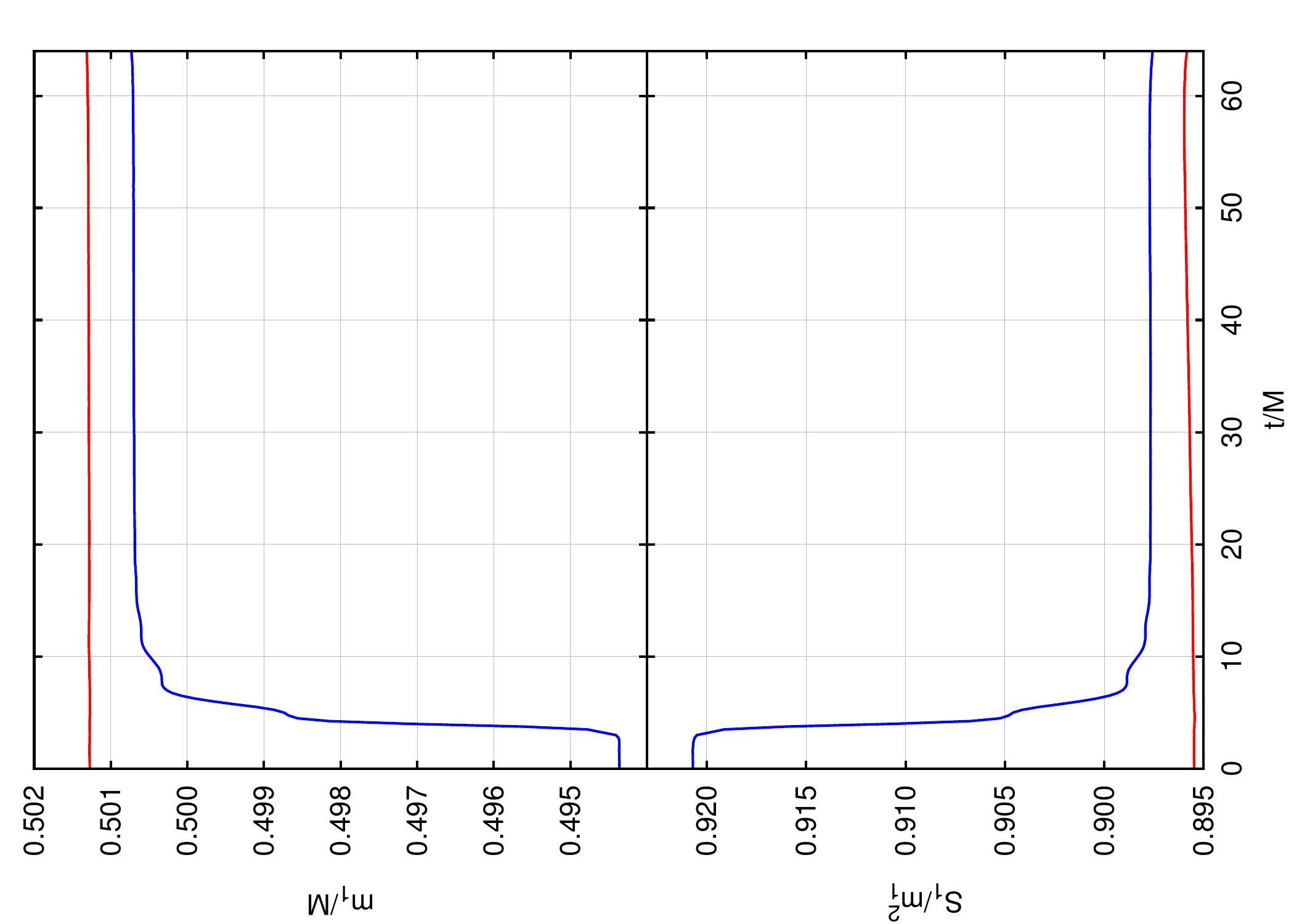}
  \includegraphics[angle=270,width=0.49\columnwidth]{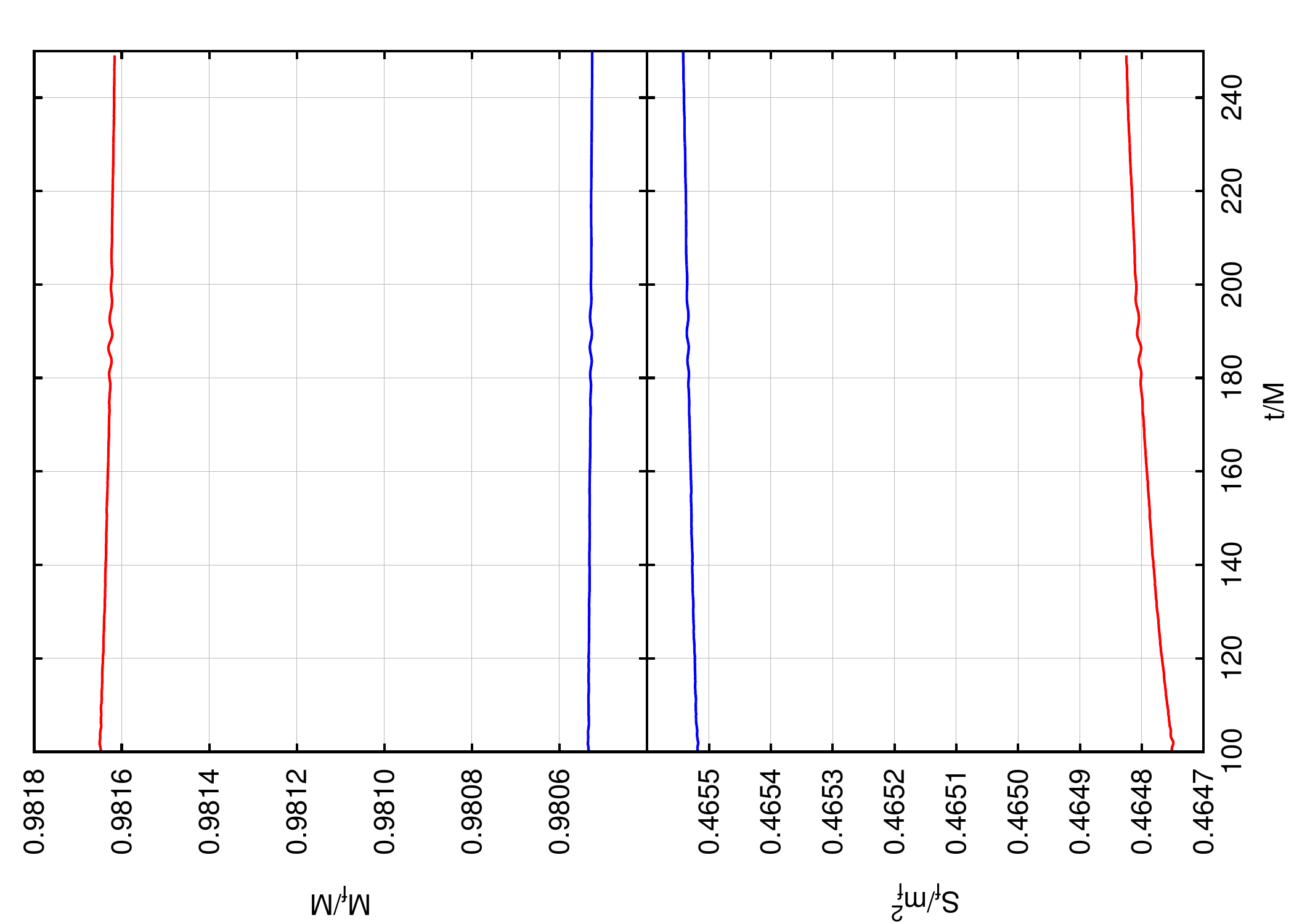}
  \caption{Masses of the individual BHs from the start of the
simulation until merger (top left). 
Spins of the individual BHs until merger (bottom left).
Masses of the remnant BH (top right).
Spin of the remnant BH (bottom right).
BY, blue; \hispid, red.}
  \label{fig:initfinal_by}
\end{figure}

As done in~\cite{Hannam:2006zt}, we compare the new data to Bowen-York
(conformally flat) initial data with the same spin parameters.
For given
binary separations and spin parameters, the horizon masses and spins
for the \hispid and BY data
are not identical, as shown in Fig.\@~\ref{fig:initfinal_by}, since
the initial radiation content and distortions are not the same.
However, they are close enough for comparisons of physical 
quantities such as the gravitational waveforms.

\begin{table}
\caption{Initial data parameters for the equal-mass, head-on 
configurations. The punctures are initially at rest and 
are located at $(\pm b,0,0)$ with spins 
$S$ aligned or antialigned with the $z$ direction, mass parameters
$m_{\text{p}}$, horizon (Christodoulou) masses $m_{1} = m_{2} = m_{\text{H}}$, total ADM mass
$M_{\rm ADM}$, and the dimensionless spins $a/m_{\text{H}} = S/m_{\text{H}}^2$, where $a_{1} = a_{2} = a$.
The Bowen-York configurations are denoted by BY, and the \hispid by HS.  Finally, UU or UD denote the direction of
the two spins, both  either aligned (UU) or antialigned (UD).
}\label{tab:ID}
\begin{ruledtabular}
\begin{tabular}{lcccccc}
Configuration  & $b/M$ & $m_{\text{p}}$ & $S/M^2$ & $a/m_{\text{H}}$ & $m_{\text{H}}$ & $M_{\rm ADM}/M$\\
\hline
BY90UU & 6 & 0.191475 & 0.225 & 0.8977& 0.500702& 0.982362\\
HS90UU & 6 & 0.5      & 0.225 & 0.8958& 0.501287& 0.982353\\
BY90UD & 6 & 0.191475 & 0.225 & 0.8977& 0.500702& 0.982396\\
HS90UD & 6 & 0.5      & 0.225 & 0.8955& 0.501208& 0.982388\\
\hline
HS99UU & 6 & 0.5      & 0.2475& 0.9896& 0.500162& 0.980124\\
HS99UD & 6 & 0.5      & 0.2475& 0.9887& 0.499981& 0.980163\\
\end{tabular}
\end{ruledtabular}
\end{table}

We study a few test cases of equal-mass BHB configurations starting from rest with spins aligned (UU) or counteraligned (UD) with each other, and perpendicular to the line joining the
BHs.
We evolve both BHBs with the \hispid data and the standard BY choice (for
spins within the BY limit). We also evolve BHBs with near-maximal
spin, $\chi=0.99$, a regime
unreachable for BY initial data.
Table \ref{tab:ID} gives the initial data parameters of these BHB configurations.

\begin{figure}
  \includegraphics[angle=270,width=0.9\columnwidth]{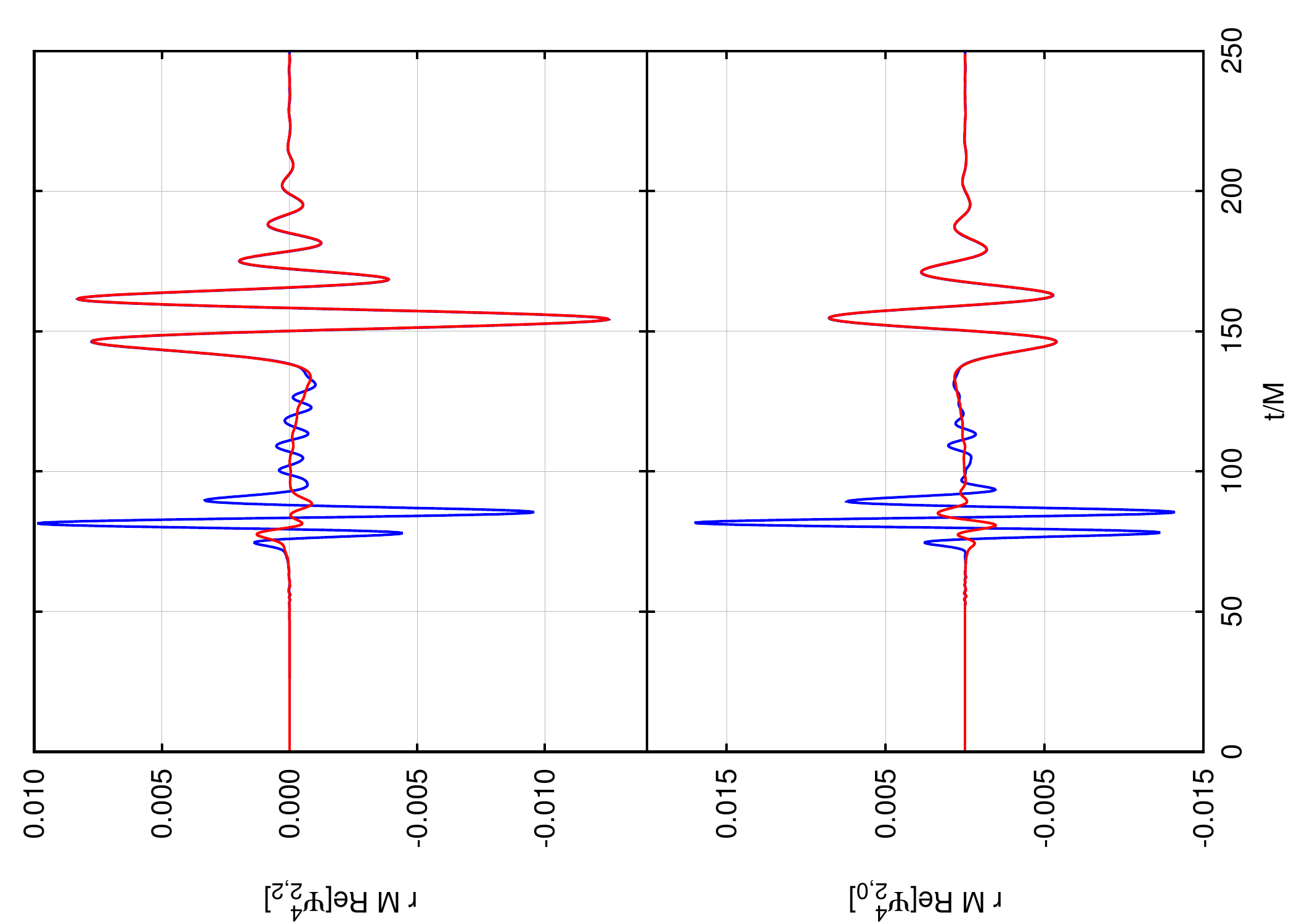}
  \caption{$(\ell=2,m=2)$ mode of $\Psi_4$ at $r=75M$ (top panel). 
(2,0) mode of $\Psi_4$ at $r=75M$ (bottom panel) for spinning binaries
with $\chi=0.9$. BY data, blue (showing a much larger initial burst);
\hispid, red.}
  \label{fig:y220_by}
\end{figure}

Figure\@~\ref{fig:y220_by} shows a comparison of waveforms 
$r\psi_4$ extracted at an observer location $r=75M$. We clearly
see that the initial radiation content (located around $t\sim80M$)
of the BY data for
equal-mass spinning BHBs with $\chi=0.9$ has an amplitude comparable
to that of the physical merger. On the other hand, the \hispid
initial data has greatly reduced initial radiation content (one order
of magnitude smaller).
Although not apparent in these plots, the much lower initial radiation
content is not only more physical, but also leads to more accurate
computations of waveforms. This initial pulse reflects from
the refinement boundaries (since they are not perfectly transmissive)
leading to high frequency errors and convergence issues when looking
at much finer details of the waveform phase 
\cite{Zlochower:2012fk,Etienne:2014tia}.

The evolution of BHs with $\chi=0.99$ requires high
resolution, particularly during the first $10M$ of evolution, but
otherwise proceeds with the standard \MP setup 
\cite{Campanelli:2005dd}.  A summary of the
properties of the final merger remnant BH are listed in
Table~\ref{tab:spinerad}.
Even though the initial data has no
initial orbital angular momentum, UU configurations radiate angular momentum due to mutual
frame dragging effects in the opposite direction, as observed in 
\cite{Campanelli:2006fg}.
The UD configurations, on the other hand, do not radiate angular
momentum, but do recoil.
Here we see that both the \by and \hispid data for black holes in a UD
configuration with spins $\chi = 0.9$ lead to recoils of
$35.95\pm0.05\ \KMS$. The much more extreme case of a UD configuration
with spins $\chi =0.99$ yields a recoil of $38.4\pm0.09\ \KMS$. 
The recoil
can be modeled as~\cite{Lousto:2012gt}
\begin{equation*}
V_{\text{recoil}}=\sum_{j=1,3,5...}k_j\Delta^j
\end{equation*}
where $\Delta=(\chi_{(+)}- \chi_{(-)})/2$
and
$k_j$ are fitting constants (this form applies only to equal-mass binaries with vanishing total spin). With only two data points,
one can only reliably fit the first constant. We find $k_1 =
39.62\pm0.36\ \KMS$. 

\begin{table}
\caption{The final mass, final remnant spin, and recoil velocity for each configuration. 
}\label{tab:spinerad}
\begin{ruledtabular}
\begin{tabular}{lllll}
Configuration & $M_{\text{rem}}/M$ & $\chi_{\mathrm{rem}}$ & $V$\\
\hline
BY90UU & 0.98053& 0.46554& 0\\
HS90UU & 0.98162& 0.46483& 0\\
BY90UD & 0.98073& 0      & 35.90\\
HS90UD & 0.98181& 0      & 36.01\\
\hline
HS99UU & 0.97971& 0.52501& 0\\
HS99UD & 0.97900& 0      & 38.40\\
\end{tabular}
\end{ruledtabular}
\end{table}

\begin{figure}
  \includegraphics[angle=270,width=\columnwidth]{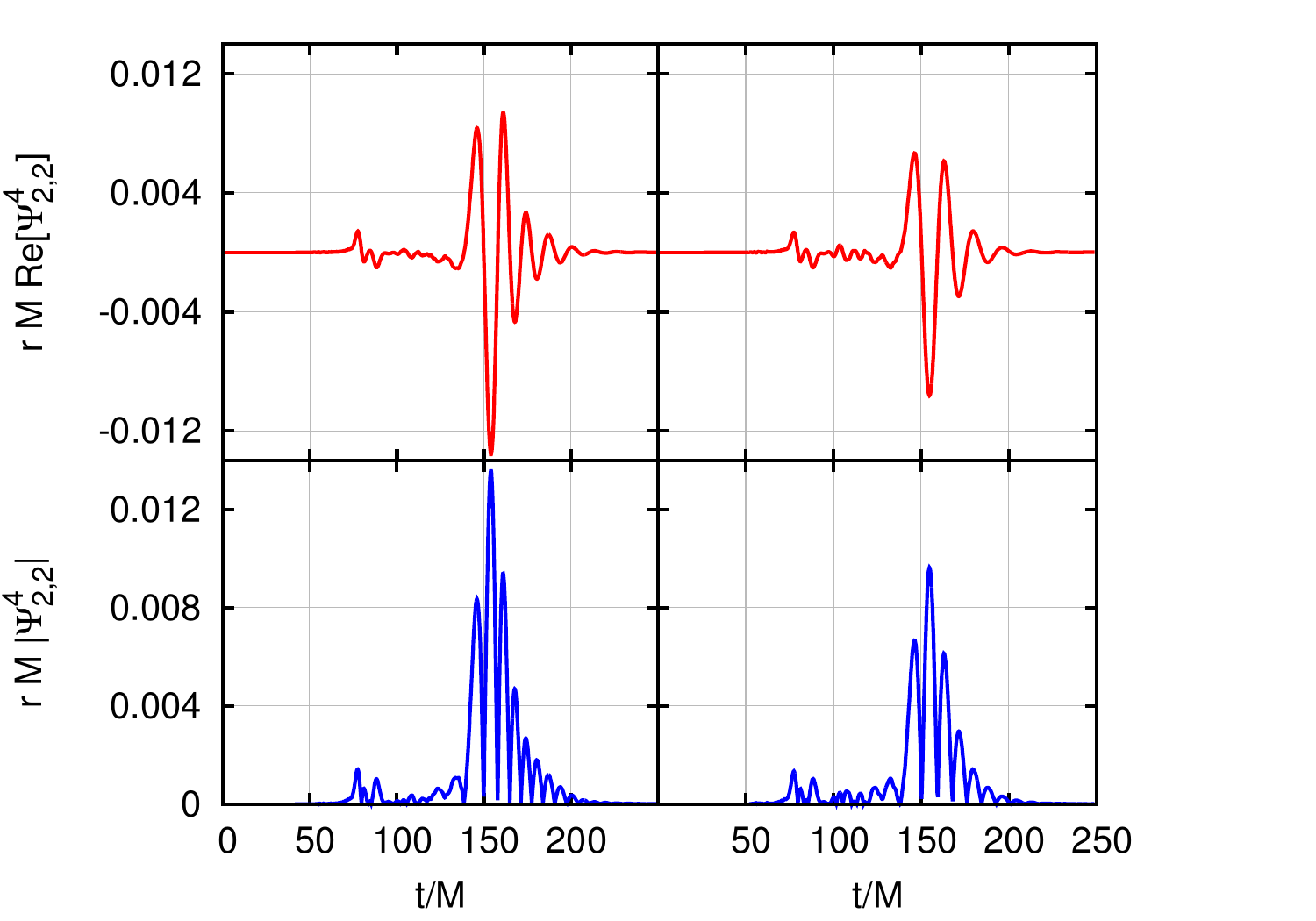}
  \caption{Waveforms and waveform magnitudes of the UU (left panels)
    and UD (right panels) configurations
with highly spinning BHs $\chi=0.99$. Note the small 
amplitude of the initial data radiation content at around $t=75M$, compared
to the merger signal after $t=130M$.}
  \label{fig:WM0.99}
\end{figure}

In Fig.\@~\ref{fig:WM0.99}, we  show the waveforms for the UU and UD cases for highly spinning BHs. The initial radiation 
content has a much smaller amplitude than the merger waveform---even
in the head-on case---significantly reducing contamination of
the physical signals by unresolved high-frequency reflections.

\subsection{Quasicircular Nonspinning Black-Hole Binaries}\label{sec:BoostedBHs}

One of the most astrophysically important applications of
numerical relativity is the evolution of BHBs in quasicircular orbits.
As such, it is critical that \hispid data are able to reproduce the
quasicircular binaries that can be generated by \by techniques. In
this section, we will concentrate on nonspinning binaries, while in
the next section, we will consider binaries with both moderate and
extreme spins.

\begin{figure}[!ht]
  \includegraphics[angle=270,width=\columnwidth]{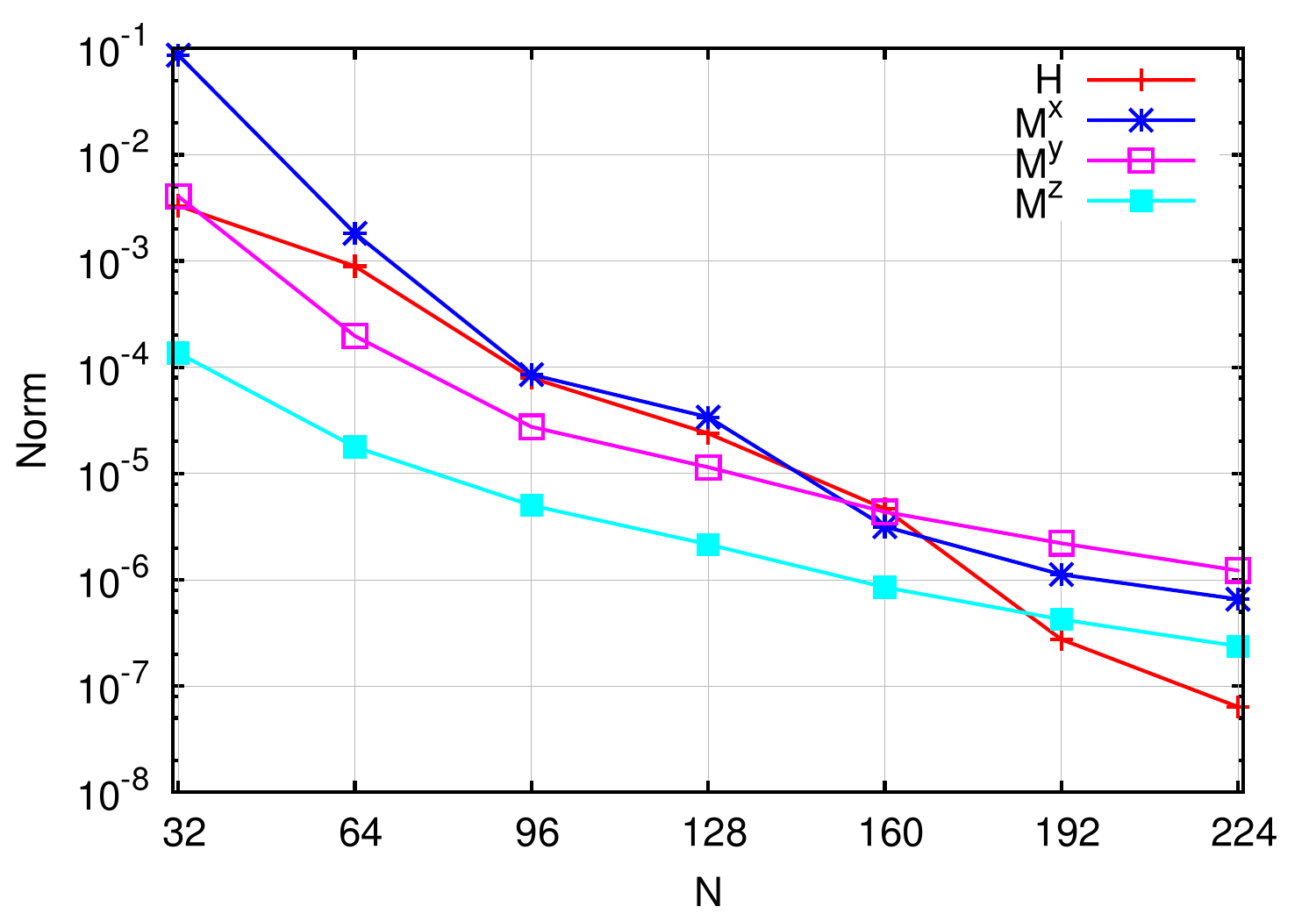}
  \includegraphics[angle=270,width=\columnwidth]{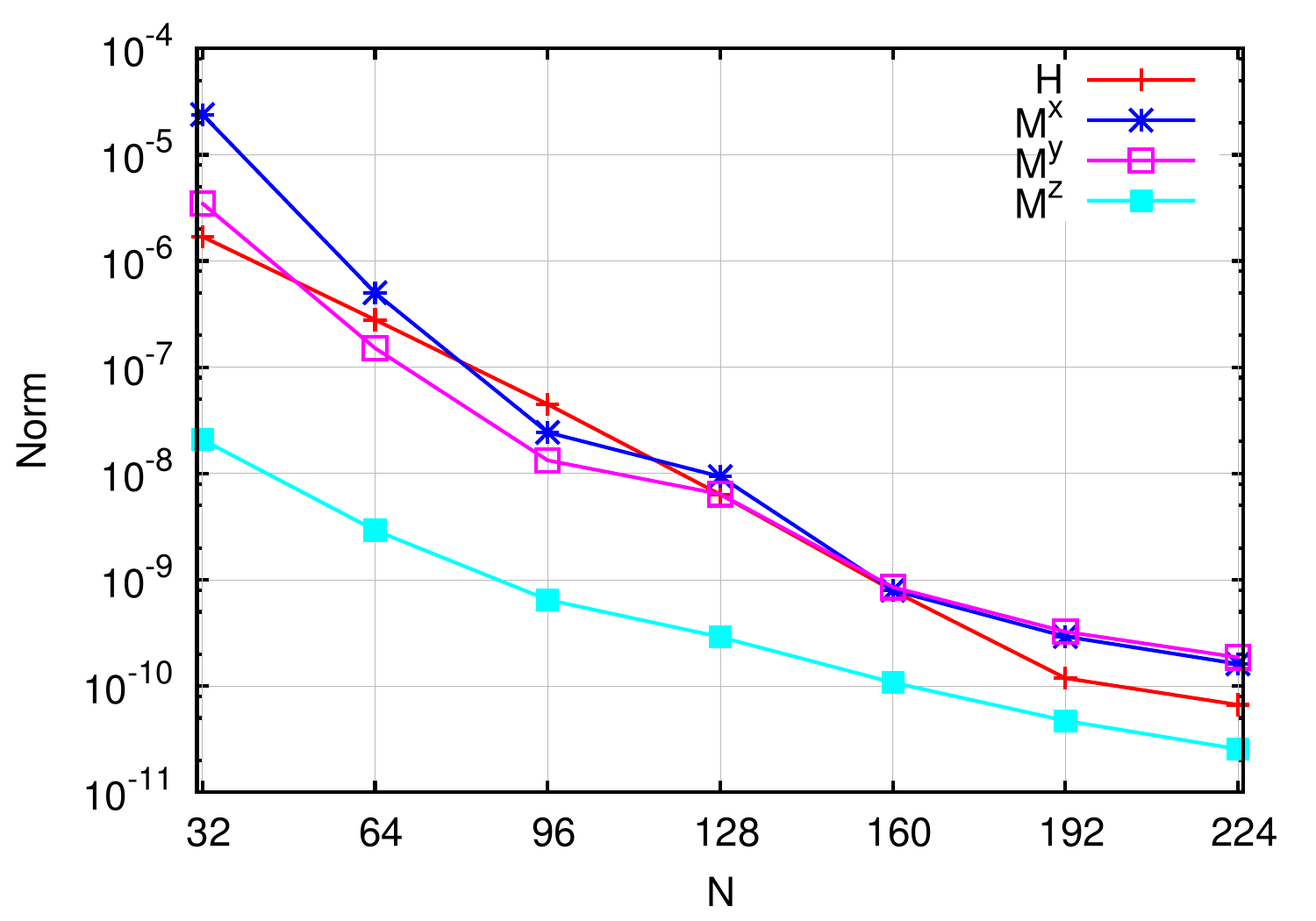}
  \caption{Convergence of the residuals of the Hamiltonian and momentum constraints 
versus the number of collocation points $N$ for Schwarzschild BHBs in a quasicircular orbit
with exponential attenuation parameters $\omega_{(\pm)}=1.0$ and $p=6$. 
Orbital parameters $P^y=0.0848M$, $d=12M$.  
(Top panel) for a  small grid along $x$-axis
($4.75 \leq x \leq 5.25$, $0.1 \leq y \leq 0.6$, and $0.5 \leq z \leq 1.0$).
(Bottom panel) for the  full numerical evolution grid.
}
  \label{fig:convQC}
\end{figure}

Figure\@~\ref{fig:convQC} shows the convergence
rate for nonspinning initial data solution with
the number of collocation points for a typical set of orbital
parameters. Hamiltonian and momentum constraint residuals reach levels
below ${\cal O} (10^{-6})$ near the horizon and ${\cal O}(10^{-10})$
in the bulk. Once again, these measures are for $L^2$ norms of the
constraints in a small volume just outside the horizon and in the bulk
of the simulation domain.

\begin{figure}[!ht]
  \includegraphics[angle=270,width=\columnwidth]{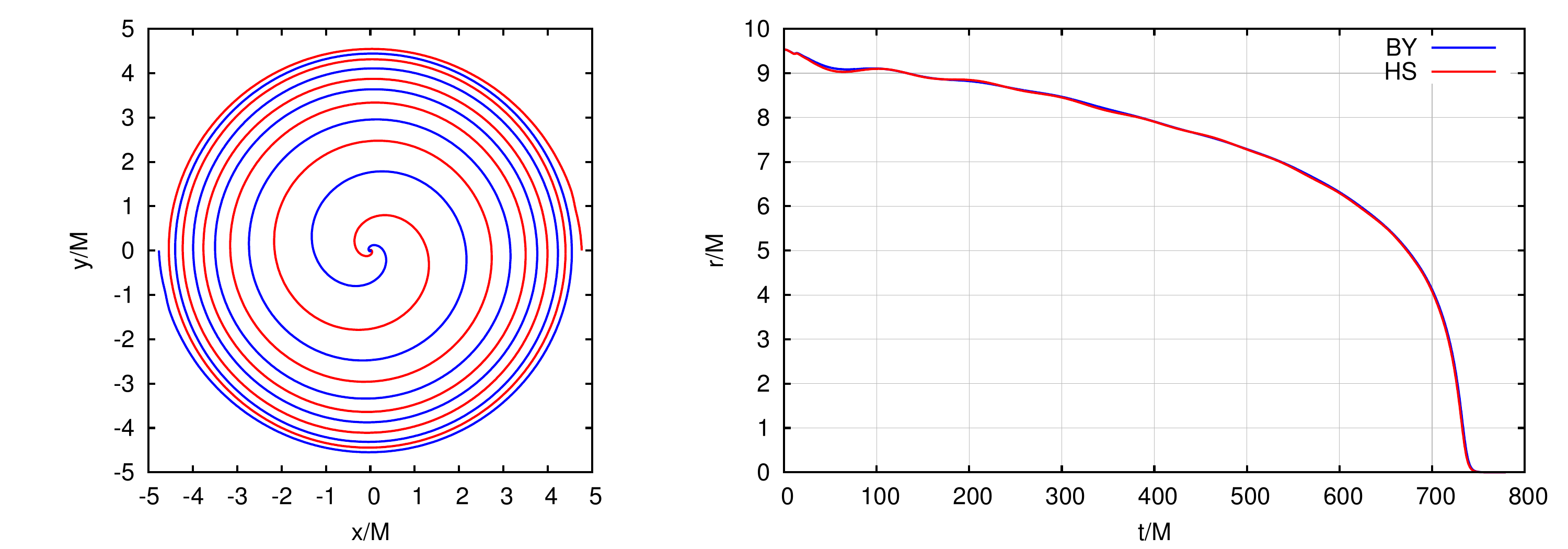}
  \caption{The orbital trajectories of the binary and a comparative
radial decay of the Lorentz boost and BY initial data. Note the near
perfect agreement of the \by and \hispid data evident in the right
panel. (Left panel) The trajectories of the two black
holes for \hispid data only.
}
  \label{fig:QCtracks}
\end{figure}

In order to evaluate the effectiveness  of the \hispid approach for
generating binary data,  we perform a numerical
evolution of a binary in the merger regime and compare our Lorentz
boost data with the traditional BY solution. We chose
initial parameters with low  eccentricity for each set of data,
as given by Table~\ref{tab:lbID}.
The BHs orbit nearly five times before merging 
(see Fig.\@~\ref{fig:QCtracks}), and at
$t\sim700M$, merge to a spinning remnant BH with
the properties given in Table~\ref{tab:final}.
Note the near perfect agreement of the orbital decay between \by and
\hispid data evident in the right panel of Fig.~\ref{fig:QCtracks}.

\begin{table}[!ht]
\caption{Initial data parameters for the equal-mass, boosted 
configurations. The punctures are initially at rest and
are located at $(\pm b,0,0)$ with momentum
$\vec{P} = P_{x} \hat{x} + P_{y} \hat{y}$, mass parameters
$m_{\text{p}}$, horizon (Christodoulou) masses $m_{\text{H}}$, 
and the total ADM mass $M_{\rm ADM}$.
The configurations are Bowen-York (BY) or 
Lorentz boosted (HS) initial data.  
}\label{tab:lbID}
\begin{ruledtabular}
\begin{tabular}{lcccccc}
Configuration   & $b/M$ & $m_{\text{p}}$ & $P_{x}$ & $P_{y}$ & $m_{\text{H}}$ & $M_{\rm ADM}/M$\\
\hline
BYQC & 4.7666 & 0.48523 & -0.001153 & 0.09932 & 0.5 & 0.98931 \\
HSQC & 4.7666 & 0.48745 & -0.001138 & 0.09794 &  0.5 & 0.98914 \\
\end{tabular}
\end{ruledtabular}
\end{table}

\begin{table}[!ht]
\caption{The final mass and spin for each configuration.}\label{tab:final}
\begin{ruledtabular}
\begin{tabular}{llll}
Configuration & $M_{\text{rem}}/M$ & $\chi_{\mathrm{rem}}$\\
\hline
BYQC & 0.95162 & 0.68643\\
LBQC & 0.95155 & 0.68646\\
\end{tabular}
\end{ruledtabular}
\end{table}

\begin{figure}[!ht]
  \includegraphics[angle=270,width=\columnwidth]{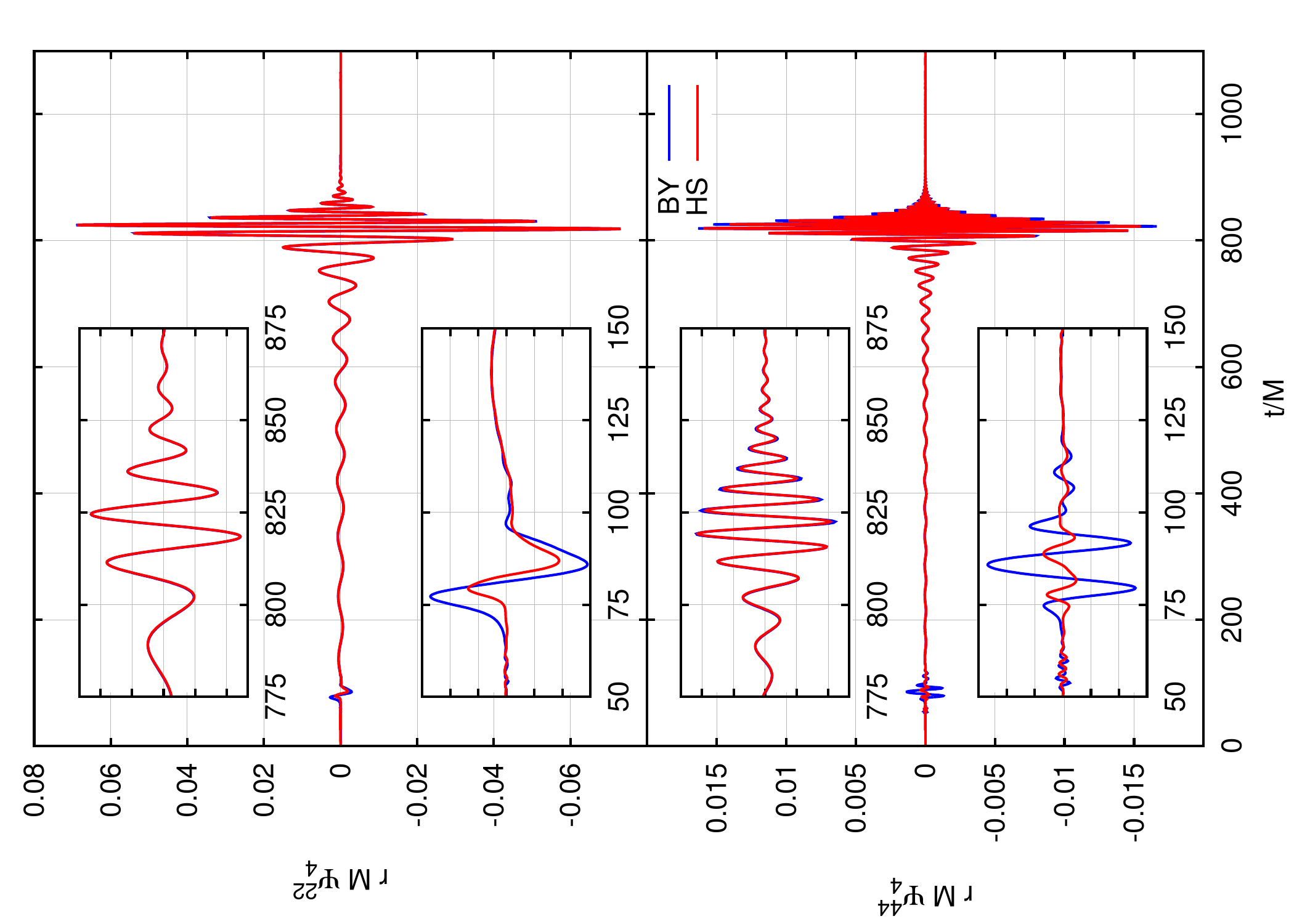}
  \caption{Comparison of the waveforms generated from
  the \hispid and BY initial data for the modes
$(\ell,m) = (2,2)$ and $(\ell, m) = (4,4)$ for nonspinning,
quasicircular binaries. Note the difference in initial radiation
content at around $t=75M$.
}
  \label{fig:QCwaveforms}
\end{figure}

Of course, the primary output from these simulations is the
gravitational waveform.
Figure\@~\ref{fig:QCwaveforms} shows the $(\ell=2, m=2)$ and
$(\ell=4,m=4)$ modes of $\psi_4$ for the \hispid and \by
binaries.
 While the two waveforms superpose for most of the simulation,
they differ substantially in the initial bursts
(located at around $t=75M$). The BY data have nearly a
factor  of 2 larger amplitude for the initial burst
relative to the \hispid data for the leading $(2,2)$ mode, 
and this ratio grows for the $(4,4)$ mode to a
factor $\sim 5$. This burst of initial radiation may have consequences
in the cases when high accuracy of the waveforms is needed (in
particular on the phase at late times) as it generates errors reflecting
on the refinement boundaries of the grid~\cite{Zlochower:2012fk}. 
Because the initial burst is much smaller, the \hispid data 
have the added benefit that the initial burst does not affect the
subsequent dynamics of the binary to the extent it does for \by.
Hence the input parameters (BH mass and momentum) matching more closely with 
the actual parameters of the binary after the initial burst
dissipates.
We also note that there is a phase and amplitude mismatch among
\hispid and BY
waveforms at merger. This may be due to a combination of slightly different
initial orbital parameters and the above mentioned disparity in the initial
radiation content.

\subsection{Quasicircular Spinning Black-Hole Binaries }\label{sec:spinningorbiting}

To assess how accurately the \hispid approach produces spinning
binaries, we again compare evolutions of \hispid and \by data.
Here, we study a few test cases of unequal-mass black-hole-binary
configurations starting in quasicircular orbits with antiparallel
(UD) spins, perpendicular to the line joining the black holes. We
evolve both black-hole binaries with the \hispid data and the standard
BY choice (for spins within the BY limit).

We also evolve black hole
binaries with nearly extremal parallel (UU) spins, $\chi=0.95$, a
regime unreachable for BY initial data. Table~\ref{tab:ch5_ID} gives
the initial data parameters of these black-hole binary configurations.
Antiparallel spins result in antisymmetric emission of gravitational
radiation, leading to a recoil in the merger remnant.

\begin{widetext}

\begin{table}
\caption{Initial data parameters for spinning, orbital configurations.
The punctures are initially located at $(\pm b,0,0)$, having mass
ratio $q = m_{(+)} / m_{(-)}$, with spins aligned or antialigned with
the $z$ direction, mass parameters $m_{(\pm)}$, total ADM mass $M_{\rm
ADM}$, and dimensionless spins $\chi$. The linear momenta of the holes,
$P_{(+)}^{y}$ and $P_{(-)}^{y}$ are initially purely in the $y-$direction
The Bowen-York configurations
are denoted by BY, and the \hispid by HS.  Finally, UU or UD denote
the direction of the two spins, both either aligned (UU) or antialigned (UD).} 
\label{tab:ch5_ID}
\begin{tabular}{lcccccccc}
\hline
\hline
Configuration & $b/M$ & $q$ & $m_{(+)}$ & $m_{(-)}$ & $\chi$ & $P_{(+)}^{y}$ & $P_{(-)}^{y}$ & $M_{\rm ADM}/M$ \\
\hline
BY80UD & 5.4489 & 2 & 0.4078 & 0.1998 & 0.80 & 0.081882 & -0.081882 & 0.991789 \\
HS80UD & 5.4489 & 2 & 0.6667 & 0.3333 & 0.80 & 0.086168 & -0.083831 & 0.996224 \\
\hline
HS95UU-A & 4 & 1 & 0.5000 & 0.5000 & 0.95 & 0.103289 & -0.103289 & 0.986854 \\
HS95UU-B & 5 & 1 & 0.5056 & 0.5056 & 0.95 & 0.092251 & -0.092251 & 0.988631 \\
\hline
\hline
\end{tabular}
\end{table}
\end{widetext}

\begin{figure}[!ht]
  \centering
  \includegraphics[angle=270,width=0.49\columnwidth]{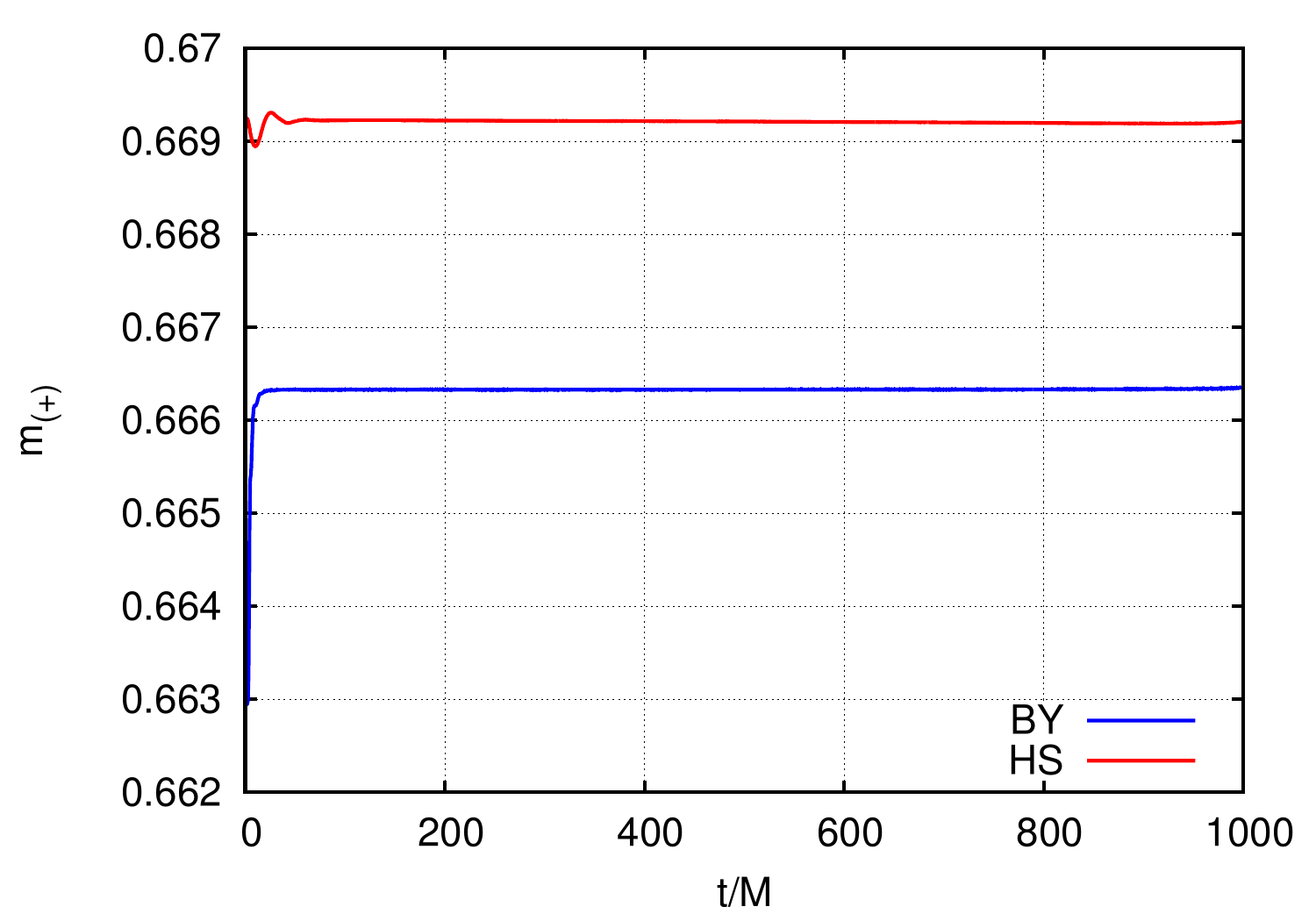}
  \includegraphics[angle=270,width=0.49\columnwidth]{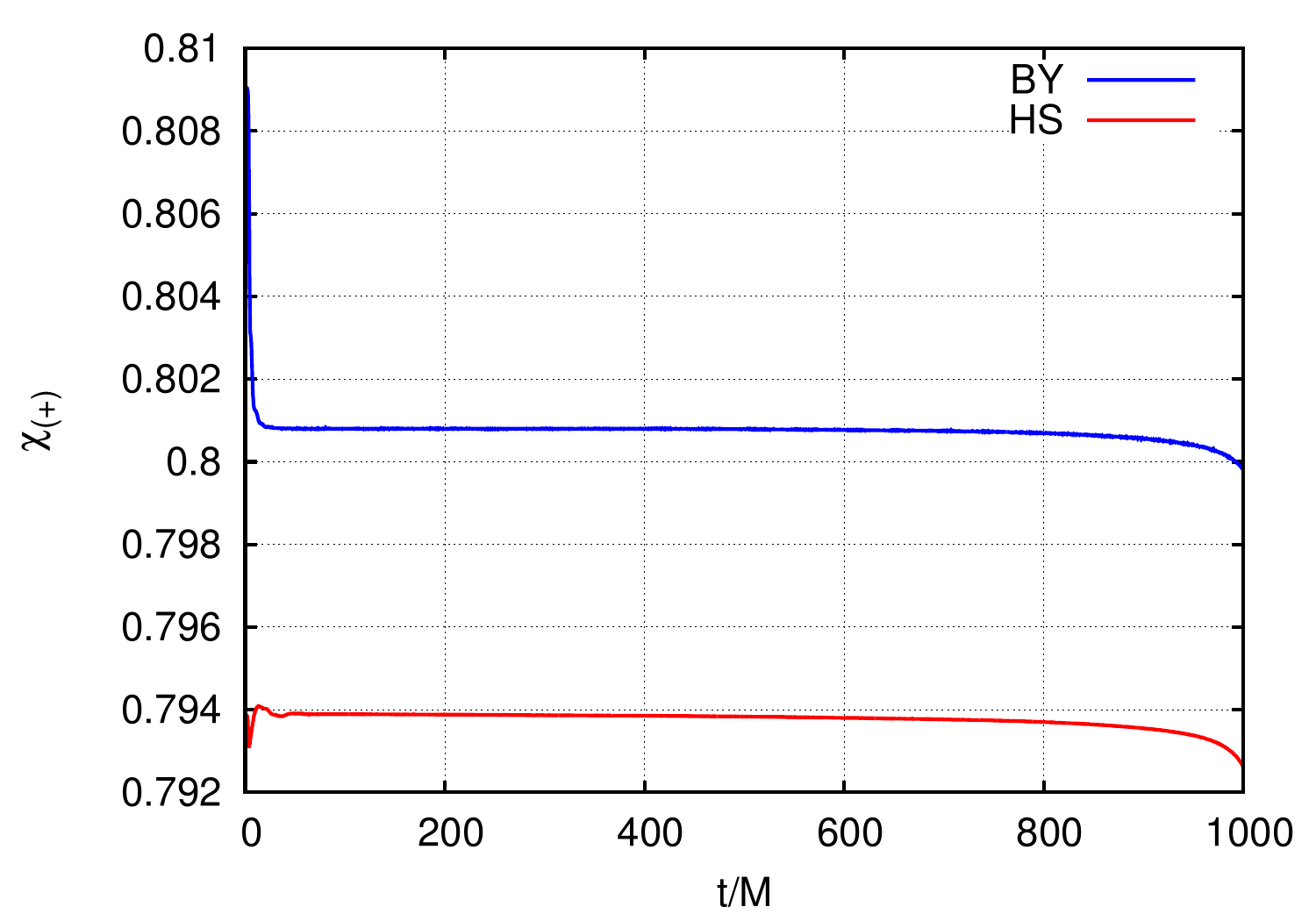}
  \includegraphics[angle=270,width=0.49\columnwidth]{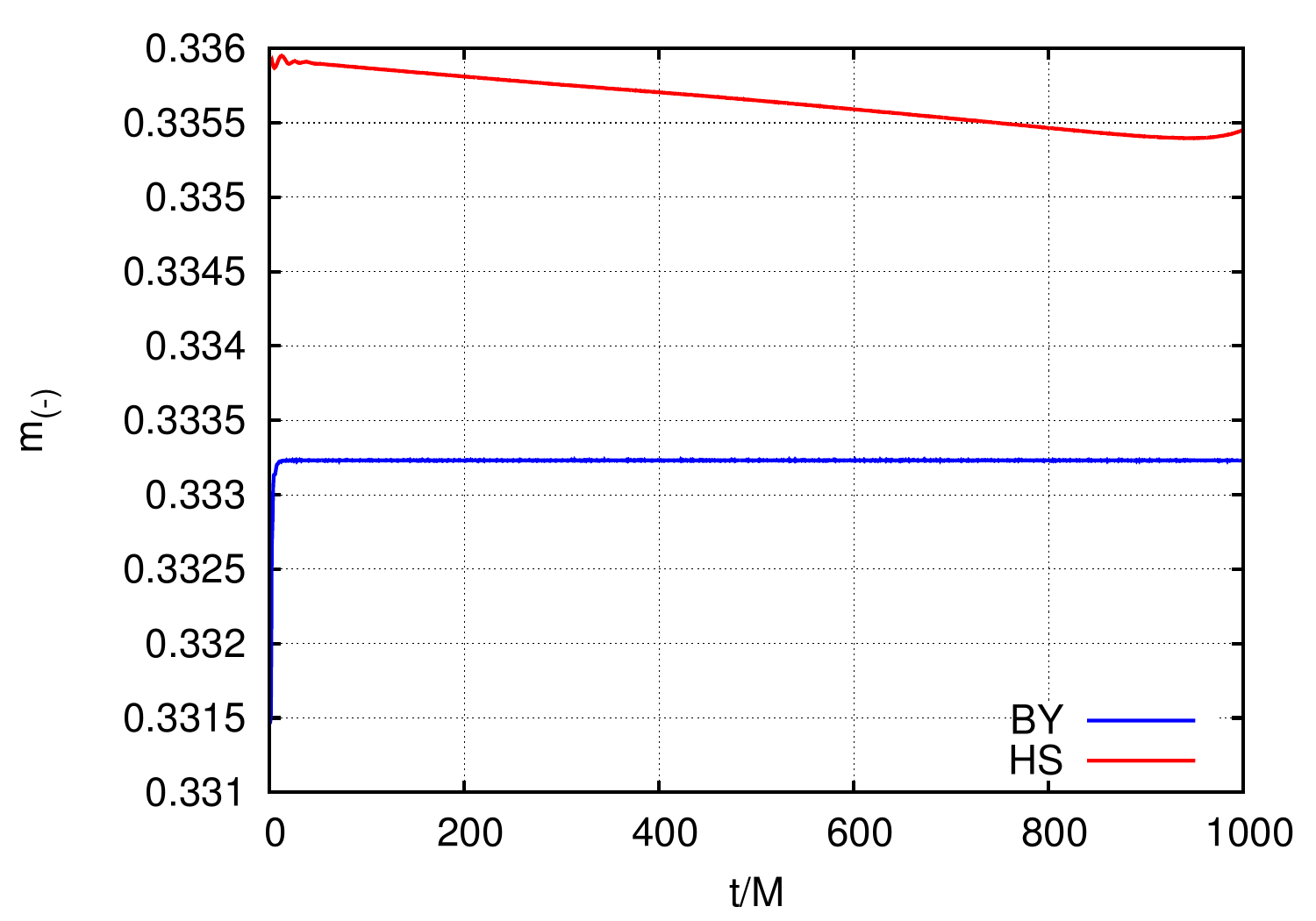}
  \includegraphics[angle=270,width=0.49\columnwidth]{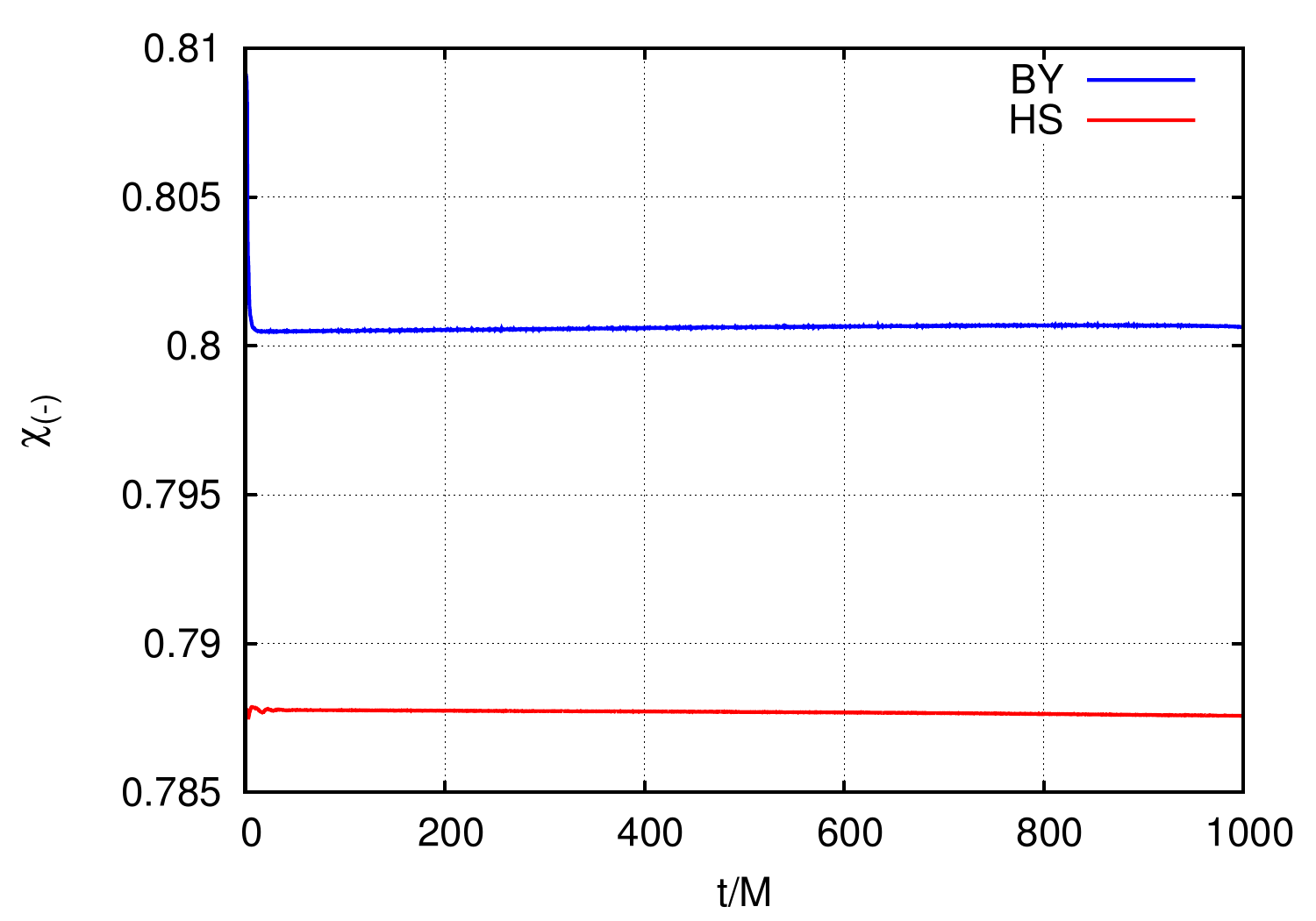}
  \caption{The evolution of the irreducible mass and dimensionless spin of the individual black holes with $q = 2$ and $\chi = 0.8$, comparing Lorentz-boosted Kerr and Bowen-York initial data. The top panels show the larger black hole, and the bottom panels show the smaller black hole.}
\label{fig:q2_evolution}
\end{figure}

\begin{figure}[!ht]
  \centering
  \includegraphics[angle=270,width=0.8\columnwidth]{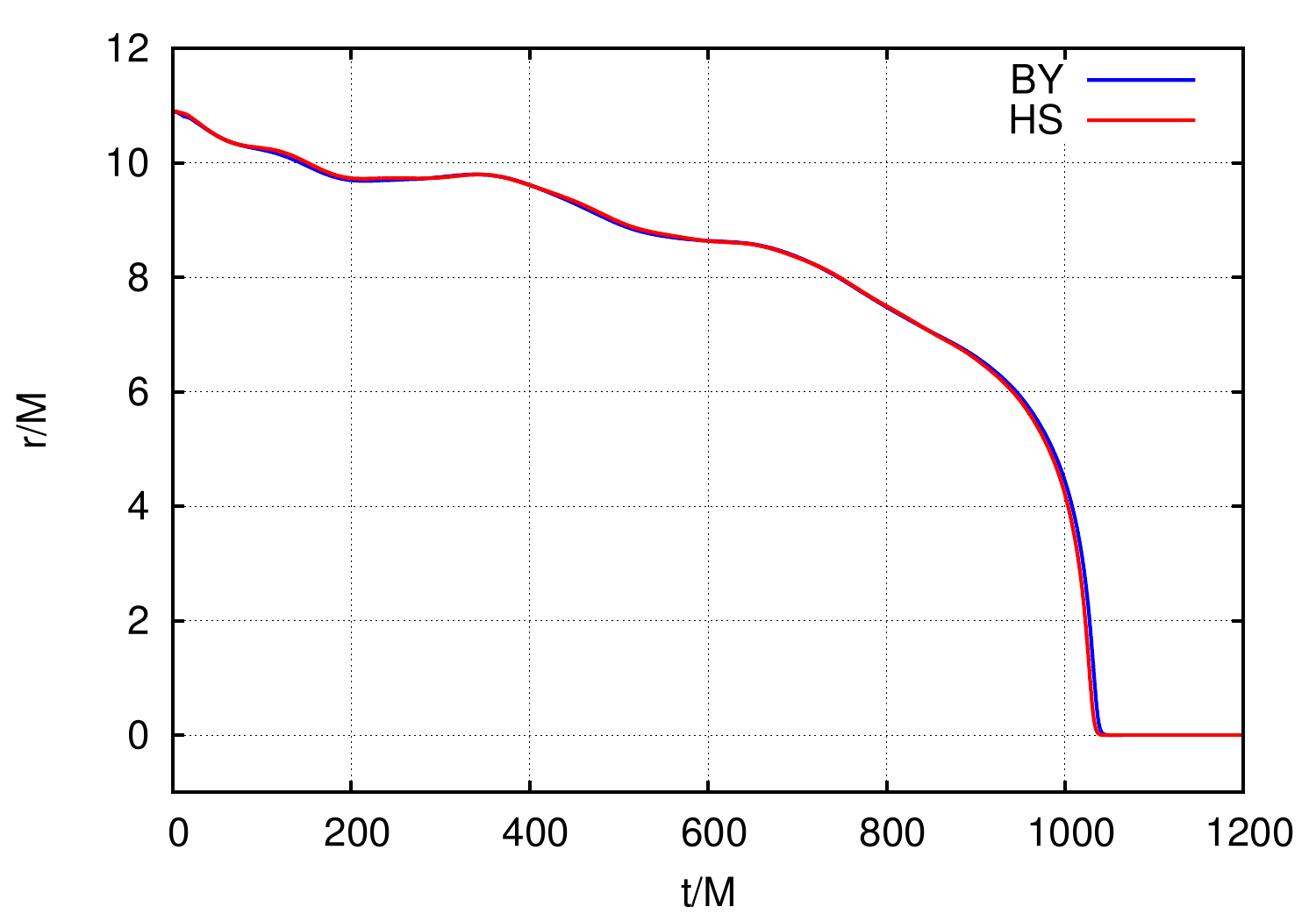}
  \caption{The evolution of the coordinate separation between the black holes in an orbiting binary with $q = 2$ and $\chi = 0.8$, comparing Lorentz-boosted Kerr and Bowen-York initial data.}
\label{fig:q2_separation}
\end{figure}

\begin{figure}[!ht]
  \centering
  \includegraphics[angle=270,width=0.8\columnwidth]{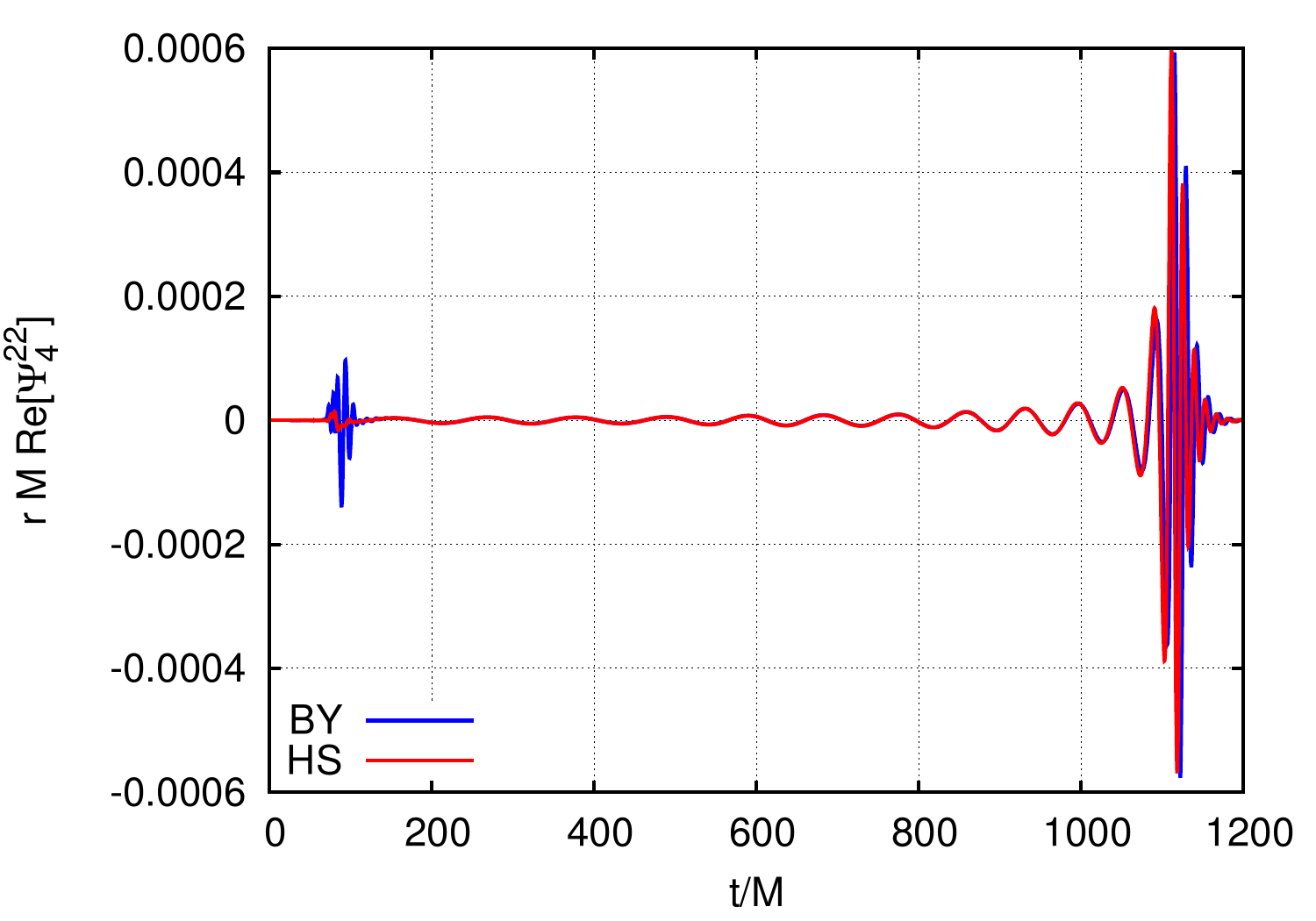}
  \includegraphics[angle=270,width=0.8\columnwidth]{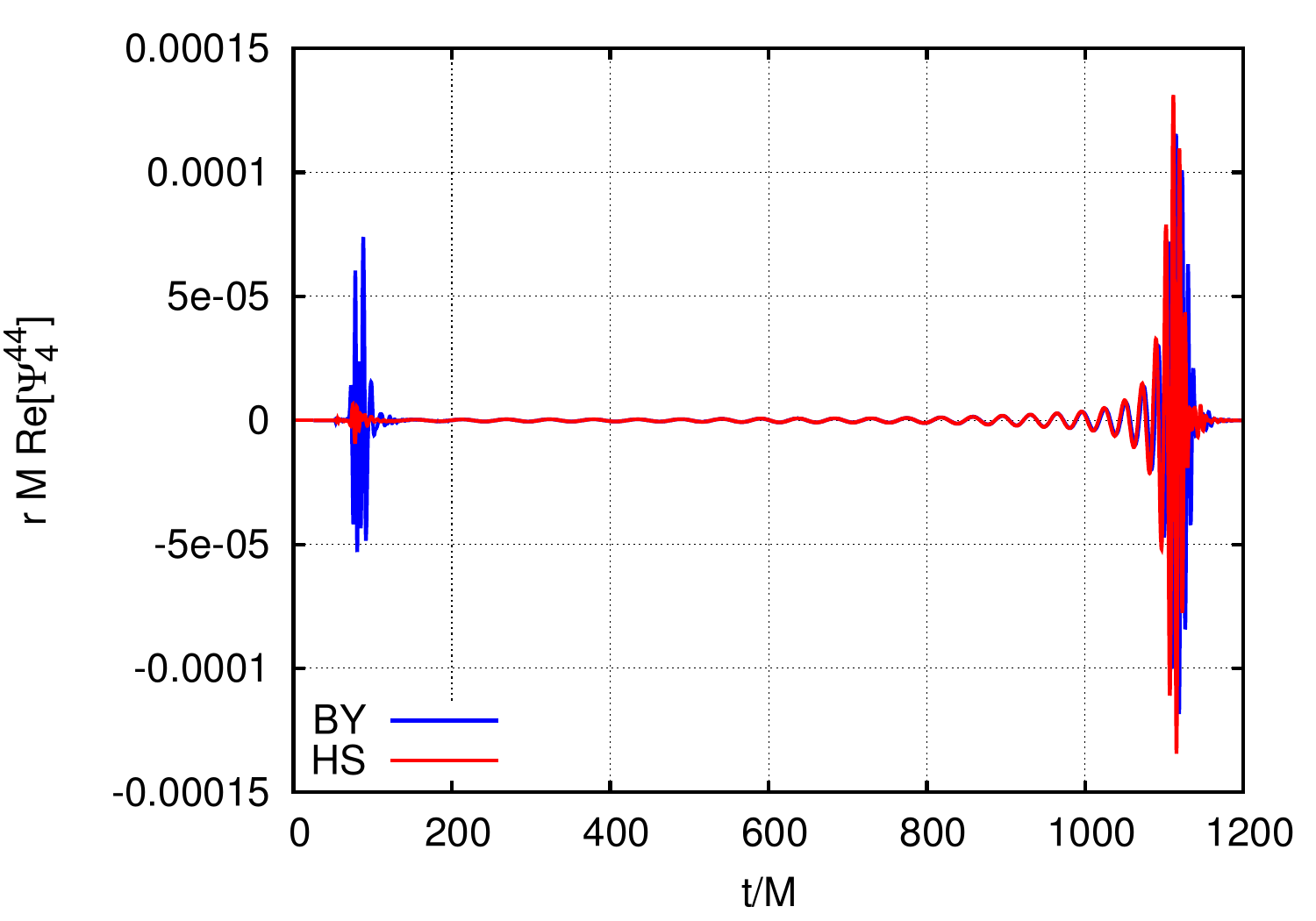}
  \caption{Comparison of the waveforms generated from the Lorentz-boosted Kerr and Bowen-York initial data for the modes $(\ell,m) = (2,2)$ and $(\ell, m) = (4,4)$. Note the difference in initial radiation content at around $t=75M$.}
\label{fig:q2_waveforms}
\end{figure}

\begin{figure}[!ht]
  \centering
  \includegraphics[angle=270,width=0.49\columnwidth]{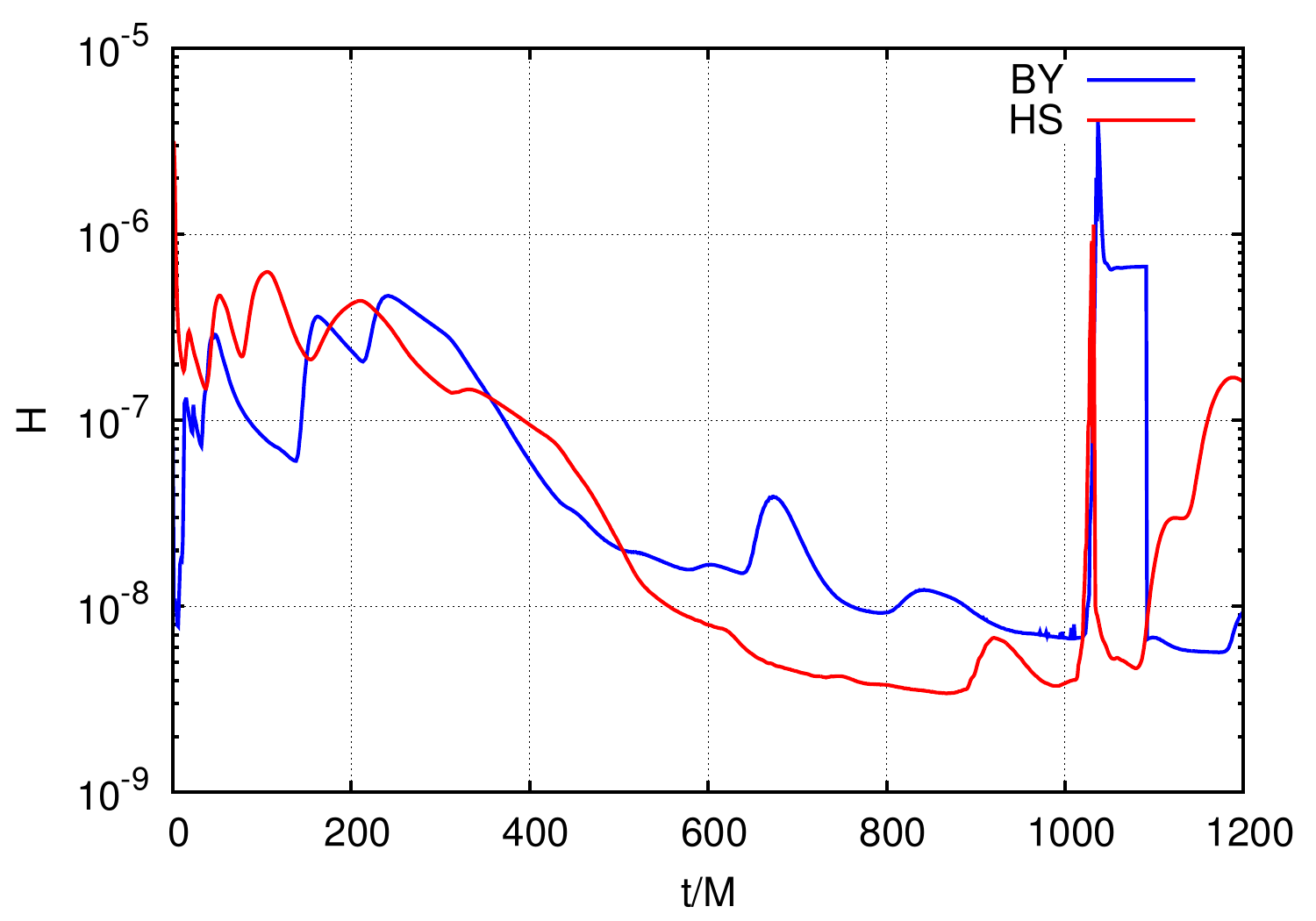}
  \includegraphics[angle=270,width=0.49\columnwidth]{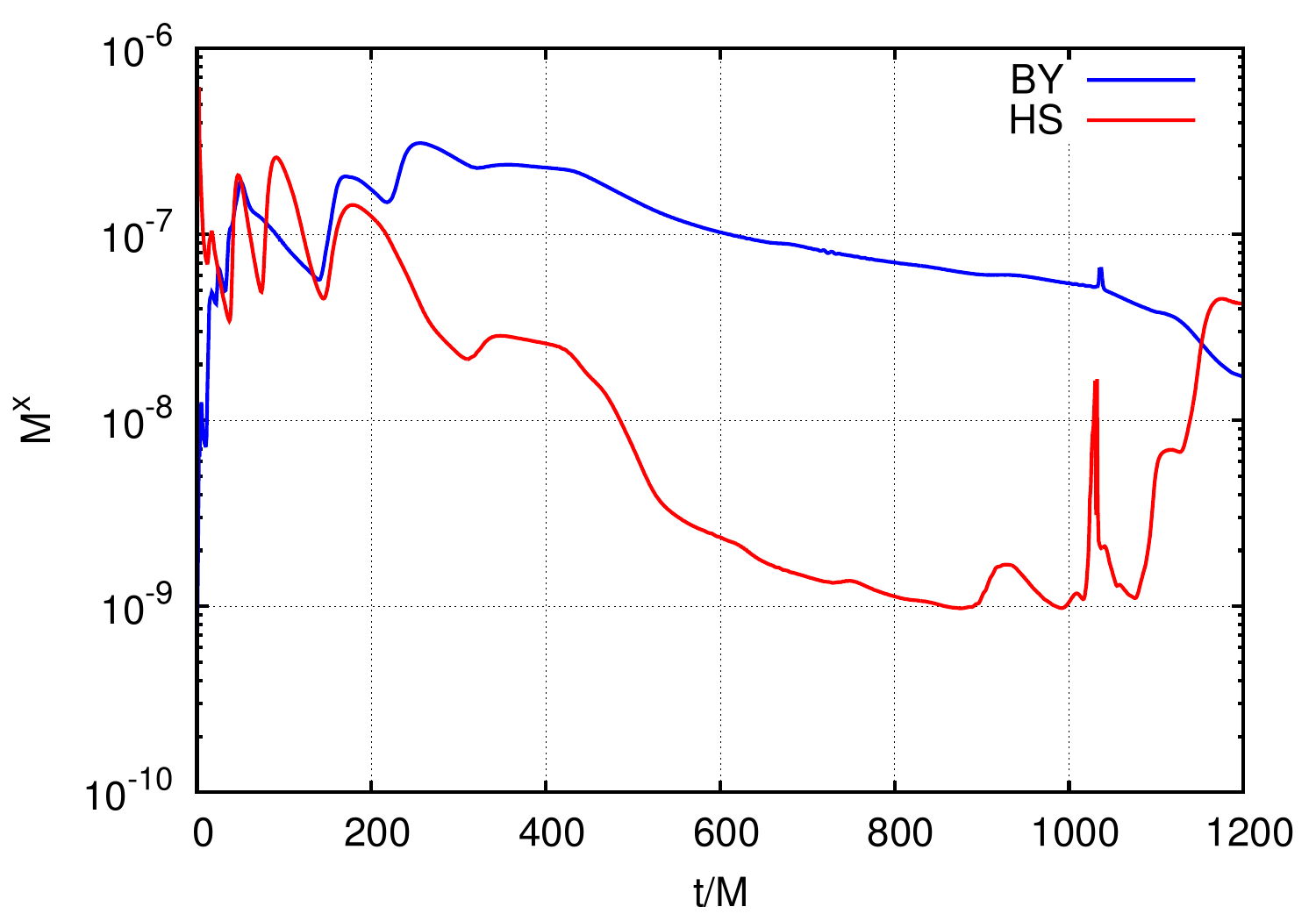}
  \includegraphics[angle=270,width=0.49\columnwidth]{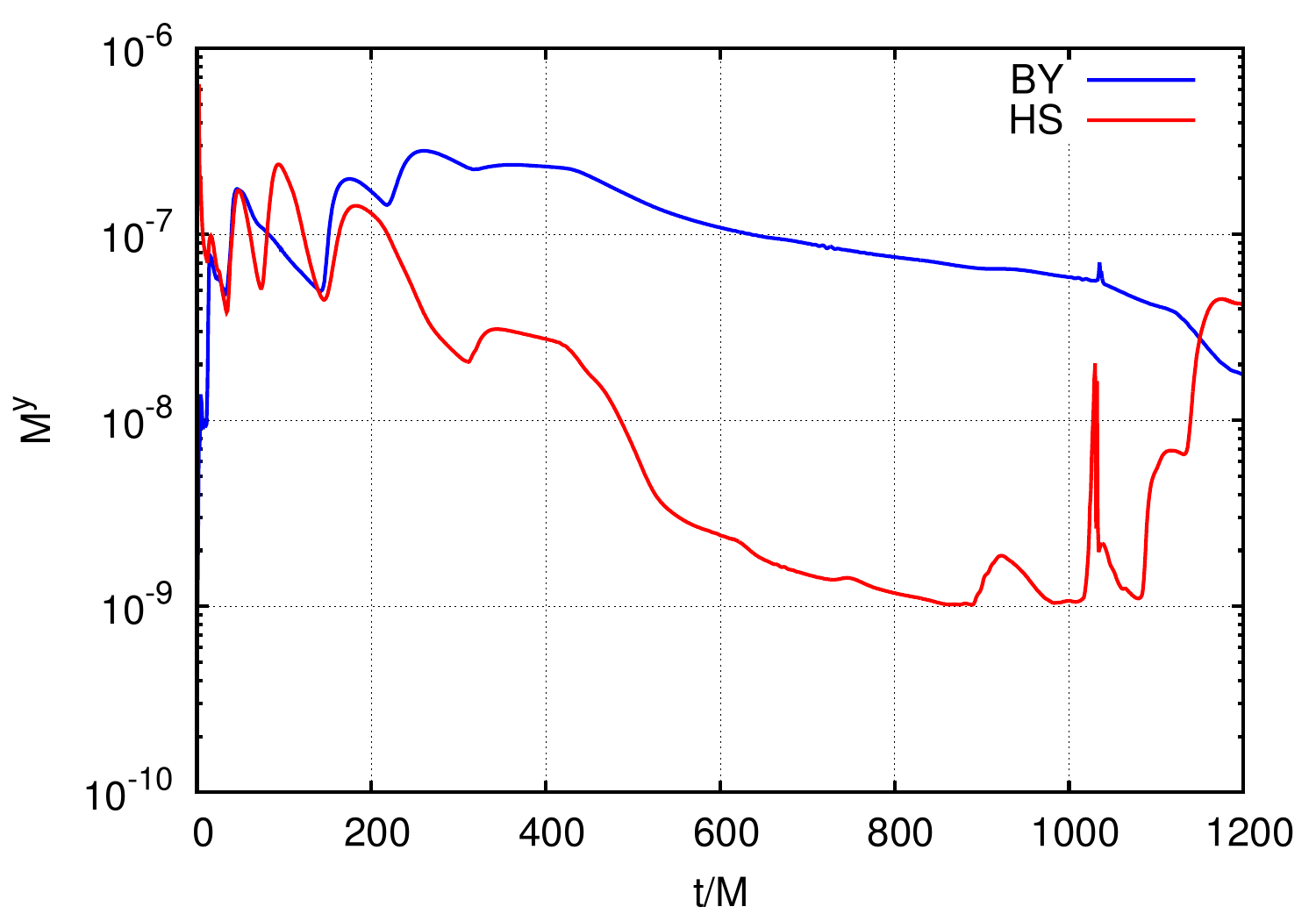}
  \includegraphics[angle=270,width=0.49\columnwidth]{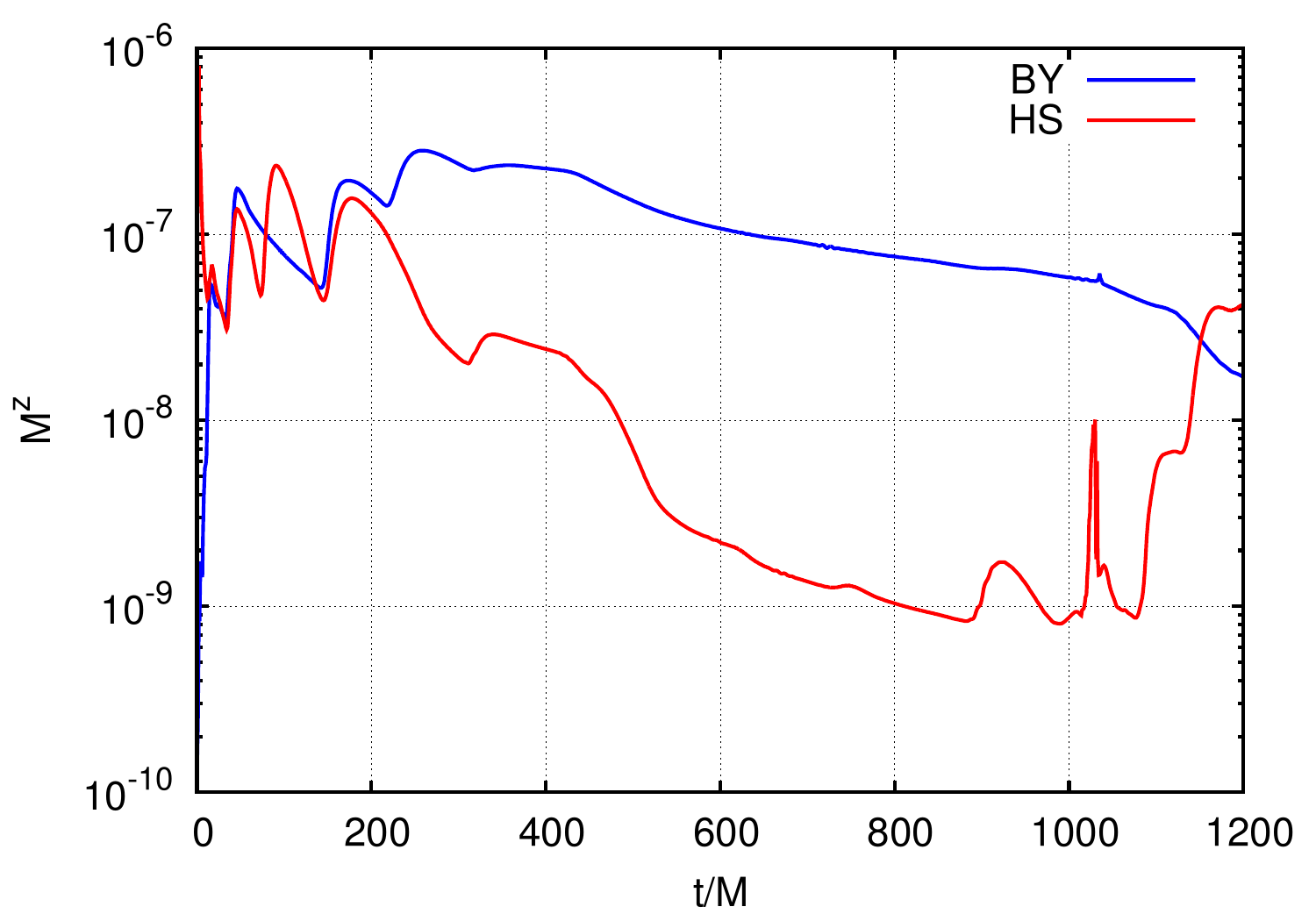}
  \caption{The evolution of the Hamiltonian ($H$) and momentum ($M^x$, $M^y$, and $M^z$) constraints for a black-hole binary with $q = 2$ and $\chi = 0.8$, comparing Lorentz-boosted Kerr and Bowen-York initial data.}
\label{fig:q2_constraints}
\end{figure}

For given binary separations, the \hispid and \by parameters (spin,
momentum, horizon masses)  are not identical, as shown
in Fig.~\ref{fig:q2_evolution}, since the initial radiation content
and distortions are not the same. However, they are  close
enough for comparisons of physical quantities such as the puncture
separations shown in Fig.~\ref{fig:q2_separation} and the
gravitational waveforms shown in Fig.~\ref{fig:q2_waveforms}.
Figure~\ref{fig:q2_constraints} shows that the conformally curved
initial data yield better evolved constraint satisfaction than the
conformally flat case. The final measured parameters are shown in
Table~\ref{tab:ch5_final}.

\begin{table*}
\centering
\caption{The relaxed mass ratio and initial spins and final remnant mass, spin, and recoil velocity. 
The final two columns give the analytic fits for the final mass and spin given the initial parameters.
}
\label{tab:ch5_final}
\begin{tabular}{lccccccccc}
\hline
\hline
Configuration & $q^{\text{relax}}$ & $\chi^{\text{relax}}_{1}$ & $\chi^{\text{relax}}_{2}$ & $M_{\text{rem}}/M$ & $\chi_{\text{rem}}$ & $V$ [$\text{km s}^{-1}$] & $M^{\text{fit}}_{\text{rem}}/M$ & $\chi^{\text{fit}}_{\text{rem}}$ & $V^{\text{fit}}$\\
\hline
BY80UD   & 0.50010 & 0.80053 & -0.80080 & 0.96831 & 0.41037 & $420 \pm 2$ & 0.96844 & 0.40941 & 421.86\\
HS80UD   & 0.50185 & 0.78776 & -0.79389 & 0.96815 & 0.41008 & $414 \pm 7$ & 0.96833 & 0.41258 & 419.08\\
\hline
HS95UU-A & 1.00000 & 0.9465 & 0.9465 & 0.8942  & 0.9402 & 0 & 0.8940 & 0.9403 & 0\\
SXS\#157 & 1.00000 & 0.9496 & 0.9496 & 0.8937  & 0.9409 & 0 & 0.8936 & 0.9410 & 0\\
HS95UU-B & 1.00000 & 0.9520 & 0.9520 & 0.8925  & 0.9413 & 0 & 0.8933 & 0.9415 & 0\\
\hline
\hline
\end{tabular}
\end{table*}

\begin{figure}[!ht]
  \centering
  \includegraphics[angle=270,width=0.8\columnwidth]{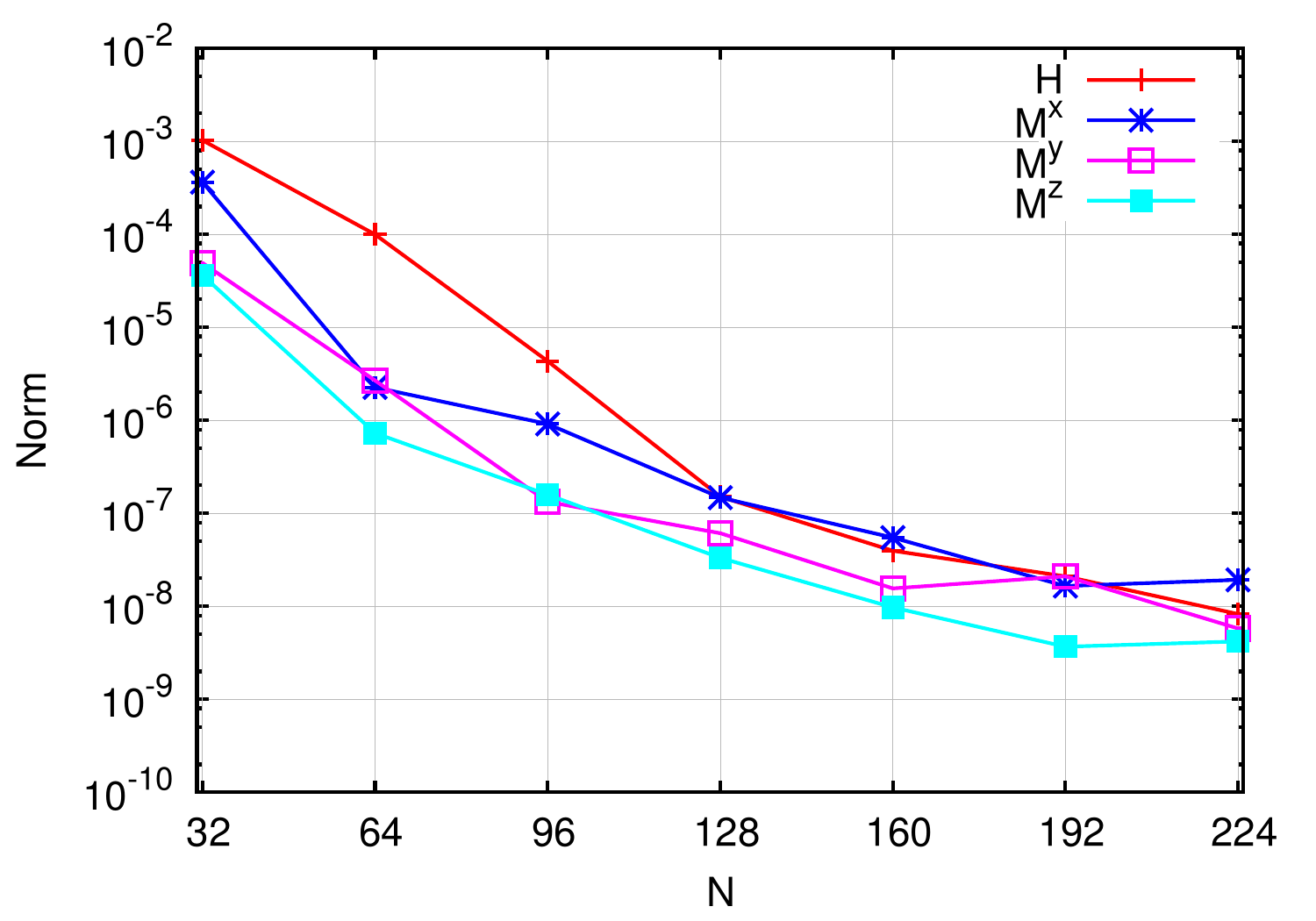}
  \includegraphics[angle=270,width=0.8\columnwidth]{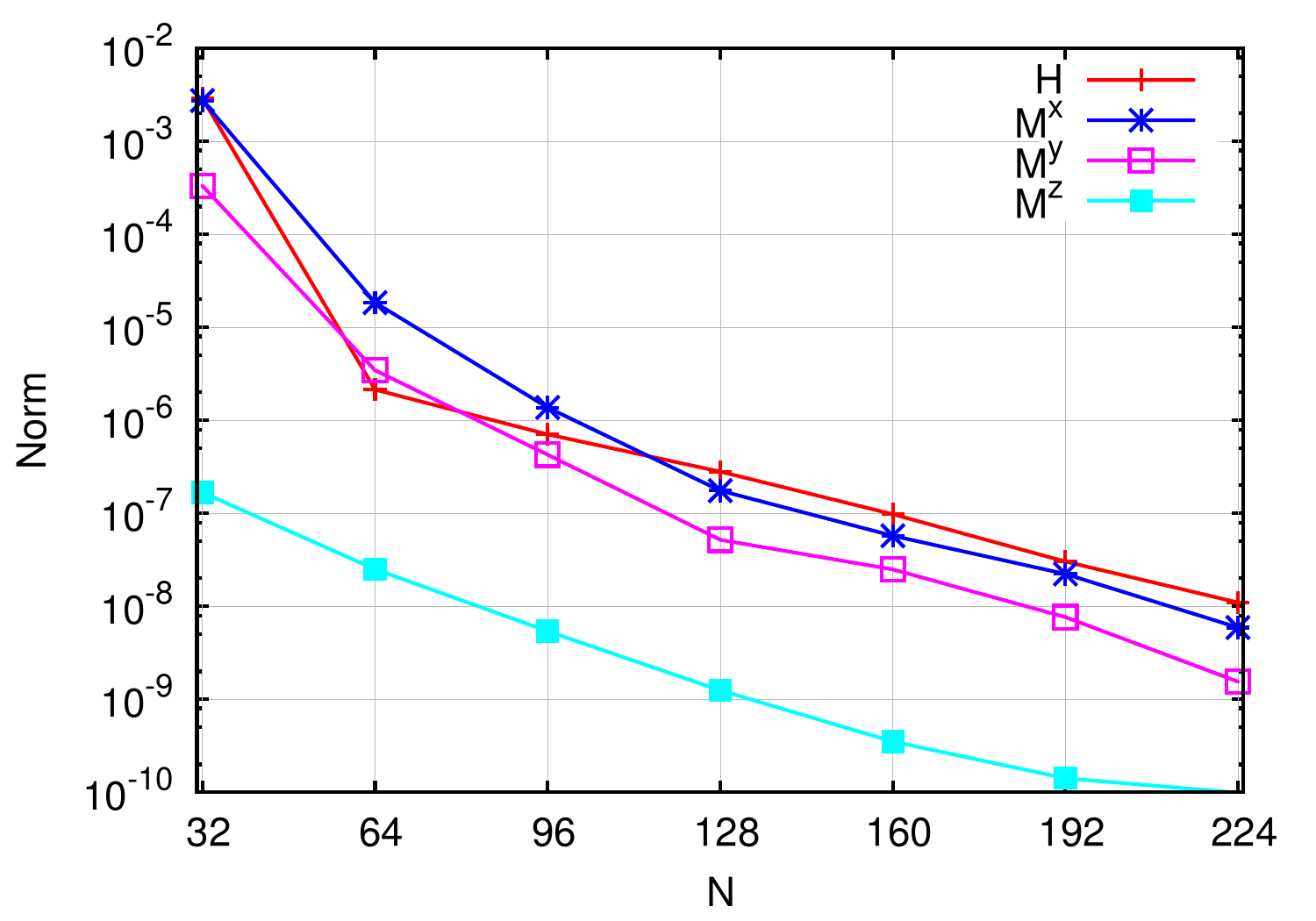}
  \caption{Convergence of the Hamiltonian and momentum constraint 
     residuals on the full numerical grid with increasing number of 
     collocation points. This example shows an equal-mass Kerr black-hole binary in a quasicircular orbit with spin $\chi = 0.95$, 
     momentum $P_{(\pm)}^{y} = \pm 0.09225$, and separation $d = 10 M$ using 
     the LES coordinates.  (Top panel) For a small grid near puncture 
     ($4.75 \leq x \leq 5.25$, $0.1 \leq y \leq 0.6$, and $0.5 \leq z \leq 1.0$).
     (Bottom panel) On the full numerical evolution grid.
   }
\label{fig:BK_convergence}
\end{figure}

One of our main motivations to study a new set of initial data is to
be able to simulate highly spinning black holes, beyond the BY (or
conformally flat) limit, $\chi \approx
0.93$~\cite{Cook90a,Dain:2002ee,Lousto:2012es}. In
Fig.~\ref{fig:BK_convergence} we show the level of satisfaction of the
constraints for our new initial data for spinning black-hole binaries
with equal masses and spin parameters $\chi = 0.95$. The $L^2$ norm of
the constraints
converge to a level of ${\cal O}(10^{-7}) - {\cal O}(10^{-6})$ near the horizons, and down
to ${\cal O}(10^{-10})$ in the bulk.  We
do not consider points interior to the horizons in our $L^2$
calculations. If one requires greater satisfaction of the constraints,
one can fine-tune the attenuation functions to that end.

\begin{figure}[!ht]
  \centering
  \includegraphics[angle=270,width=0.9\columnwidth]{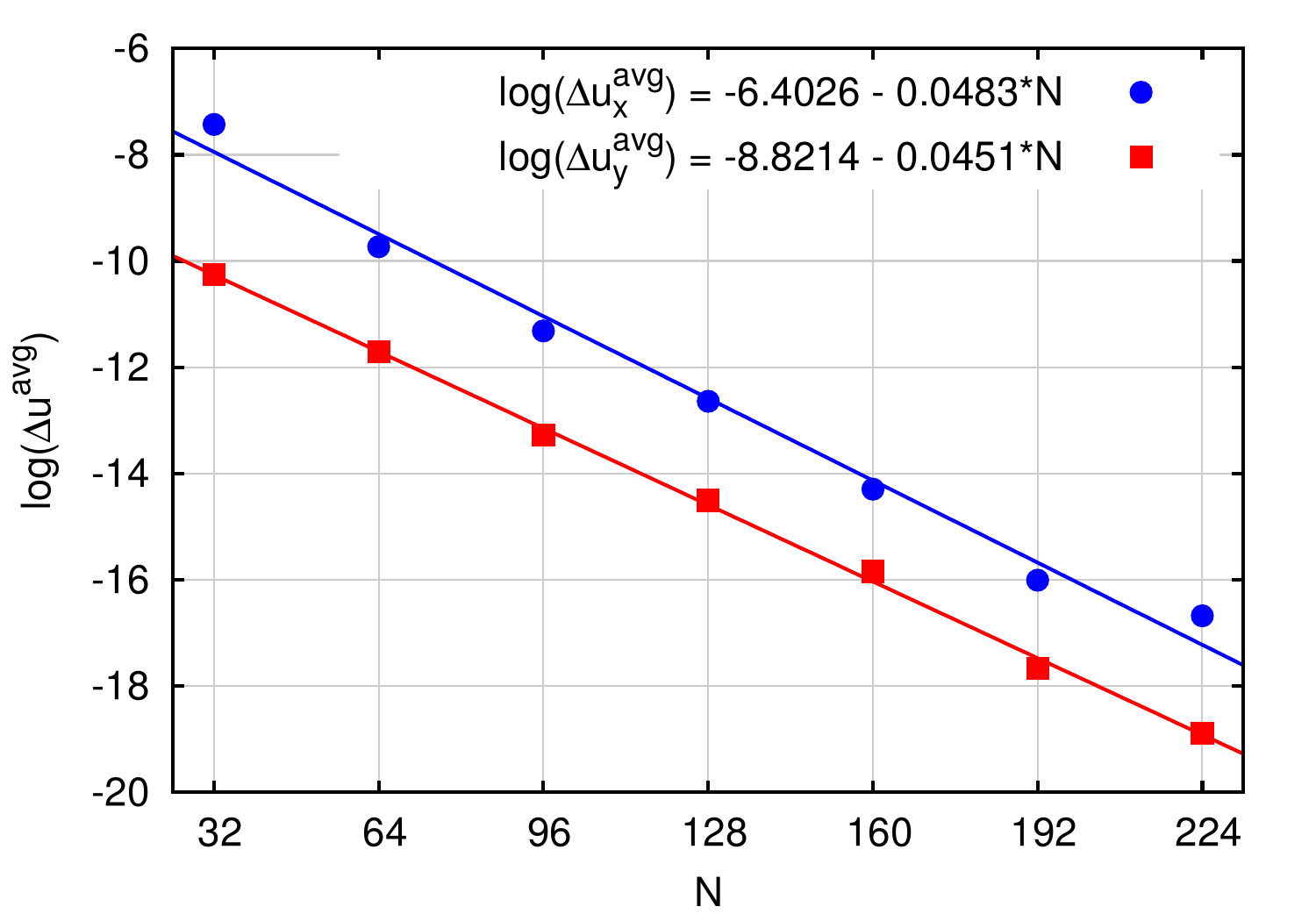}
  \caption{Exponential convergence for the same data in Fig.~\ref{fig:BK_convergence}
     along the $x$ and $y$ axes  of the 
     averaged quantity [See Eq. (\ref{eq:psi})] $\Delta u = u_{256} - u_N$, where N is the 
     number of collocation points.
     The average is taken over $0M$ to $50M$ along $x$ or $y$.
     Since there is higher resolution near the BHs, these averages are weighted more towards the
     regions around the BHs.  The solid lines show a best least-squares fit assuming 
     exponential convergence.
   }
\label{fig:BK_convergence_exp}
\end{figure}

It is important to note that the actual metric functions generated by
solving Eq.~(\ref{eq:constraints}) converge even faster than the
constraints themselves. In Fig.~\ref{fig:BK_convergence_exp}, we show
how the function $u$ converges with the number of collocation points.
To do this, we compare the $u$ generated with $N=256$ collocation points
with the values obtained using fewer collocation points. We do this
both along the $x$-axis (where the black holes are located) and along
the $y$-axis (in between the holes). In both cases, we find exponential
convergence.

\begin{figure}[!ht]
  \centering
  \includegraphics[angle=270,width=0.8\columnwidth]{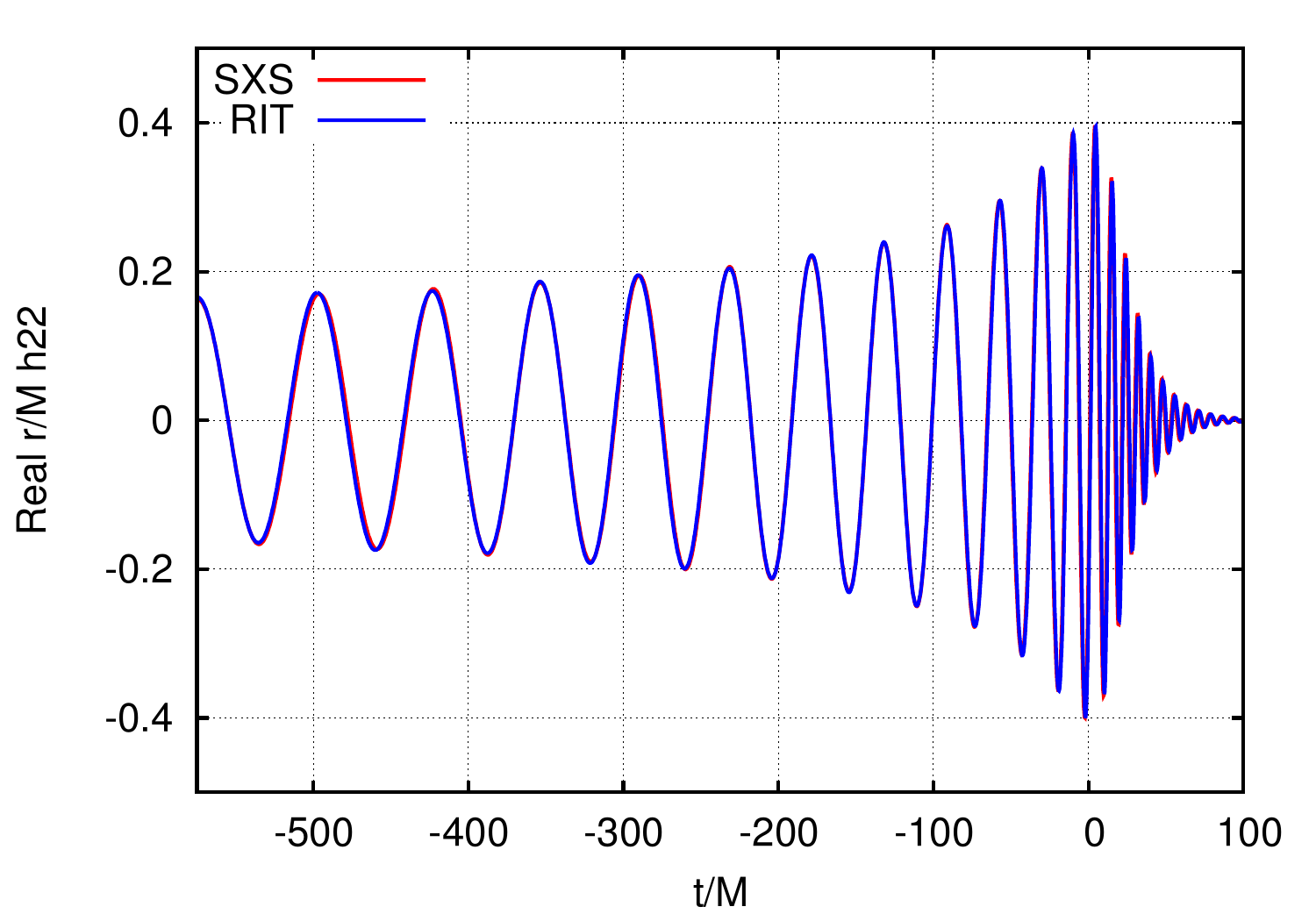}
  \includegraphics[angle=270,width=0.8\columnwidth]{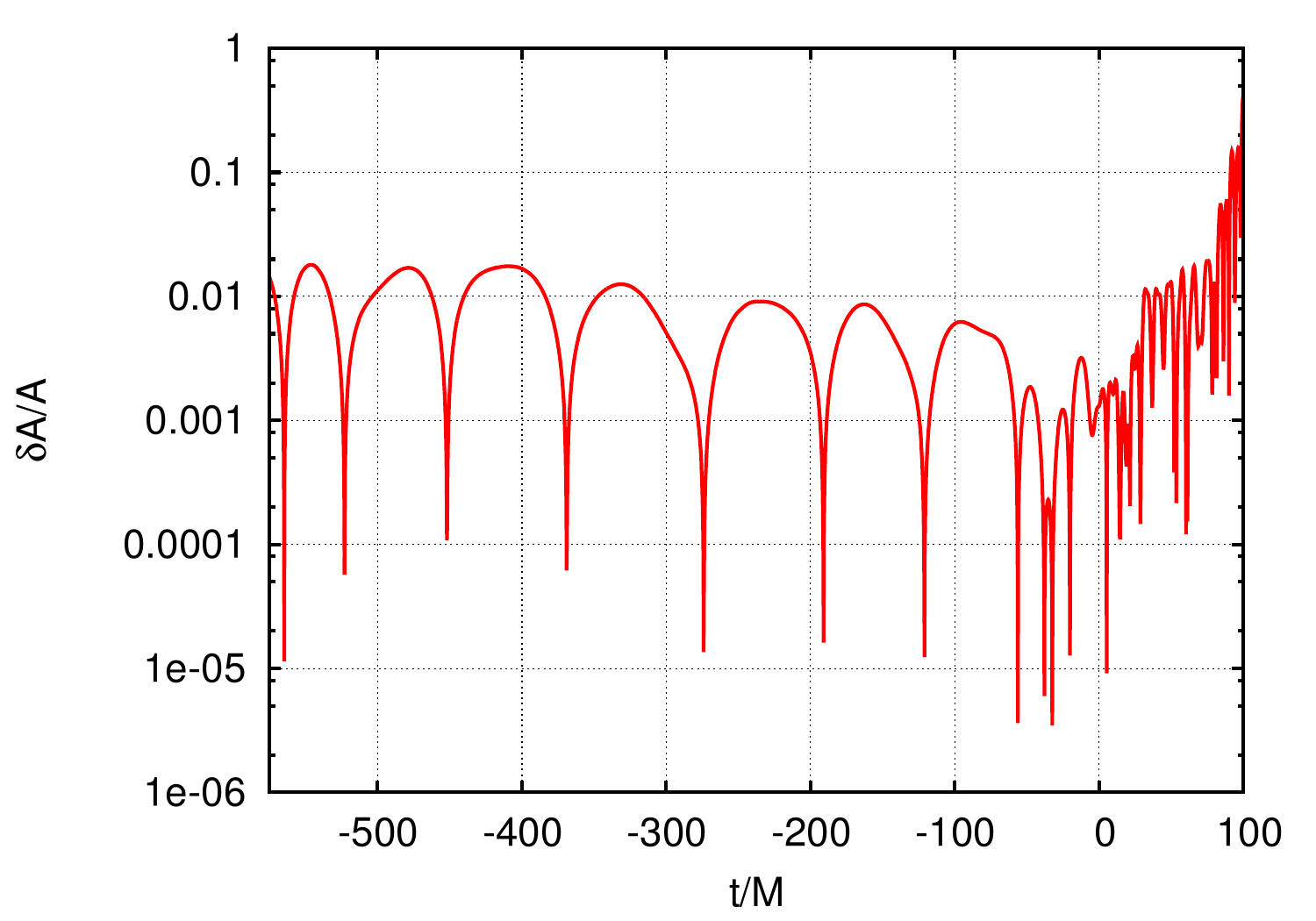}
  \includegraphics[angle=270,width=0.8\columnwidth]{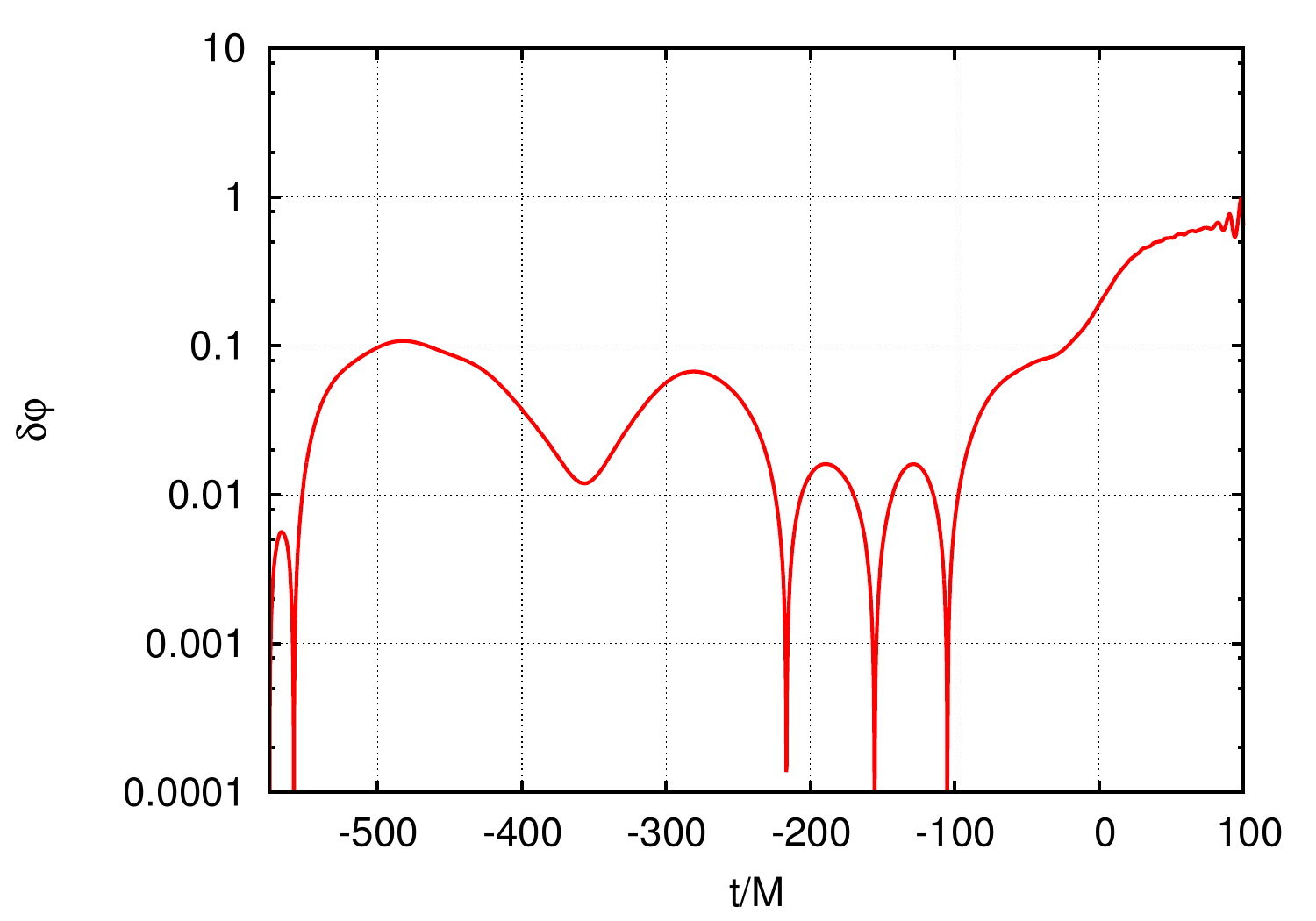}
  \caption{ The 22 mode of the gravitational waveform generated by an
  equal-mass, equal-spin (with spins $\chi = 0.95$) BHB (HS95UU-A) generated using isotropic coordinates compared to SXS catalog \#157.  Middle and bottom panels show the percent difference between the two for the amplitude and phase, respectively.}
\label{fig:uu95-jim-wf}
\end{figure}

\begin{figure}[!ht]
  \centering
  \includegraphics[angle=270,width=0.8\columnwidth]{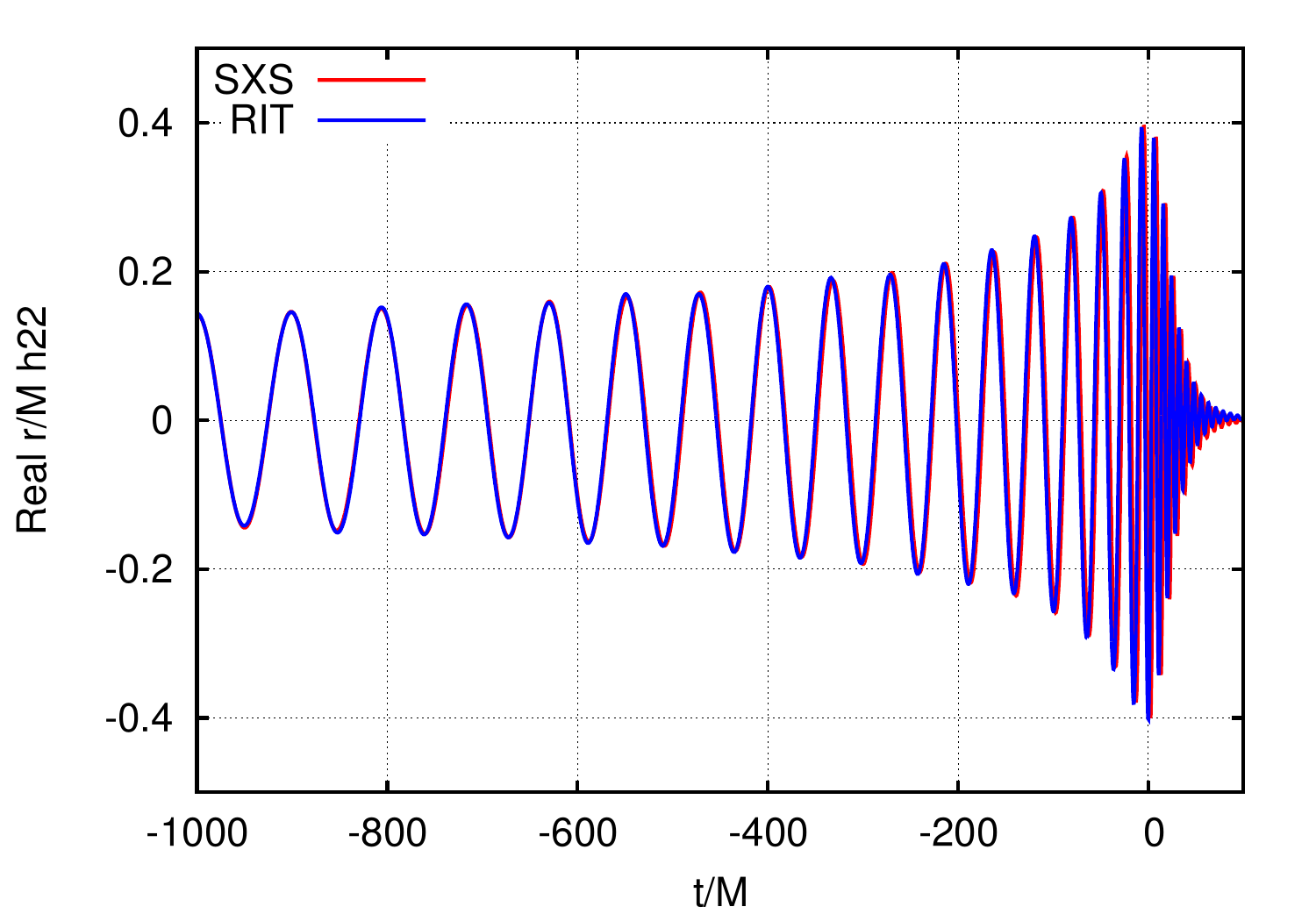}
  \includegraphics[angle=270,width=0.8\columnwidth]{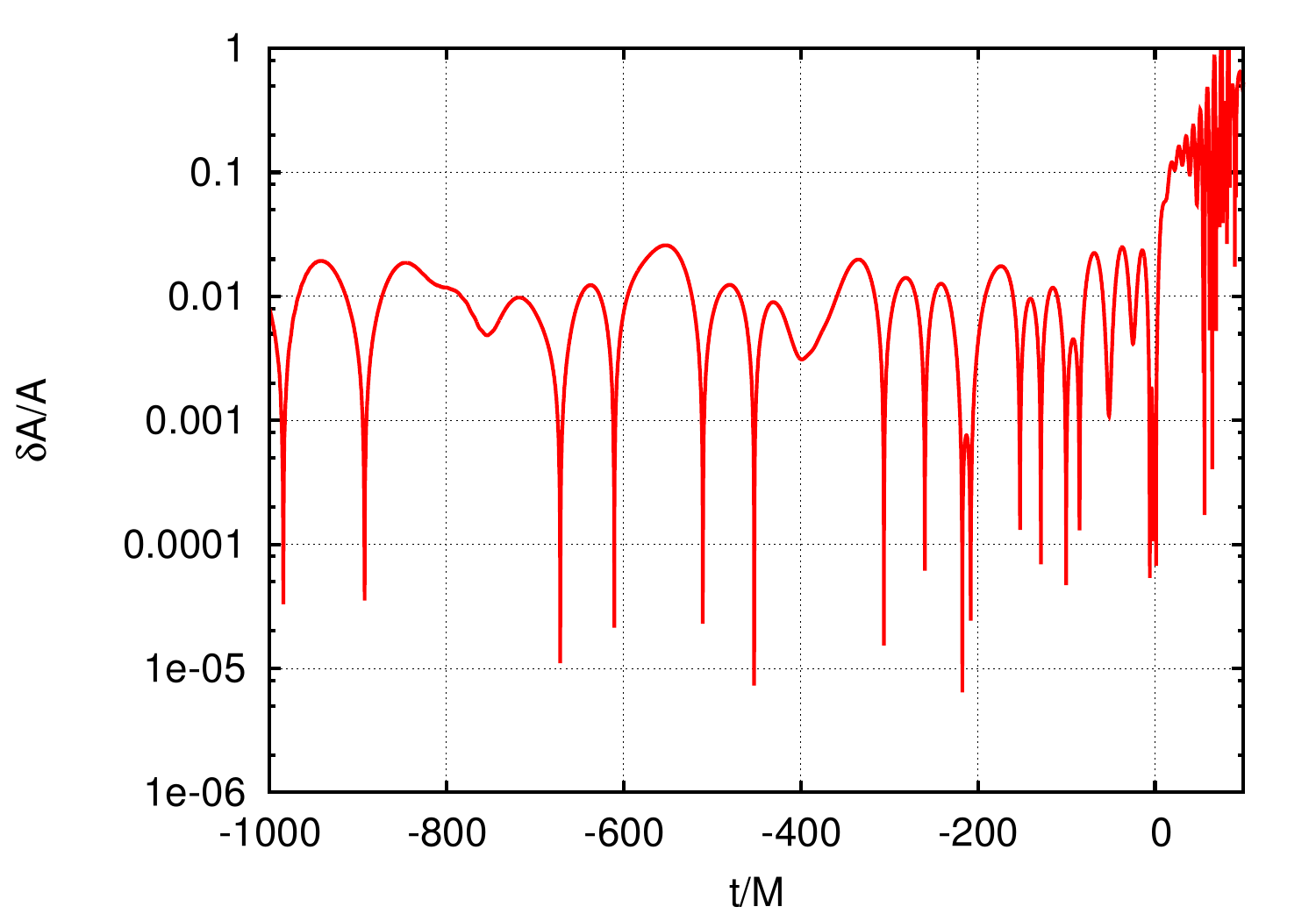}
  \includegraphics[angle=270,width=0.8\columnwidth]{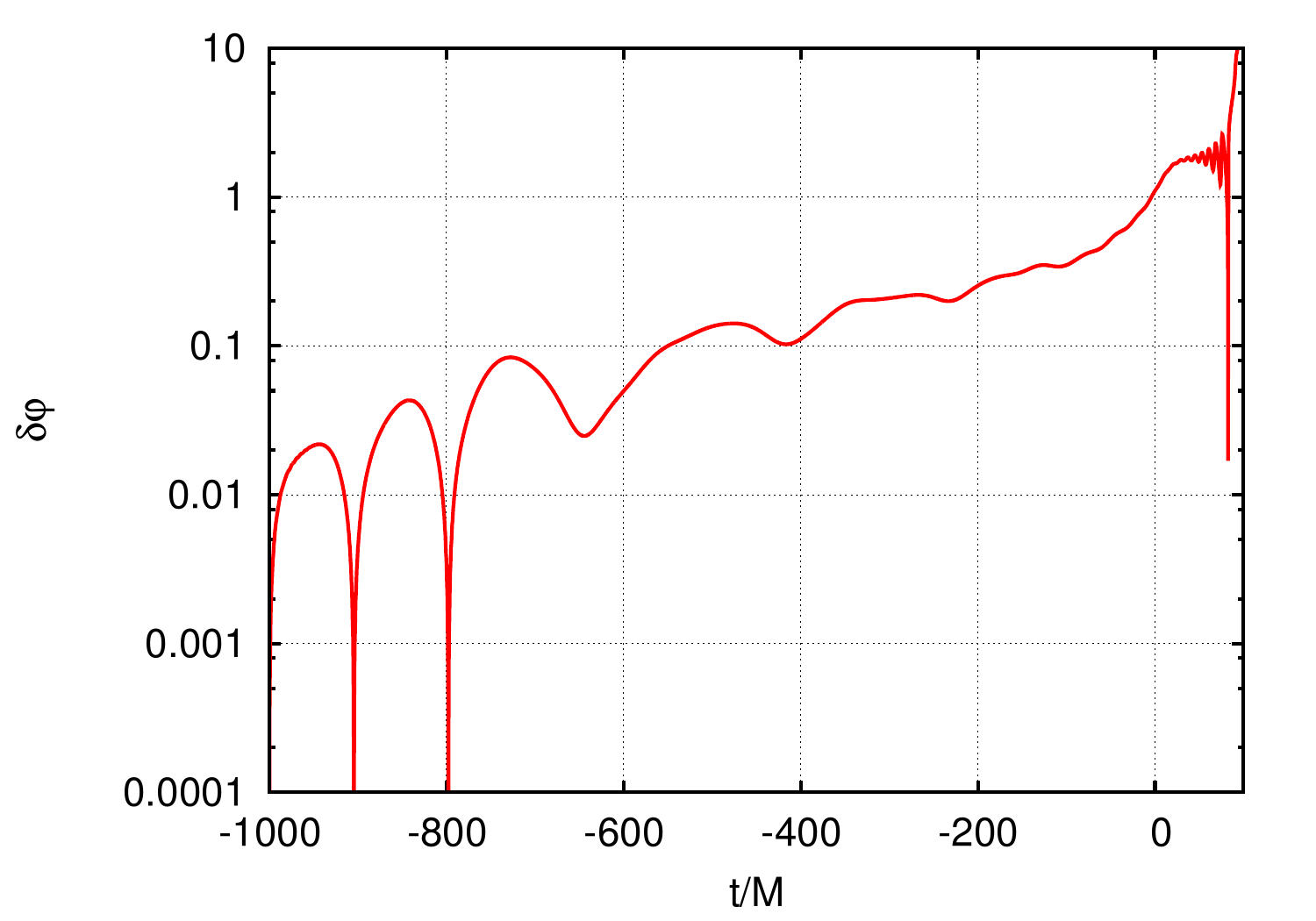}
  \caption{ The 22 mode of the gravitational waveform generated by an
  equal-mass, equal spins of 0.95 BHB  (HS95UU-B) generated using LES coordinates compared to SXS catalog \#157.  Middle and bottom panels show the percent difference between the two for the amplitude and phase, respectively.}
\label{fig:uu95-yosef-wf}
\end{figure}

In Figs.~\ref{fig:uu95-jim-wf} and \ref{fig:uu95-yosef-wf}, we compare the $(\ell=2, m=2)$ mode of
$\psi_4$ for the two UU95 runs with the corresponding SXS waveform from the
SXS catalog~\cite{SXS:catalog}.
For most of the waveform, we observe relative errors (compared to
SXS) of $2\%$ for the amplitude of the waveform. The phase differences
between the two \hispid runs and the SXS run are between $0.1$ and
$0.2$ rad.\ for most of the waveform. The HS95UU-A configuration
agrees with the SXS simulation to a higher degree, with a phase
difference of under 0.2 rad.\ through merger. 

\begin{figure}[!ht]
  \centering
  \includegraphics[angle=270,width=0.49\columnwidth]{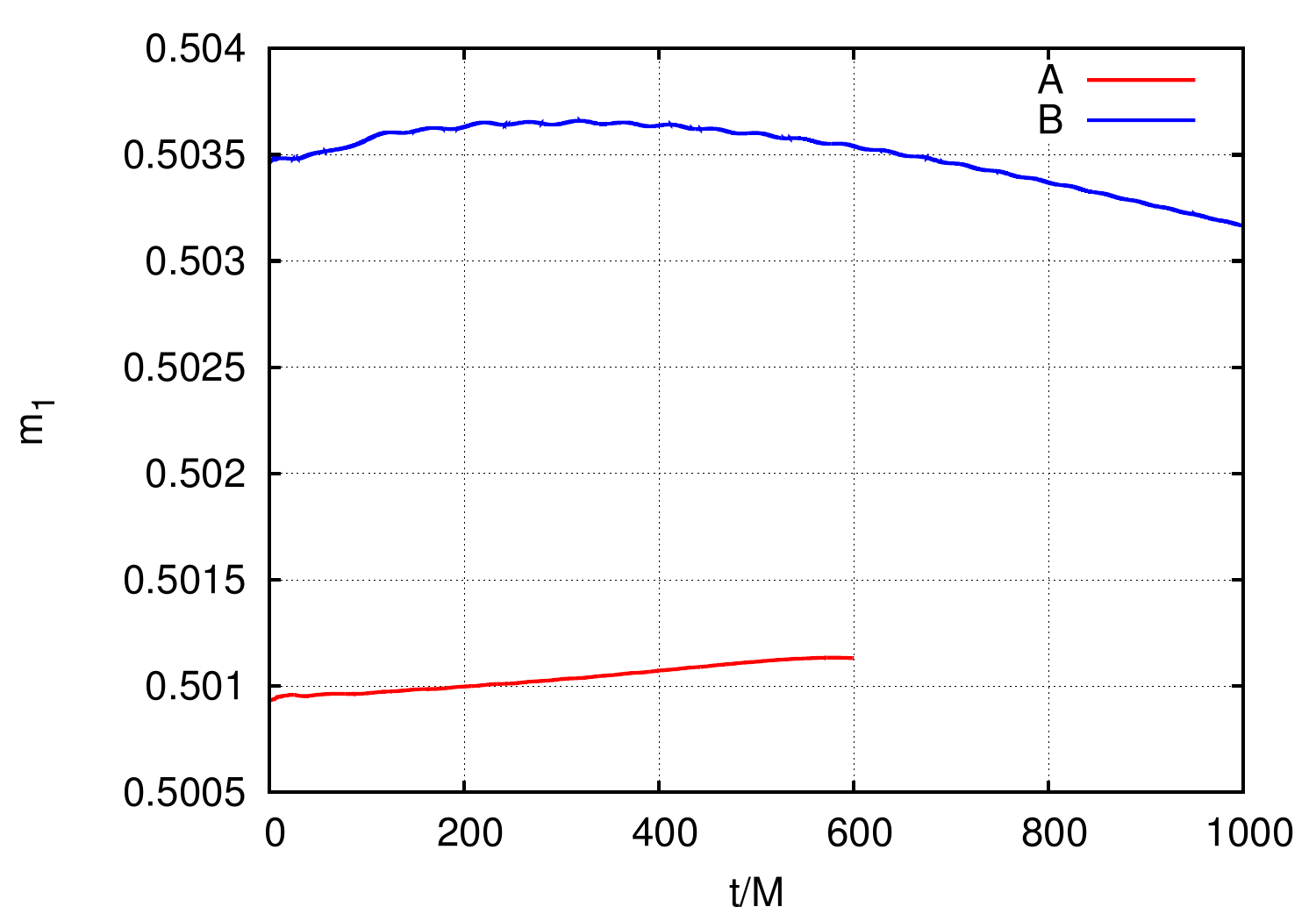}
  \includegraphics[angle=270,width=0.49\columnwidth]{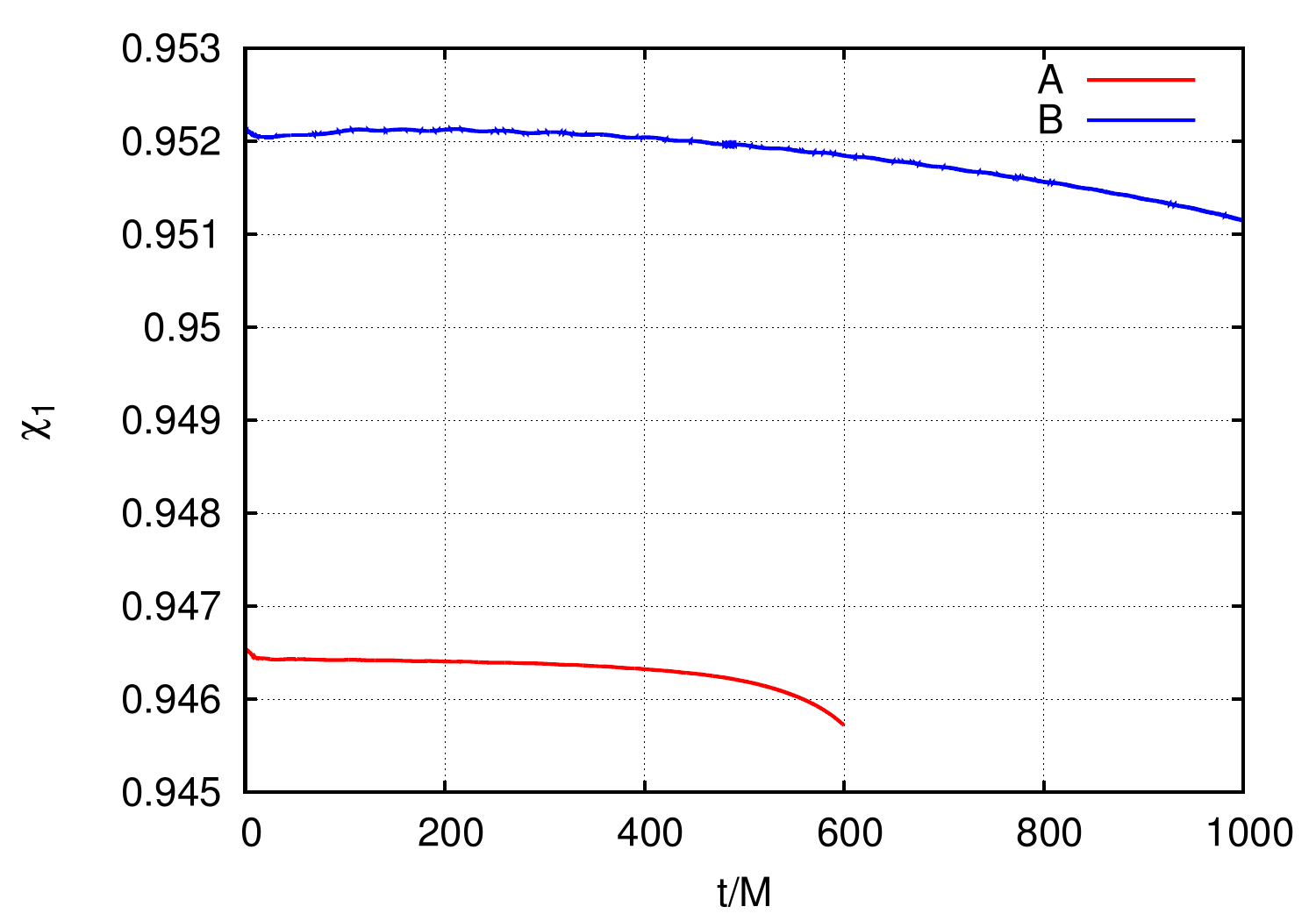}
  \caption{The evolution of the horizon mass and dimensionless spin of one of the individual black holes with $q = 1$ and $\chi = 0.95$. 
}
\label{fig:BK_chi95}
\end{figure}

An evolution of the mass and spin parameters for an equal-mass binary
with $\chi = 0.95$ is shown in Fig.~\ref{fig:BK_chi95}. The initial
and final parameters for this run are listed in the final rows of
Tables~\ref{tab:ch5_ID} and~\ref{tab:ch5_final}, respectively. To
check the validity of these results, we compare the final mass and
spin, calculated during the numerical simulation using the apparent
horizon and isolated horizon formalism, to an analytic fitting formula
(the last three columns in
Table~\ref{tab:ch5_final}~\cite{Healy:2014yta}). These analytic
fitting formulas were developed using a set of 37 aligned and
antialigned spinning unequal-mass systems, as well as an additional
38 simulations from the SXS catalog~\cite{SXS:catalog}, which included
aligned systems with spins up to $\chi = 0.98$.  The fitting formulas
give $M_{\text{rem}}/M = 0.8940$ and $\chi_{\text{rem}} = 0.9403$ for
case A, and $M_{\text{rem}}/M = 0.8933$ and $\chi_{\text{rem}} =
0.9415$ for case B, differing from our measured results by about
$0.01\%$ to $0.09\%$. We also compare the remnant values with the
SXS\#157 run which falls right in between our A and B runs.

\begin{figure}[!ht]
  \centering
  \includegraphics[angle=270,width=0.8\columnwidth]{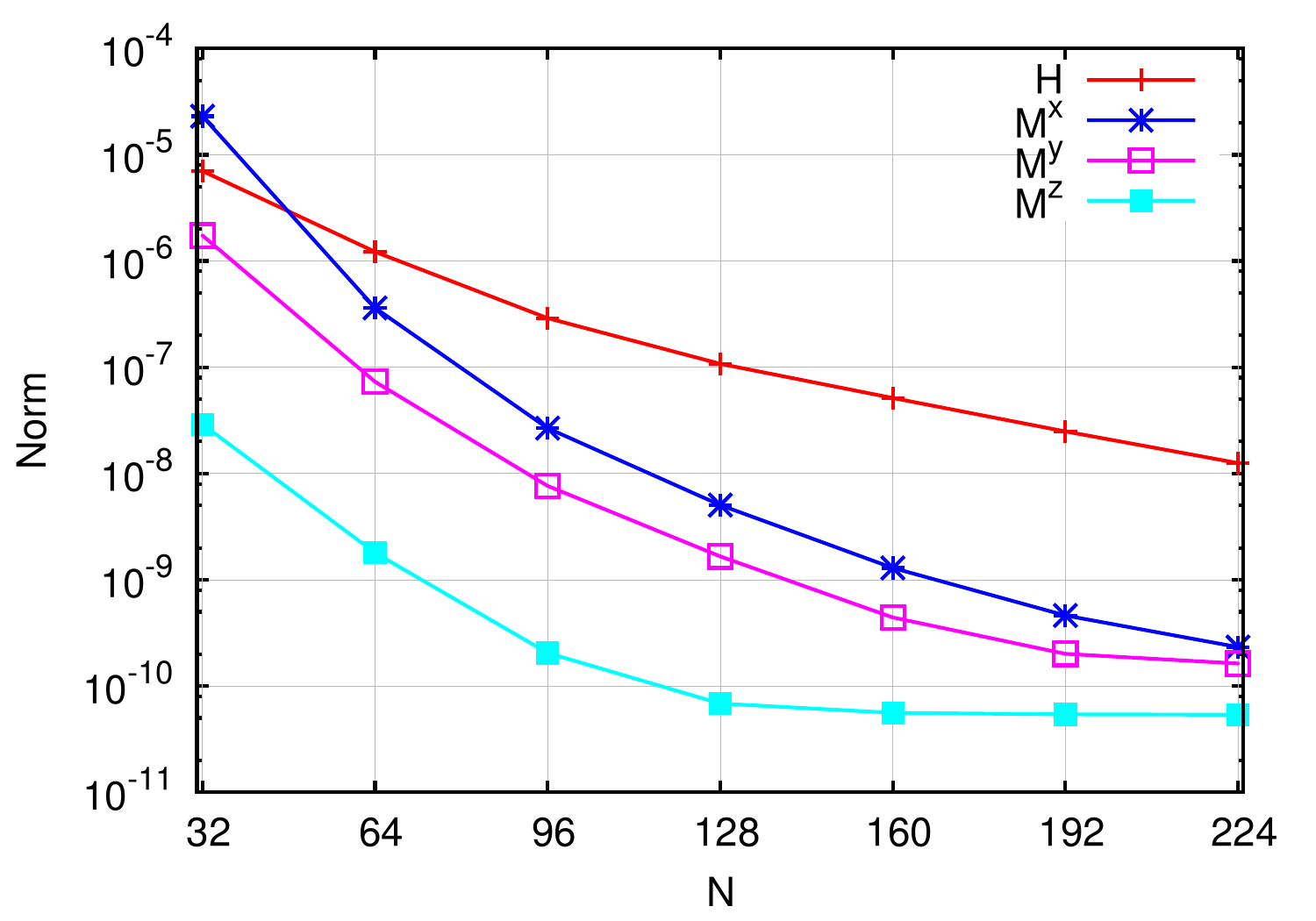}
  \caption{Convergence of the residuals of the Hamiltonian and
  momentum constraints versus the number of collocation points $N$ for
a quasicircular binary in LES coordinates with $\chi=0.974$ in the UU configuration.}
\label{fig:LES_convergence}
\end{figure}

The \hispid data using LES coordinates is not limited to $\chi\leq
0.95$. To demonstrate this, we solved the initial data for a
quasicircular binary with spins $\chi=0.974$ and a separation of
$d=10M$. As shown in Fig.~\ref{fig:LES_convergence}, the constraints
converge to ${\cal O}(10^{-10})$ (the Hamiltonian is larger by a
factor of 100, but it is still converging essentially exponentially).

To summarize, 
in this section we have shown that we are able to implement puncture initial data for highly spinning, orbiting black-hole binaries by attenuated superposition of conformal Lorentz-boosted Kerr metrics. We modified the \textsc{TwoPunctures} thorn, in the \textsc{Cactus}/\textsc{Einstein Toolkit} framework, to solve the Hamiltonian and momentum constraint equations simultaneously for highly spinning black holes. We verified the validity of the data by showing convergence of the Hamiltonian and momentum constraint residuals with the number of collocation points in the spectral solver. We then showed, by evolving this data, that the radiation content of these initial data was much lower than the standard conformally flat choice at spins $\chi = 0.8$. This produced a more accurate and realistic computation of gravitational radiation waveforms. We go on to simulate a black-hole binary on quasicircular orbit with $\chi = 0.95$, beyond the Bowen-York limit. The mass and spin of the resulting remnant black hole agrees with the analytic fitting function estimates to between $0.01\%$ and $0.09\%$.

\section{Conclusions and Discussion}\label{sec:discussion}

In this paper we were able to implement puncture initial data for
highly spinning and highly boosted BHBs by the attenuated
superposition of conformal Kerr and Lorentz-boosted Schwarzschild metrics
and for the boosted Kerr case. We verified
the validity of the data by showing convergence of the Hamiltonian
and momentum constraint residuals with the number of collocation points
in the spectral solver.
We then showed, by evolving this data, that the spurious radiation content
of these initial data was much lower than the standard conformally
flat choice. This produced a more accurate and realistic computation
of gravitational radiation waveforms that we compare with both, the 
Bowen-York initial data evolution for lower spins ($\chi=0.90$) and the SXS's 
waveforms for higher spins with $\chi=0.95$. 
This represents the first \MP evolution
of highly spinning black holes beyond the conformally-flat ansatz
limit \cite{Lousto:2012es} of $\chi=0.935$.
These cleaner initial data allowed us to explore different choices of 
the \MP gauge (initial lapse and shift) in Appendix \ref{sec:gauge},
as well as alternative/additional
evolution variables as introduced by CCZ4 \cite{Alic:2011gg}.

This initial data implementation will allow for simulation of 
extremely boosted and highly spinning orbiting BHBs to explore the corners 
of the BHB parameter space, in a regime 
of theoretical and astrophysical interest.
The high boost case was recently used in \cite{Healy:2015mla} to
study the head-on high-energy collision of nonspinning black holes.
It also allows for revisiting some of the most interesting
spin dynamic effects in BHBs, such as the
hangup~\cite{Campanelli:2006uy}, 
flip-flops~\cite{Lousto:2014ida,Lousto:2015uwa,Lousto:2016nlp}, 
and large recoils~\cite{Lousto:2011kp,Healy:2014yta},
as well as extreme BHB collisions~\cite{Healy:2015mla}.

\acknowledgments 
The authors thank M.\ Campanelli, G.~Lovelace, 
and M.\ Zilh\~ao for discussions.

The authors gratefully acknowledge the NSF for financial support from Grants
 NSF Grants  No. PHY-1607520, No. ACI-1550436 , No. AST-1516150, No. ACI-1516125, No. PHY-1305730, No.
 PHY-1212426, No. PHY-1229173, No. AST-1028087,
 No. OCI-0725070 (PRAC subcontract  2077-01077-26), No. OCI-0832606.
Computational resources were provided by XSEDE allocation
TG-PHY060027N, and by NewHorizons and BlueSky Clusters 
at Rochester Institute of Technology, which were supported
by NSF grant No. PHY-0722703, DMS-0820923, AST-1028087, and PHY-1229173.
This work was supported in part by the 2013--2014 Astrophysical Sciences and Technology Graduate Student Fellowship,
funded in part by the New York Space Grant Consortium, administered by Cornell University.

\appendix

\section{Gauge Conditions}\label{sec:gauge}

The original \MP breakthrough formulation 
\cite{Campanelli:2005dd} remains widely used, and produces reliable BHB evolutions, as well as multi-BH systems~\cite{Lousto:2007rj}.  It also functions in the presence of matter, as in neutron-star mergers~\cite{Anderson:2007kz,Baiotti:2008ra,Kiuchi:2009jt}.
The choice of the gauges, i.e., Eq.\@~(\ref{eq:gauge}) plays a 
crucial role in stabilizing the numerical evolutions.
There is still a range of possibilities for choosing the specific form
of the gauges. While preserving the numerical
stability properties one would like to improve the accuracy 
of the simulation for a given resolution and grid structure.

Some questions about the accuracy and convergence
of the \MP method have been raised in
\cite{Zlochower:2012fk}.
Recently, Etienne {\it et al.}~\cite{Etienne:2014tia} studied
how modifications to the lapse evolution
can 
ameliorate the numerical errors that lead to
poor waveform convergence.

Here we study other choices for the initial lapse and its time evolution
to control and improve the accuracy of the numerical results for
highly spinning BHs and 
relativistic collisions of BHs generating
large amplitude gauge waves.

The Bona-Mass\`o gauge condition for the lapse evolution is~\cite{Bona94b}
\begin{equation}\label{eq:lapse}
(\partial_t - \beta^i \partial_i) \alpha = - \alpha^2\,f(\alpha)\,K,
\end{equation}
where in the original \MPA $f(\alpha)=2/\alpha$.
We also consider $f(\alpha)=1/\alpha$, with gauge speed equal to 1, 
and $f(\alpha)=8/(3\alpha(3-\alpha))$, with approximate shock avoiding
properties~\cite{Alcubierre02b}.

For the initial lapse we use
either $\alpha(t=0) =1/\psi_{\text{BL}}^{2}$
\cite{Campanelli:2005dd}  
or $\alpha(t=0)=2/(1+\psi_{\text{full}}^{4})$ 
\cite{Campanelli:2006uy} which behave like $\sim r^2$ and $\sim r^4$,
respectively near
the puncture, and have the form of the Schwarzschild lapse
in isotropic coordinates at large $r$. Here, $\psi_{\text{full}} = [\det(\gamma_{i j})]^{1/12}$.
In addition, we investigate the form
$\alpha(t=0)=1/(2\psi_{\text{BL}}-1)$.  This goes like $\sim r$ near
the puncture, similar to the `trumpet'-like behavior observed
at later evolution times when gauges settle down to a quasistationary
behavior \cite{Brugmann:2009gc}, while conserving the form of the Schwarzschild
lapse in isotropic coordinates at asymptotically large $r$.

\subsubsection{Highly spinning Black-Holes}\label{app:HSBH}

\begin{figure}
  \includegraphics[angle=270,width=0.9\columnwidth]{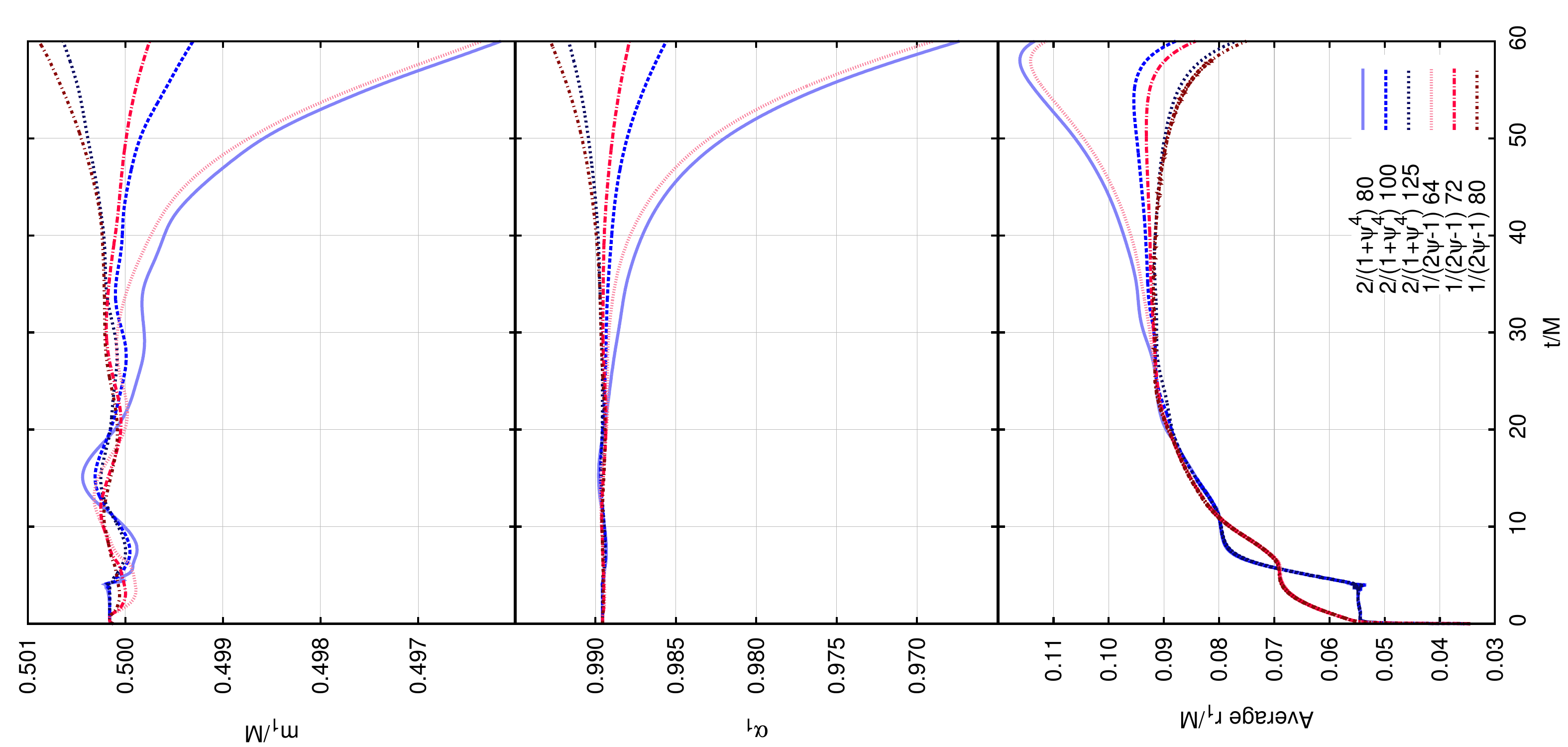}
  \caption{The effects of the choice of the initial lapse on the individual
BH masses (top panel), spins (middle panel), and coordinate radii
(bottom panel). The benefits of the new
initial lapse are evident since they follow the higher resolution behavior
with many fewer grid points by $(80/125)^3$.  For all runs $f(\alpha)=2/\alpha$.}
\label{fig:trumpinitc99}
\end{figure}

Notably, the simple choice of the initial lapse $\alpha_0=1/(2\psi_{\text{BL}}-1)$
has advantages over the other choices studied for the entire
evolution by
providing increased accuracy and computational efficiency.
Here, we display the results of the evolution from rest of BHBs with
intrinsic spin $\chi=0.99$. Figure\@~\ref{fig:trumpinitc99}
shows that we can achieve comparably accurate 
results with many fewer grid points.
The curves follow closely to each other, but with the new lapse we use
80 points per dimension compared to the 125 needed with the original 
initial lapse. This provides a speedup factor of $(125/80)^4\sim6$.

\begin{figure}
  \includegraphics[angle=270,width=0.9\columnwidth]{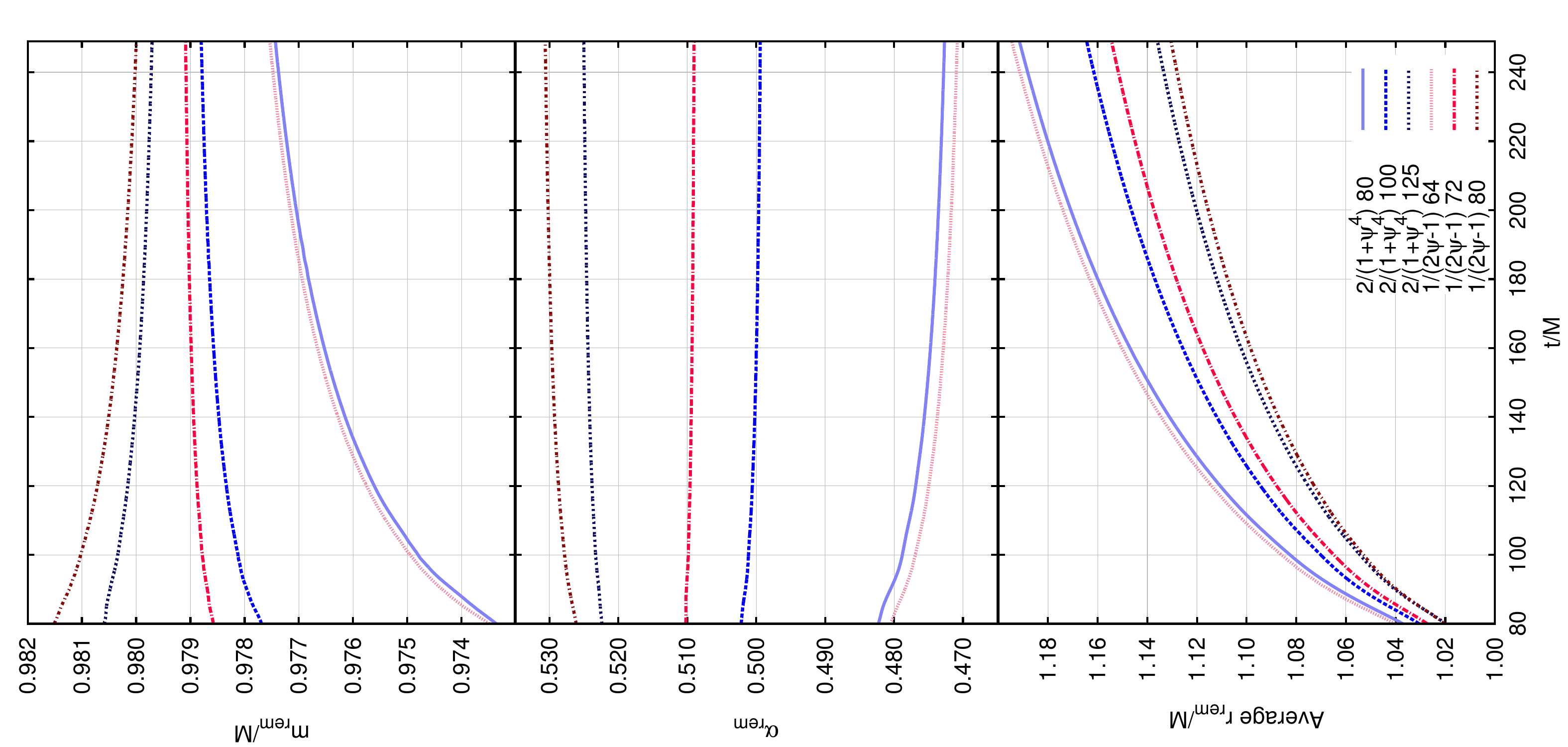}
  \caption{The effects of the choice of the initial lapse on the final
BH mass (top panel), spin (middle panel), and coordinate radius (bottom
panel). The benefits of the new
initial lapse are evident since they follow the higher resolution behavior
with many fewer grid points. For all runs $f(\alpha)=2/\alpha$.}
\label{fig:trumpfinc99}
\end{figure}

The improvement of the new initial lapse also translates into a more
accurate description of the final remnant BH, as shown in
Fig.\@~\ref{fig:trumpfinc99}. Note that even at lower resolutions we
observe a similar gain.

We interpret these results as indicating that a better choice of the initial
lapse leads to a better
coordinate evolution. The intensities of the initial gauge waves are reduced, thus allowing
a better distribution of grid points, resulting in a more efficient
numerical computation. See for instance the horizon coordinate radius
evolution in Figs.\@~\ref{fig:trumpinitc99} and \ref{fig:trumpfinc99}.

\begin{figure}
  \includegraphics[angle=270,width=0.9\columnwidth]{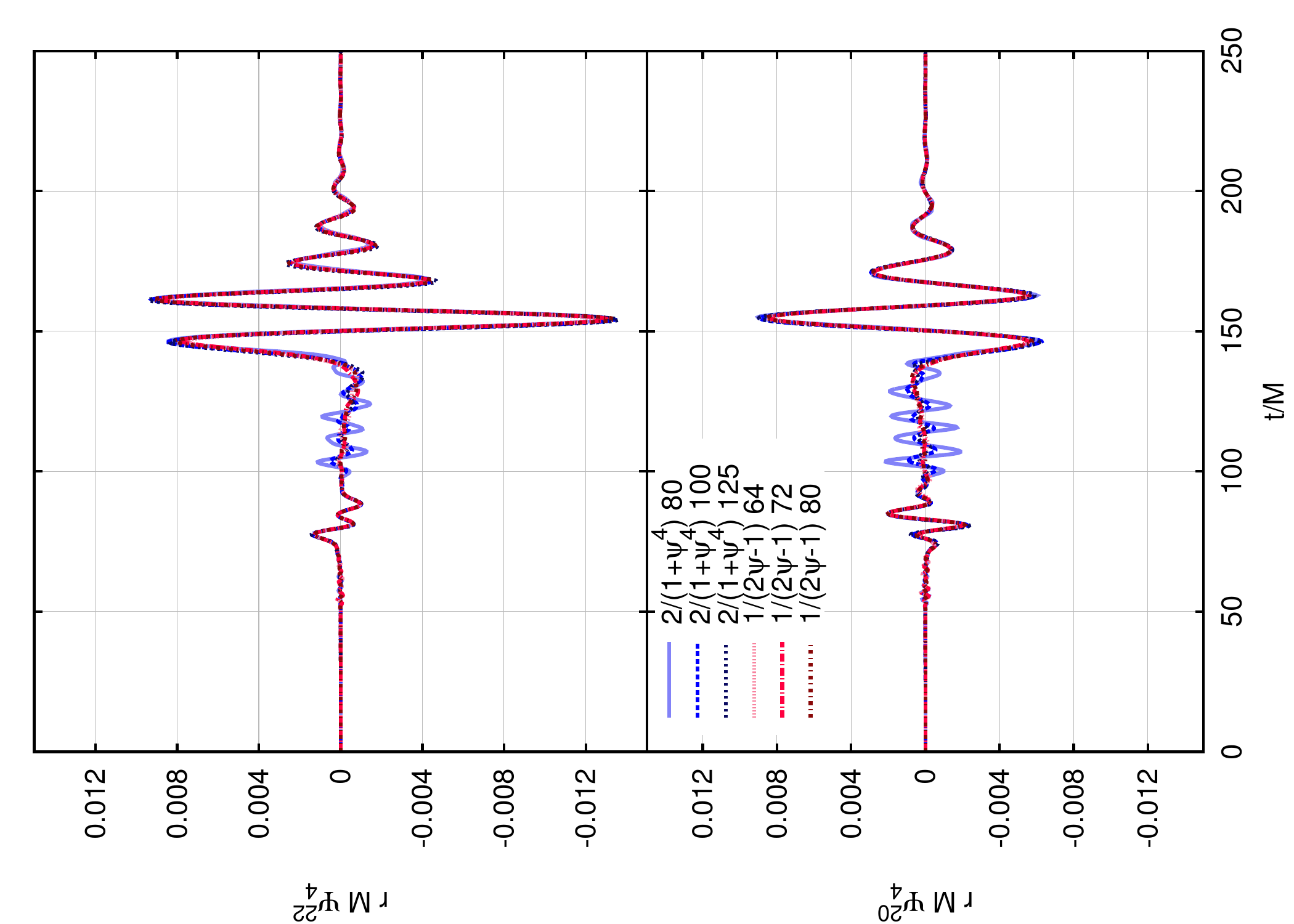}
  \caption{The effects of the choice of the initial lapse on the 
waveforms. The benefits of the new
initial lapse are evident since they show much less initial noise
before the merger.}
\label{fig:trumpre220c99}
\end{figure}

Figure\@~\ref{fig:trumpre220c99} displays the effects of the initial lapse
on the waveform. We see the notable reduction of the unphysical oscillations
premerger while reproducing accurately the physical merger waveform
for the dominant modes $(\ell,m) = (2,0)$ and $(\ell,m) = (2,2)$.
Note that this reduction of the errors due to improved gauge choices is in 
addition to and independent from the reduction of the initial burst of
radiation (with respect to the BY data) that has a physical content, 
despite being an undesirable effect.

Figures.\@~\ref{fig:trumpinitc99}, \ref{fig:trumpfinc99}, and 
\ref{fig:trumpre220c99} show three different resolutions for the 
waveforms and horizon quantities for the highly spinning $\chi=0.99$ case. 
 The convergence order was calculated for each of these quantities.  
For the individual BH spin, irreducible mass, horizon mass, and 
dimensionless spin, we find average convergence orders of 7.6, 6.2, 8.2, 
and 8.2, respectively. The eighth-order convergence is expected if the
errors are dominated by the spatial finite difference stencil.
  The same quantities for the final remnant BH 
have convergence orders between 3.3 and 4.3. We expect a fourth-order
convergence if the errors are dominated by the time integration.
For the amplitude and phase 
of the 2,2 mode, we find a convergence order between 3 and 4 
after the BHs merge. This is consistent with the time integration errors.

\subsubsection{Quasicircular orbits}\label{app:QC}

Here, we study the effects of lapse evolution choices on the case
study of equal mass, nonspinning, orbiting BHBs.
The Lorentz boost initial data has a lower radiation content than the
boost BY data and allows us to see more clearly the effects of
the initial choice and evolution of the lapse.  In this section, all 
runs studied are at the medium resolution, labeled `$100$' in Figs.
\ref{fig:trumpinitc99}--\ref{fig:trumpre220c99}.

\begin{figure}
  \includegraphics[angle=270,width=\columnwidth]{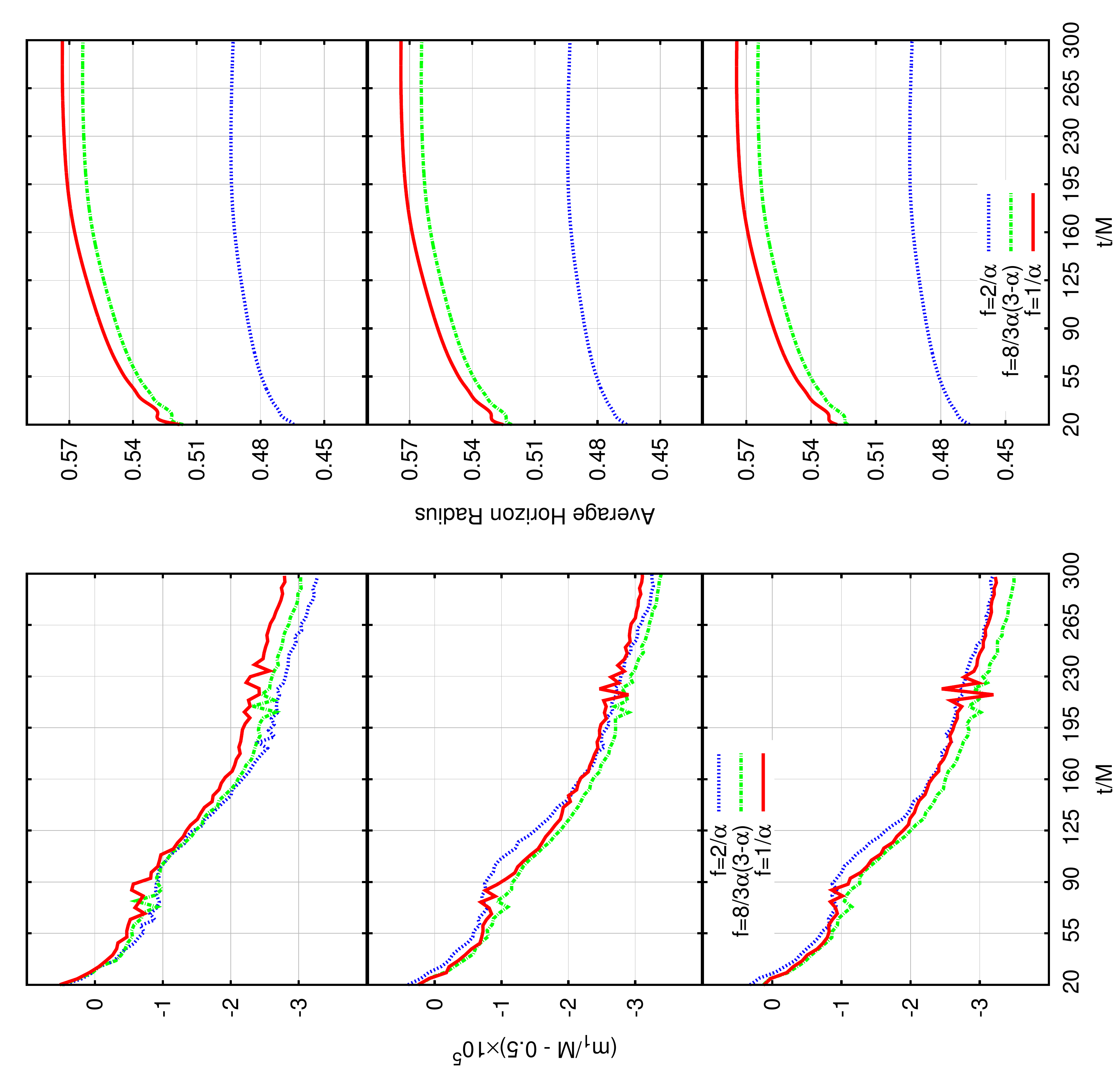}
  \caption{Individual BH horizon masses (left column) and coordinate
radii of the horizons (right column) versus time
for different evolution functions $f(\alpha)$
for the lapse. The initial lapse $\alpha_0=2/(1+\psi_{\text{full}}^4)$ (top row),
$\alpha_0=1/\psi_{\text{BL}}^2$ (middle row), $\alpha_0=1/(2\psi_{\text{BL}}-1)$ (bottom row) 
.}
  \label{fig:MHRH}
\end{figure}

Figure\@~\ref{fig:MHRH} displays the effects of gauge versus resolution
on physical quantities like
the horizon mass (left column) and horizon radius (right column). We expect the horizon mass to be essentially conserved
during the orbital period up to merger. We can see that this physical
observable varies very little with different gauge choices.
On the other hand, we observe that the coordinate
radius varies with the evolution of the lapse choice, but not as much
with the initial lapse. After a sudden growth, typical of a gauge settling, 
the horizon radius reaches a constant value.
 The original \MP
choice, $f(\alpha)=2/\alpha$, keeps the value of the horizon coordinate
closer to its original value, which could be beneficial for setting up
the initial mesh refinement levels.

\begin{figure}
  \includegraphics[angle=270,width=\columnwidth]{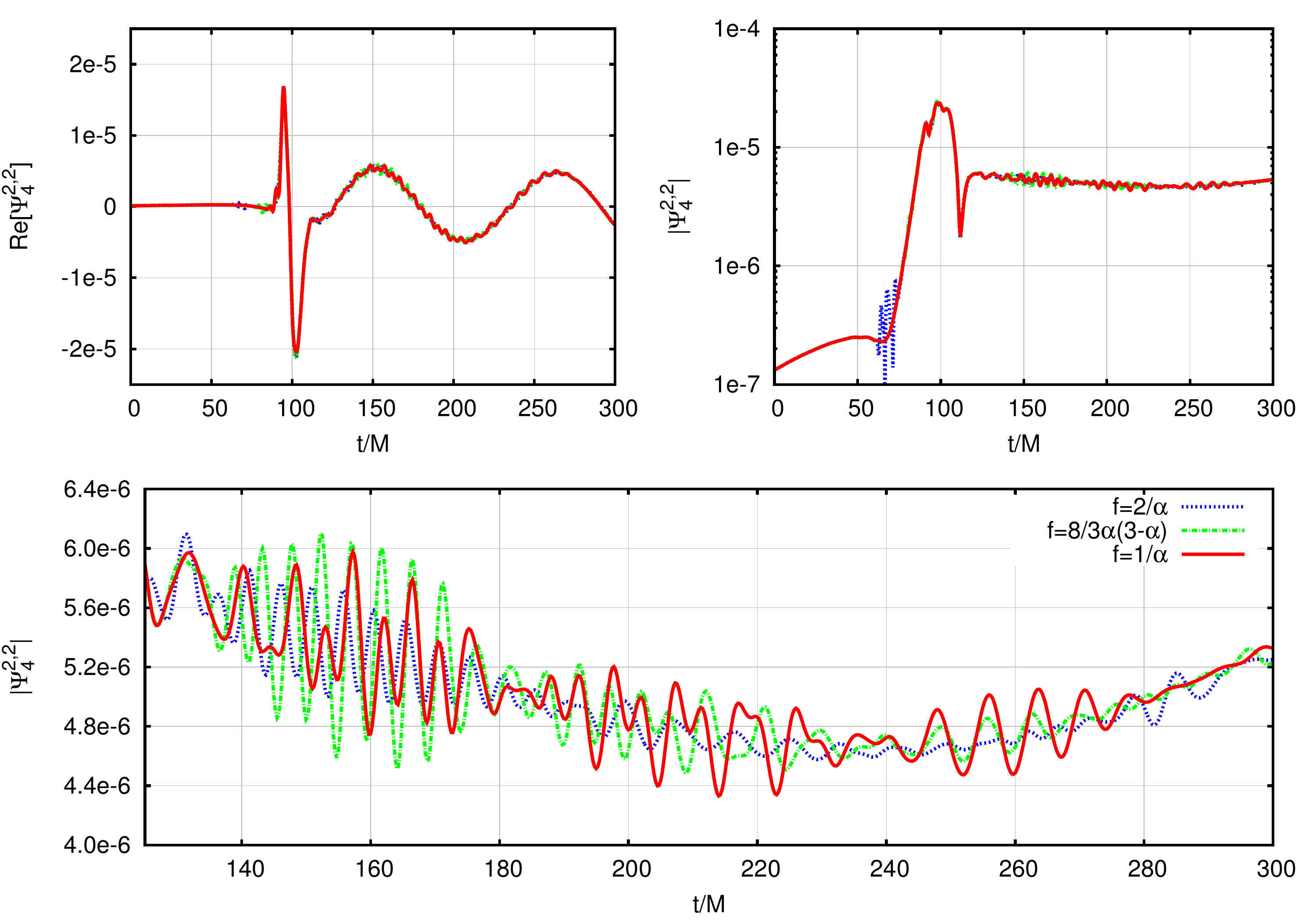}
  \caption{Waveforms extracted at an observer location $r=90M$. Real part
of $\Psi_4$ (upper left panel) and the amplitude of those waveforms
(upper right panel).
(Lower panel) An enlargement  of the amplitude oscillations
for different evolution functions $f(\alpha)$ for the lapse. 
The initial lapse here is $\alpha_0=2/(1+\psi_{\text{BL}}^4)$.}
  \label{fig:WAZ-4}
\end{figure}

Figure\@~\ref{fig:WAZ-4}
displays the waveform as seen by an observer at $r=90M$ from
the sources for different evolution functions $f(\alpha)$ for the lapse. 
The initial lapse here is $\alpha_0=2/(1+\psi_{\text{full}}^4)$. While physical
quantities like the waveform and its amplitude are essentially independent
of the gauge choices, numerical errors, which produce the high-frequency noise, are not. The bottom panel of the figure
shows a close-up view of the amplitude during the postinitial pulse period.
We observe that overall the choice $f(\alpha)=2/\alpha$ produces
a lower amplitude of this high-frequency noise.

\begin{figure}
  \includegraphics[angle=270,width=\columnwidth]{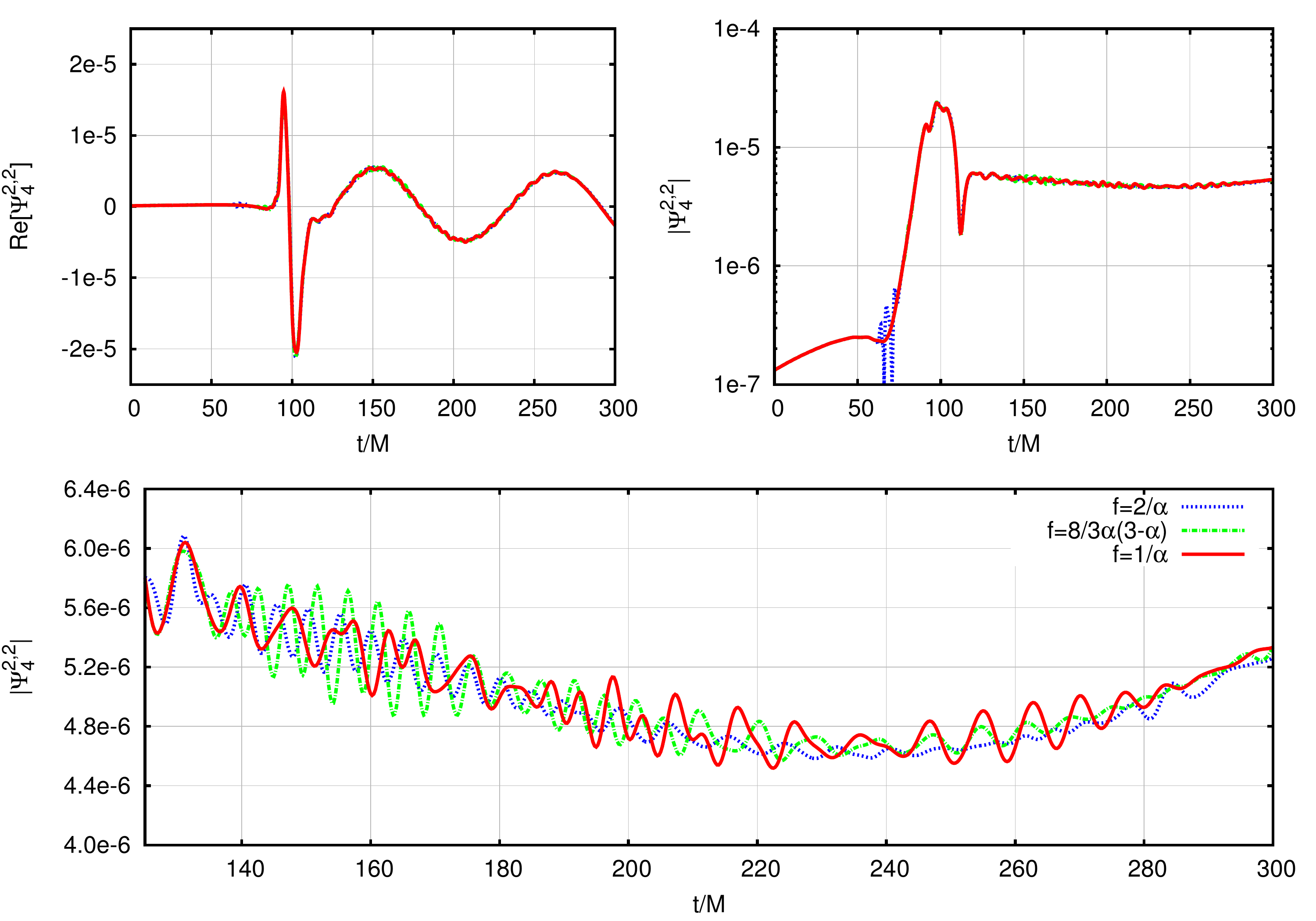}
  \caption{Waveforms extracted at an observer location $r=90M$. Real part
of $\Psi_4$ (upper left panel) and the amplitude of those waveforms
(upper right pane).
The bottom panel an enlargement of the amplitude oscillations
for different evolution functions $f(\alpha)$ for the lapse. 
The initial lapse implemented here is $\alpha_0=1/\psi_{\text{BL}}^2$.}
  \label{fig:WAZ-2}
\end{figure}

\begin{figure}
  \includegraphics[angle=270,width=\columnwidth]{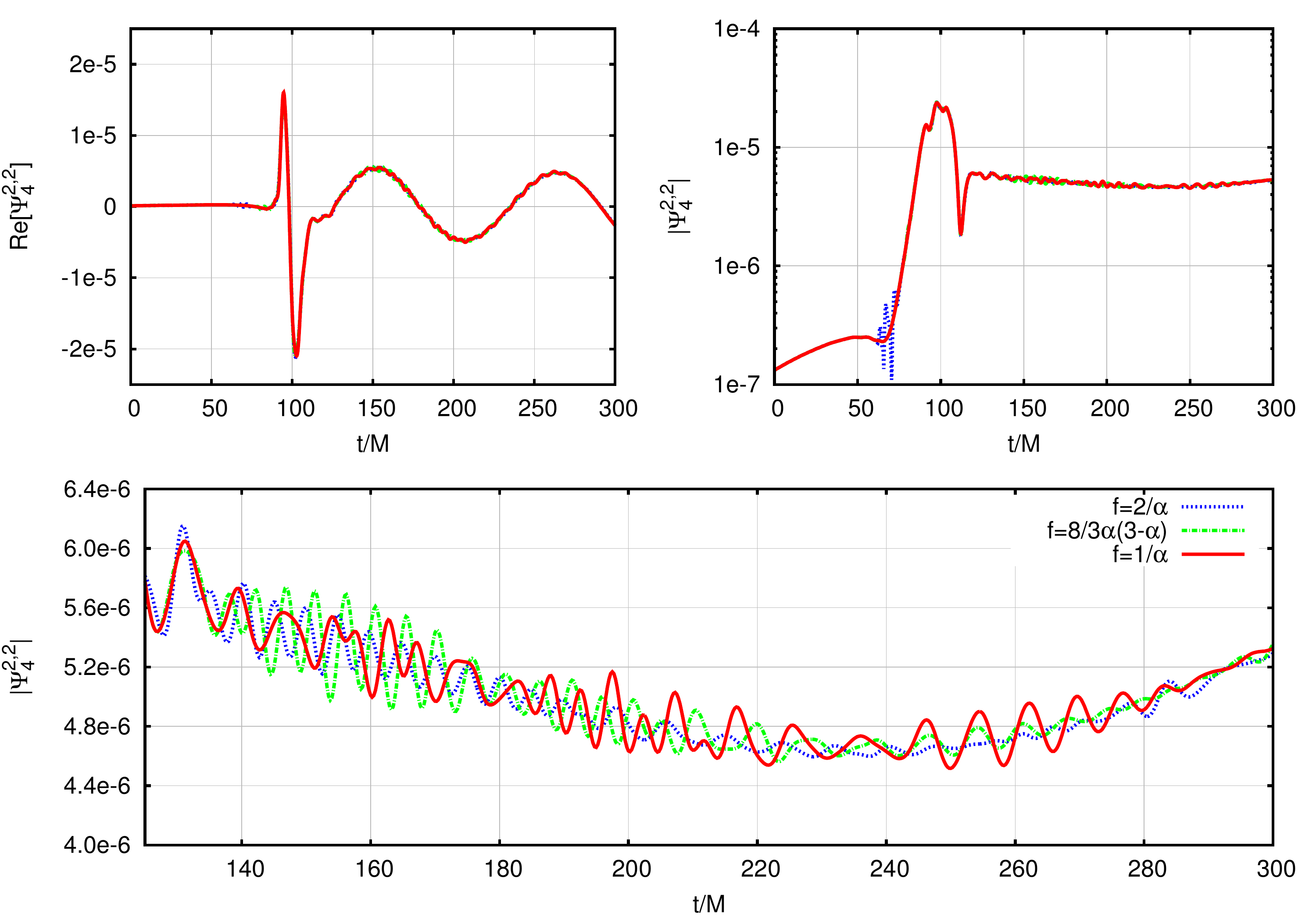}
  \caption{Waveforms extracted at an observer location $r=90M$. Real part
of $\Psi_4$ (upper left panel) and the amplitude of those waveforms
(upper right panel).
The lower panel shows an enlargement of the amplitude oscillations
for different evolution functions $f(\alpha)$ for the lapse. 
The initial lapse here is $\alpha_0=1/(2\psi_{\text{BL}}-1)$.}
  \label{fig:WAZ-1}
\end{figure}

Figures\@~\ref{fig:WAZ-2} and \ref{fig:WAZ-1} display a similar behavior
for the waveforms, but their close-up view of the noise shows a smaller 
amplitude, which suggests that the choice of the initial lapse
$\alpha_0=1/\psi_{\text{BL}}^2$ or $\alpha_0=1/(2\psi_{\text{BL}}-1)$
leads to smaller amplitude gauge waves.

Since the \MPA is a free evolution of the 
general relativistic field equations, a very important method 
to monitor its accuracy is to verify the satisfaction of the 
Hamiltonian and momentum constraints. We also monitor the 
BSSN constraints, which are on the order of $10^{-7}$ throughout 
the duration of the evolution.

\begin{figure}
  \includegraphics[angle=270,width=\columnwidth]{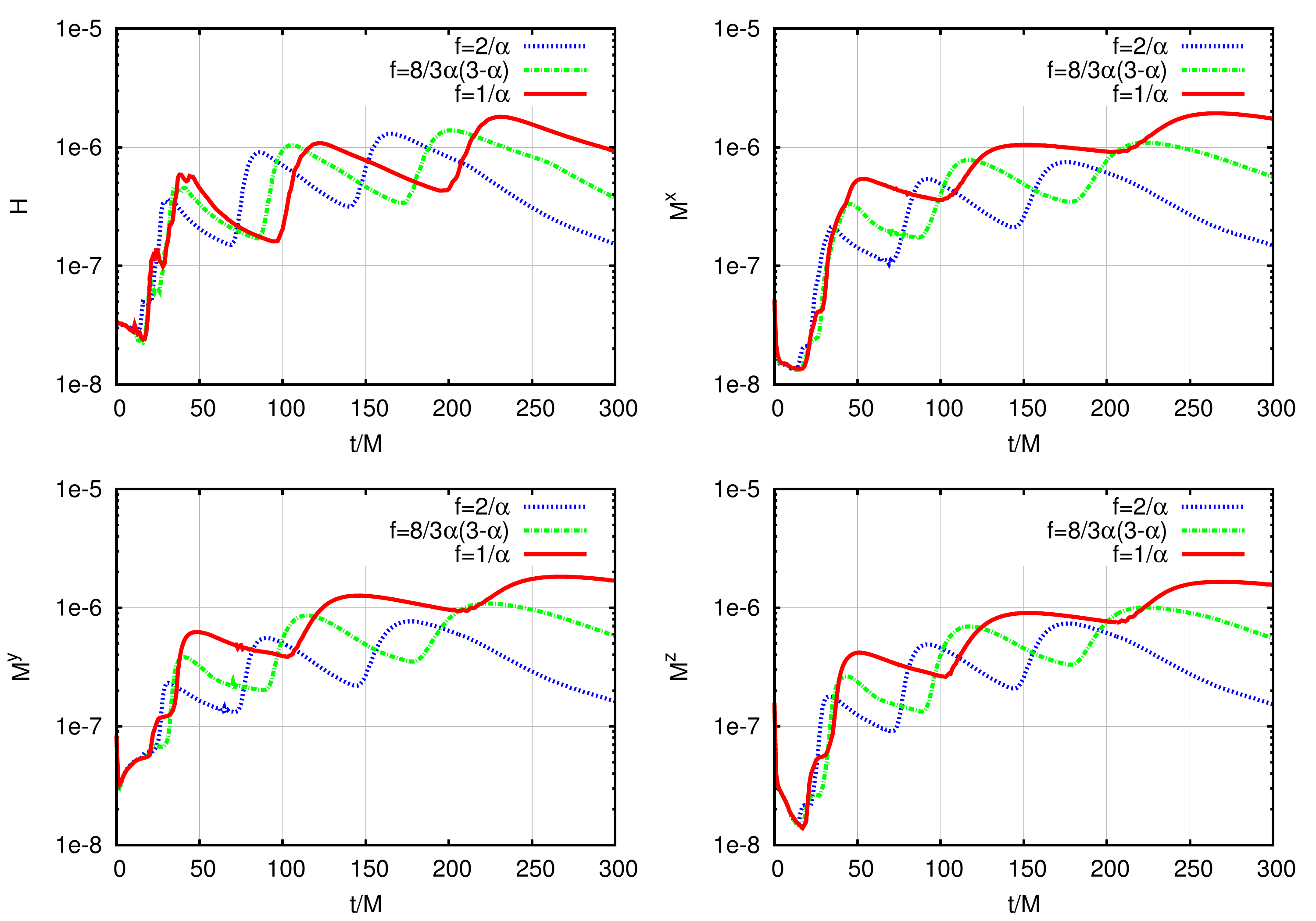}
  \caption{$L^2$-norms of the violations of the Hamiltonian and three components
of the momentum constraints versus time
for different evolution functions $f(\alpha)$ for the lapse. 
The initial lapse here is $\alpha_0=2/(1+\psi_{\text{full}}^4)$.}
  \label{fig:HM-4}
\end{figure}

\begin{figure}
  \includegraphics[angle=270,width=\columnwidth]{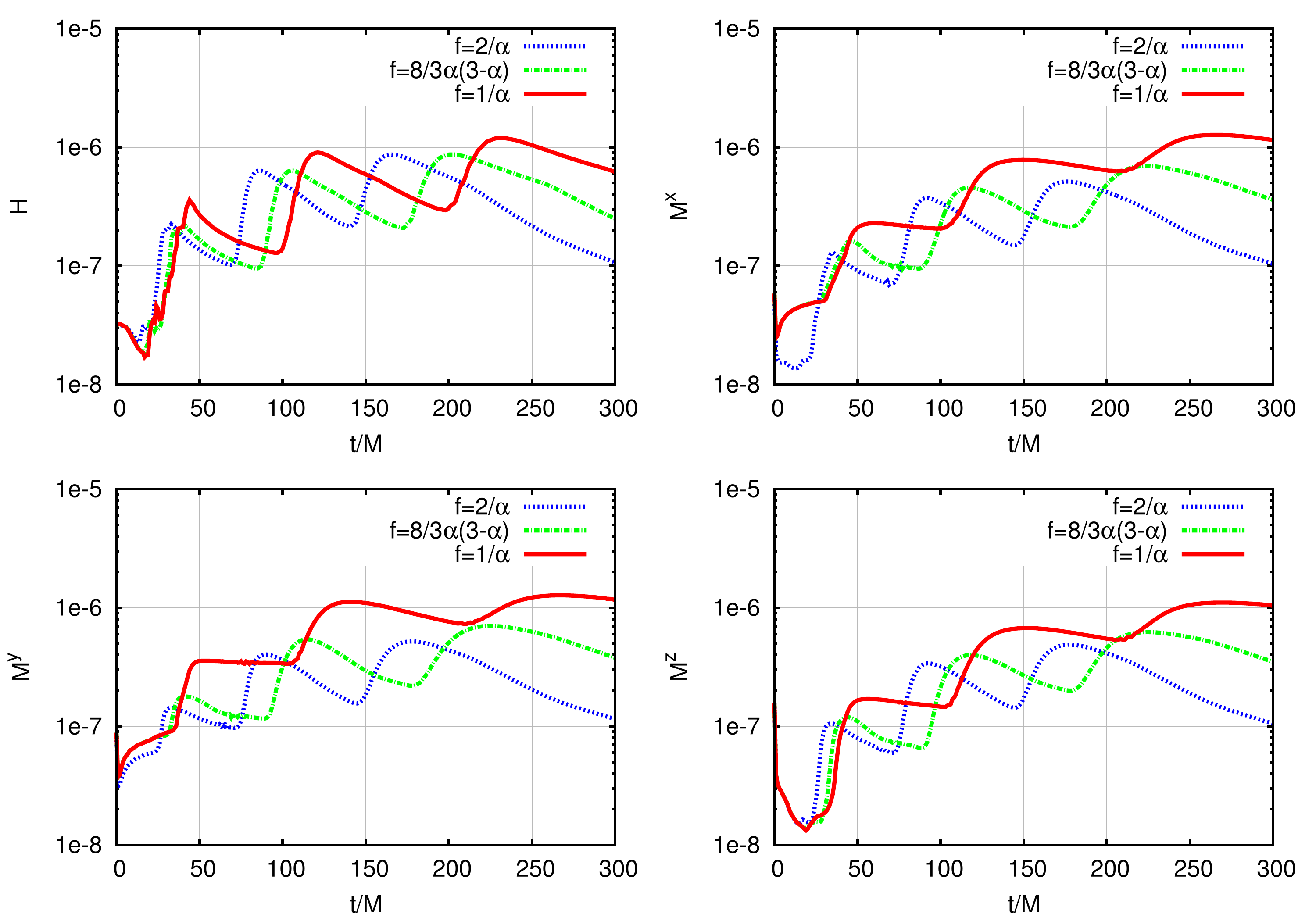}
  \caption{$L^2$-norms of the violations of the Hamiltonian and three components
of the momentum constraints versus time
for different evolution functions $f(\alpha)$ for the lapse. 
The initial lapse here is $\alpha_0=1/\psi_{\text{BL}}^2$.}
  \label{fig:HM-2}
\end{figure}

\begin{figure}
  \includegraphics[angle=270,width=\columnwidth]{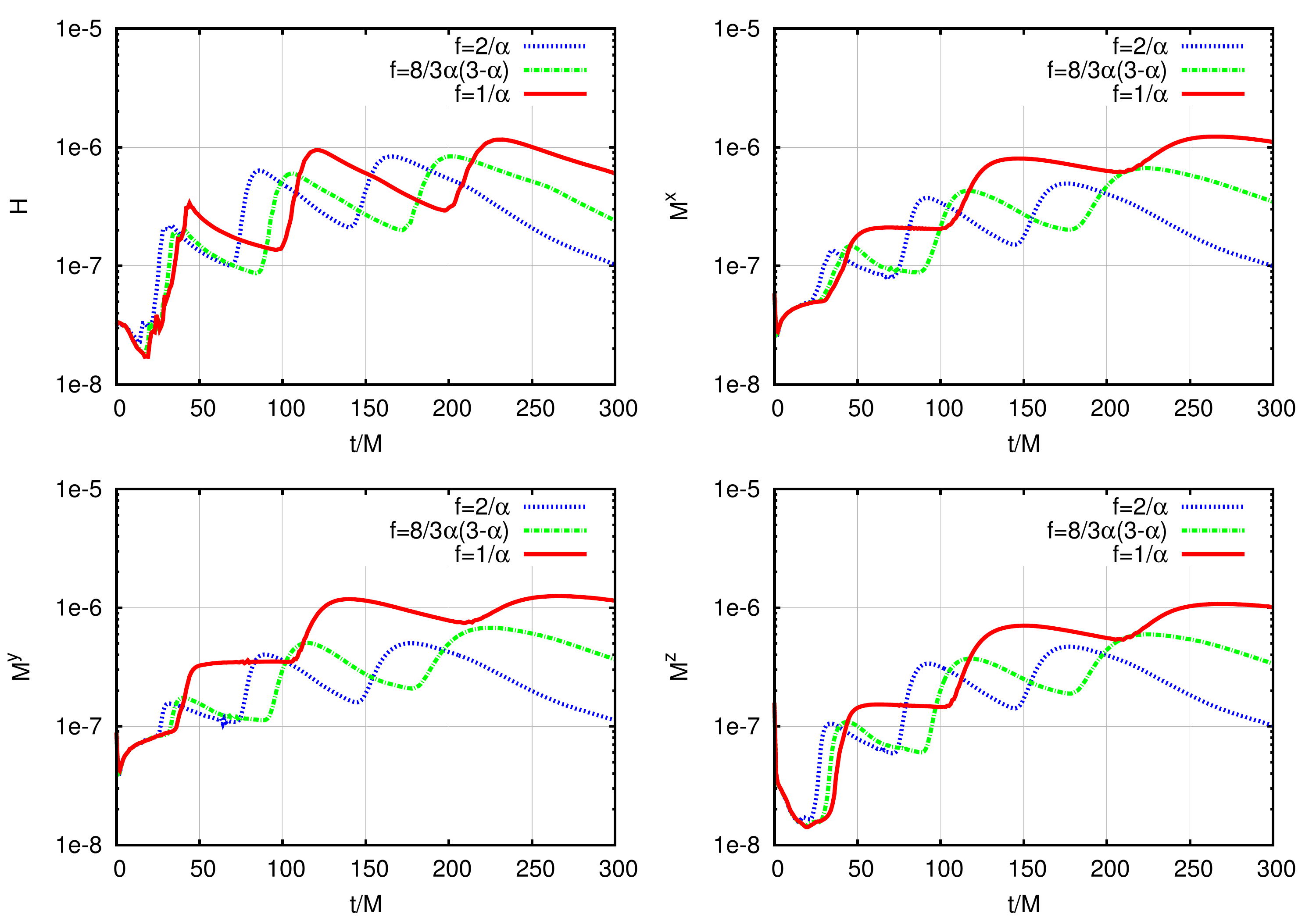}
  \caption{$L^2$-norms of the violations of the Hamiltonian and three components
of the momentum constraints versus time
for different evolution functions $f(\alpha)$ for the lapse. 
The initial lapse here is $\alpha_0=1/(2\psi_{\text{BL}}-1)$.}
  \label{fig:HM-1}
\end{figure}

Figures\@~\ref{fig:HM-4}--\ref{fig:HM-1} display the $L^2$ norm of the
nonvanishing values of the Hamiltonian and momentum
components of the constraints. We observe that the propagation
of errors travel at different speeds, associated with the gauge
velocities $\sqrt{2}$, $\sqrt{4/3}$, and $1$ for
$f(\alpha)=2/\alpha$, $8/(3\alpha(3-\alpha))$, and $1/\alpha$, respectively.
We also observe slightly larger violations for the choice 
$f(\alpha)=1/\alpha$, and $\alpha_0=2/(1+\psi_{\text{BL}}^4)$.

We thus conclude that while all three evolution choices for the lapse are
viable to evolve typical BHB simulations, the original \MP choice
$f(\alpha)=2/\alpha$ and the initial lapse 
$\alpha_0=1/\psi_{\text{BL}}^2$ or $\alpha_0=1/(2\psi_{\text{BL}}-1)$ are somewhat preferred. This study suggests there might be even more optimal choices
of $\alpha_0$ and $f(\alpha)$, as well as shift evolution gauge
conditions.
We also note that in the independent study of
Ref.\@~\cite{Etienne:2014tia},
a higher gauge velocity is preferred for the early stage of evolution.

\subsubsection{Relativistic head-on collisions}\label{app:HOC}

Since we observe a notable benefit on using the initial lapse
$\alpha_0=1/(2\psi_{\text{BL}}-1)$ in evolutions of highly spinning BHs, we
would like to explore their effect on another extreme configuration:
high-energy relativistic collisions of BHs.
The collisions were studied in
Refs.\@~\cite{Sperhake:2008ga,Shibata:2008rq,Sperhake:2009jz,Sperhake:2012me} 
with regard to potential applications
to collider-generated mini BHs. Here we will consider them
as a test case for comparing 
different gauge conditions.

\begin{figure}[!ht]
  \includegraphics[angle=270,width=0.49\columnwidth]{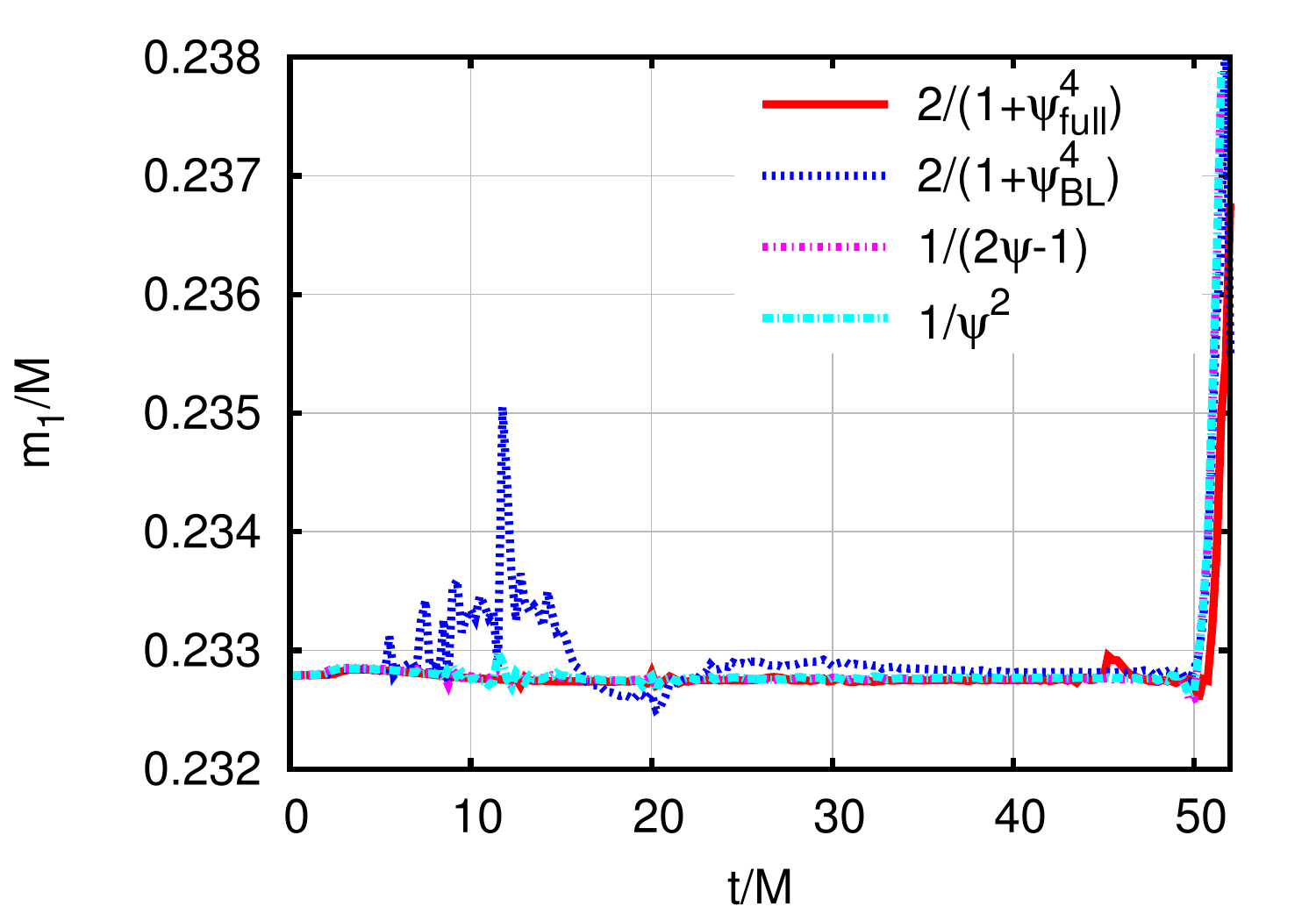}
  \includegraphics[angle=270,width=0.49\columnwidth]{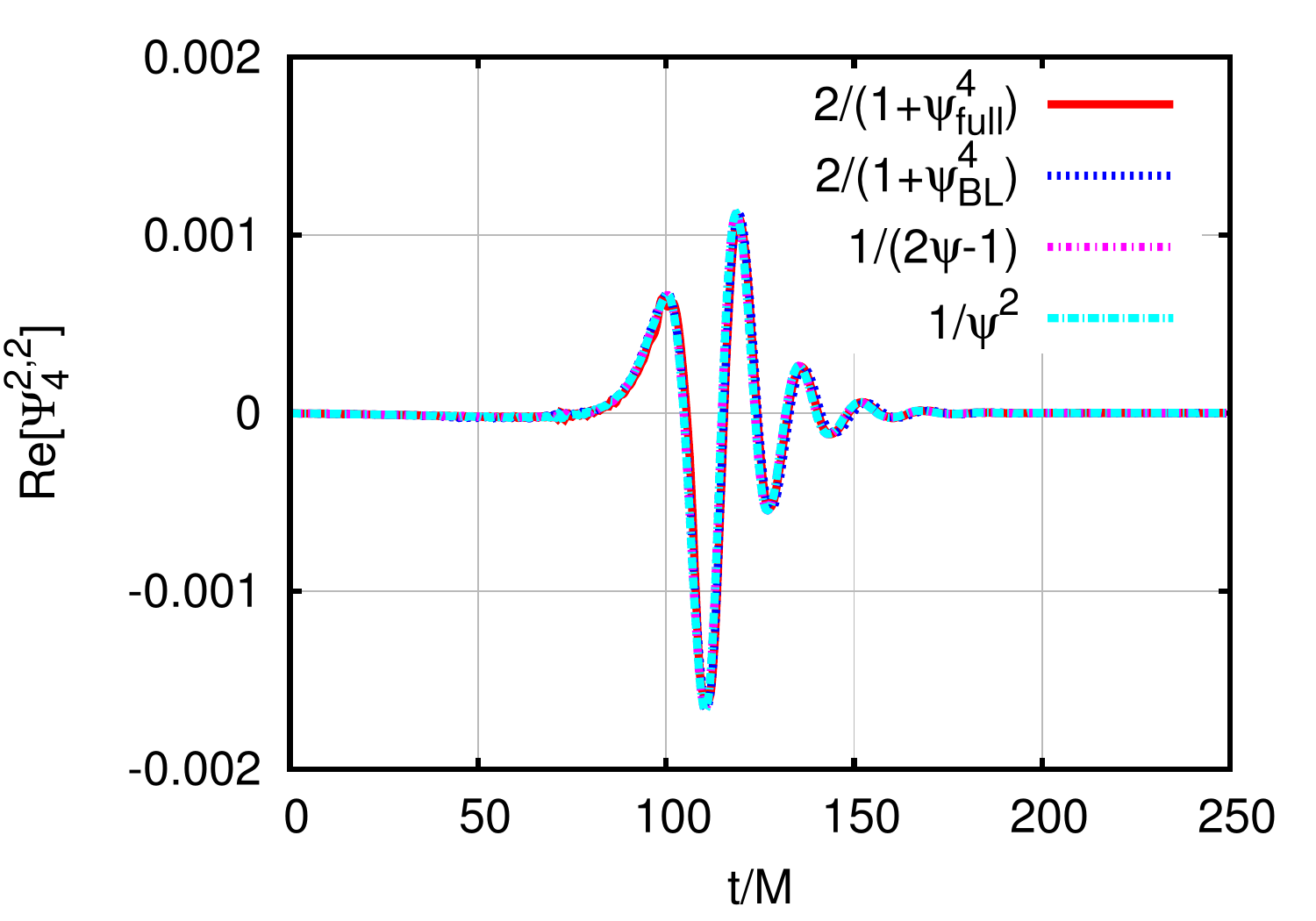}
  \caption{Horizon mass of the boosted BH with $P_x/m_{\text{H}}=\pm 2$ and
waveform after collision for different choices of the initial lapse and
evolution $f(\alpha)=2/\alpha$.
}\label{fig:HOC-MW}
\end{figure}

In Fig.\@~\ref{fig:HOC-MW} we use physical observables such as the
individual horizon masses and the gravitational radiation waveforms
as indicators of the numerical accuracy of the evolutions. We observe
that the initial lapse $\alpha_0=1/(2\psi_{\text{BL}}-1)$ gives the
best behavior for the horizons mass (i.e., the most constant)
and a waveform with reduced noise.

\begin{figure}[!ht]
  \includegraphics[angle=270,width=\columnwidth]{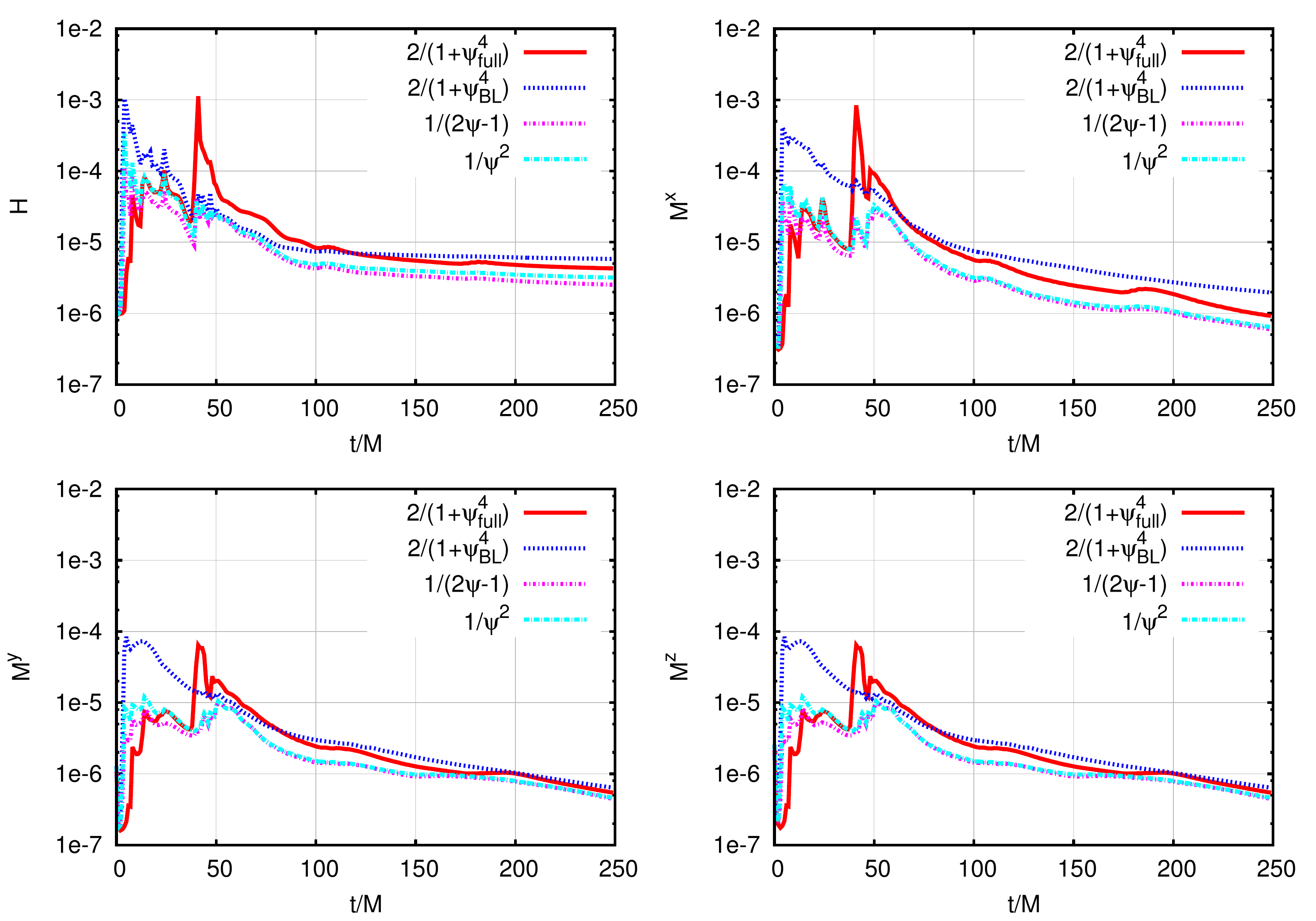}
  \caption{Hamiltonian and momentum constraints during the free evolution
for different choices of the initial lapse and
evolution $f(\alpha)=2/\alpha$.
}\label{fig:HOC-constraints}
\end{figure}

The preferred behavior of the initial lapse 
$\alpha_0=1/(2\psi_{\text{BL}}-1)$  is also confirmed with regard to the
constraint preservation as shown in Fig.\@~\ref{fig:HOC-constraints}, 
closely followed by the choice $\alpha_0=1/\psi_{\text{full}}^2$.

\begin{figure}[!ht]
  \includegraphics[angle=270,width=0.49\columnwidth]{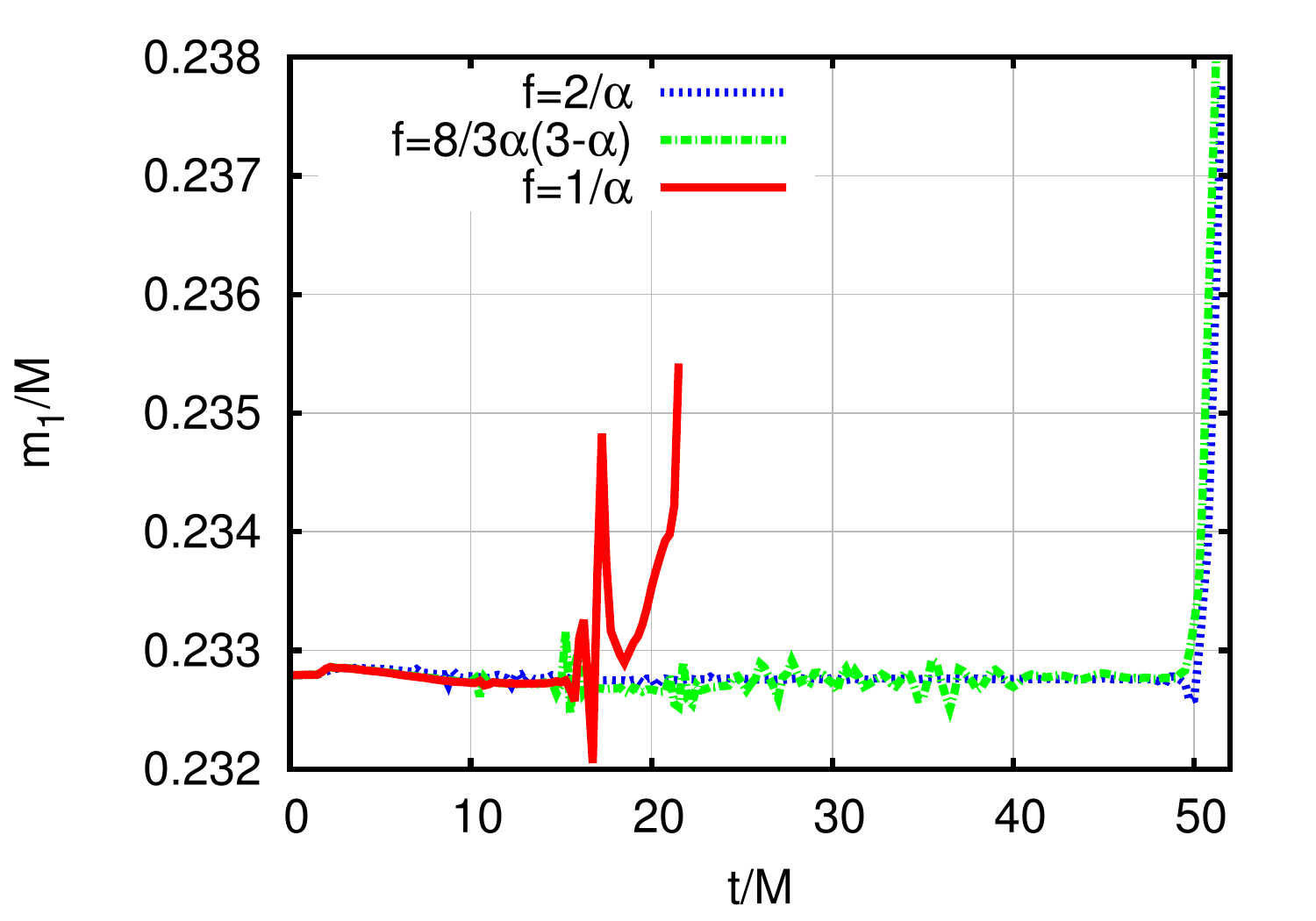}
  \includegraphics[angle=270,width=0.49\columnwidth]{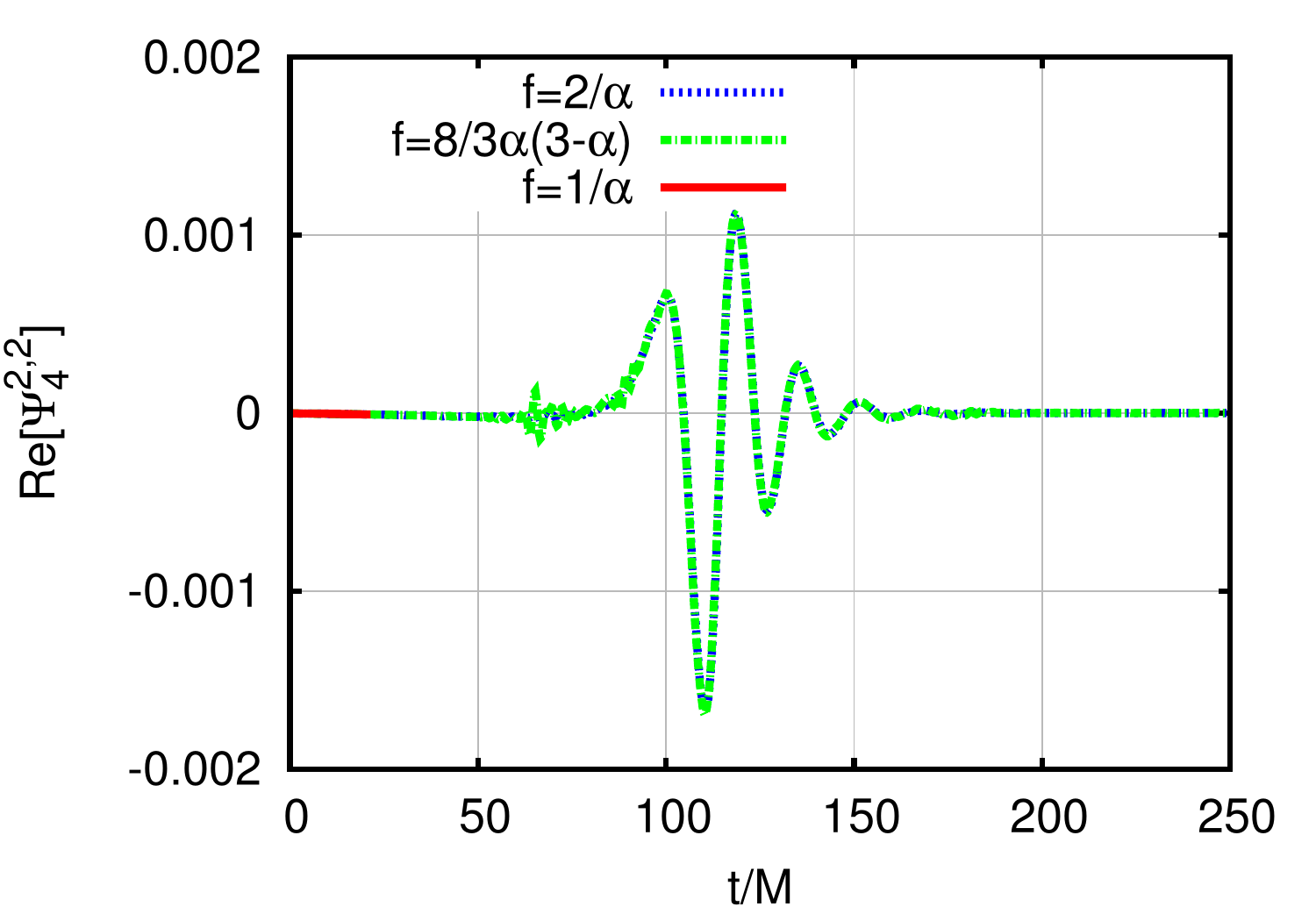}
  \caption{Horizon mass (left panel) of the boosted BH with $P_x/m_{\text{H}}=\pm 2$ and
waveform (right panel) after collision for the initial lapse $\alpha_0=1/(2\psi_{\text{BL}}-1)$ and
different choices of the evolution of the lapse.
}\label{fig:HOCle-MW}
\end{figure}

\begin{figure}[!ht]
  \includegraphics[angle=270,width=\columnwidth]{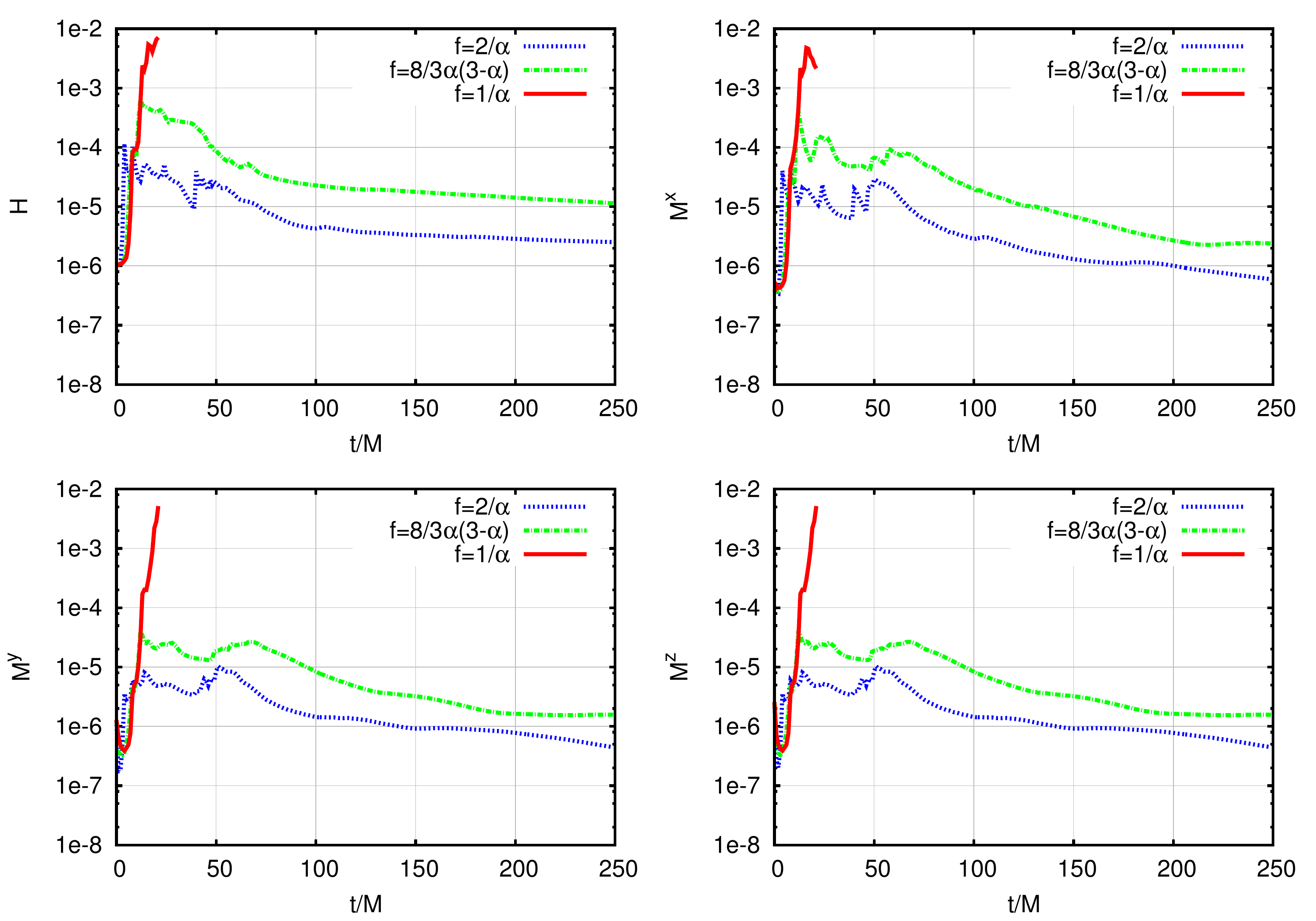}
  \caption{Hamiltonian ($H$) and momentum ($M^x$, $M^y$, and $M^z$) constraints during the free evolution
for the initial lapse $\alpha_0=1/(2\psi_{\text{BL}}-1)$ and
different choices of the evolution of the lapse.
}\label{fig:HOCle-constraints}
\end{figure}

In these evolutions we have taken the standard choice for the \MP
evolution of the lapse, $f(\alpha)=2/\alpha$ in 
Eq.\@~\eqref{eq:lapse}.
It is also worthwhile to explore alternative evolutions of
$f(\alpha)=1/\alpha$, with gauge speed equal to 1, 
and $f(\alpha)=8/(3\alpha(3-\alpha))$, with approximate shock avoiding
properties \cite{Alcubierre02b}.
The results of such evolutions are displayed in Figs.\@~\ref{fig:HOCle-MW}
and \ref{fig:HOCle-constraints}, where we have taken an initial separation
of the binary $d=66M$, $P_x/m_{\text{H}}=\pm 2$, and used the initial lapse
$\alpha_0=1/(2\psi_{\text{BL}}-1)$.

We first observe that the results of Figs.\@~\ref{fig:HOCle-MW}
and \ref{fig:HOCle-constraints} indicate that
with our numerical setup the evolution
$f(\alpha)=1/\alpha$ fails to complete (i.e., crashes) generating large errors, while
the form $f(\alpha)=8/(3\alpha(3-\alpha))$ is stable,
 but less accurate than the standard $f(\alpha)=2/\alpha$.

However, we find that for larger initial $P/m_{\text{H}}$ values,
the lapse evolution equation characterized by $f(\alpha)=2/\alpha$
fails to complete the evolution while the (approximate) shock avoiding
form $f(\alpha)=8/(3\alpha(3-\alpha))$ always succeeds.
In these cases, a large amplitude gauge wave is generated by the
high-energy collision initial data which leads to an inability for
the numerics to resolve the waves and stabilize the system.
While one can try
to fine-tune parameters of the evolution or change the evolution
equations (for instance to a Z4 type \cite{Alic:2011gg}) the form 
$f(\alpha)=8/(3\alpha(3-\alpha))$ represents a valid alternative
to the standard $f(\alpha)=2/\alpha$ evolution (which can still
be used by starting collisions further apart or slightly grazing).

\section{Calculating the ADM energy, momentum, and spin}\label{app:ADM}

For the sake of completeness we give here the explicit form of
the ADM mass, linear and angular momenta
used in the identification of the initial data parameters.

In an asymptotically flat spacetime, in asymptotically Cartesian
coordinates, the ADM mass is given by~\cite{Wald84}
\begin{equation*}
   E[h_{ab}] = \frac{1}{16 \pi} \lim_{r\to\infty} \sum_{a,b=1}^{3}
\oint \left(h_{ab,a} -h_{aa,b}\right)
\frac{x^b}{r} r^2 \, \ud \Omega
\end{equation*}
where $h_{ab} = \delta_{ab} + c_{ab}(\theta,\phi)/r + {\cal O}(1/r^2)$
is the 3-metric, $x^a = (x ,y ,z)$ are Cartesian coordinates (at
spatial infinity), and $(r,\theta,\phi)$ are the usual spherical coordinates.
The integral is over an $r={\rm const}$ sphere,
and $\ud \Omega = \sin\theta \, \ud \theta \, \ud \phi$.
Only the ${\cal O}(1/r)$ terms in the metric contribute to the ADM
mass.

For the case of superimposed boosted Schwarzschild BHs, we have

\begin{equation*}
  h_{ab} = \left(1 + \frac{m_{(+)}}{2 r_{(+)}} + \frac{m_{(-)}}{2 r_{(-)}} +
u\right)^4\left(\tilde{S}^{(+)}_{ab} + \tilde{S}^{(-)}_{ab} - \delta_{ab}\right),
\end{equation*}
where
\begin{align*}
 \left(1 + \frac{m_{(+)}}{2 r_{(+)}}\right)^4 \tilde{S}^{(+)}_{ab} &= S^{(+)}_{ab},\\
 \left(1 + \frac{m_{(-)}}{2 r_{(-)}}\right)^4 \tilde{S}^{(-)}_{ab} &= S^{(-)}_{ab},
\end{align*}
$m_{(\pm)}$ are the mass parameters of the two Schwarzschild BHs,
$r_{(\pm)}$ are ${\cal O}(r)$ with angular dependence,
and $S^{(\pm)}_{ab}$ indicates Schwarzschild metrics in boosted
coordinates.
Since $\tilde{S}_{ab} = \delta_{ab} + {\cal O}(1/r)$, the ADM mass takes
the form
\begin{align}
  E \left[h_{ab} \right] = {} & E \left[\psi^4 \tilde{S}^{(+)}_{ab} \right] + E \left[\psi^4 \tilde{S}^{(-)}_{ab} \right] - E \left[\psi^4 \delta_{ab} \right] \nonumber \\
   = {} & E \left[ \left(1+\frac{2 m_{(+)}}{r_{(+)}} \right) \tilde{S}^{(+)}_{ab} \right] +  E \left[ \left(1+\frac{2 m_{(-)}}{r_{(-)}} \right)
\tilde{S}^{(-)}_{ab} \right] \nonumber \\
 & +  E \left[ \left(\frac{2 m_{(-)}}{r_{(-)}} +  \frac{4 \hat{u}}{r} \right) \delta_{ab} \right] + E \left[ \left(\frac{2 m_{(+)}}{r_{(+)}} +
 \frac{4 \hat{u}}{r} \right)\delta_{ab} \right] \nonumber \\
& - E \left[ \left(\frac{2 m_{(+)}}{r_{(+)}} + \frac{2 m_{(-)}}{r_{(-)}} +
 \frac{4 \hat{u}}{r} \right)\delta_{ab} \right] \label{eq:adm_mass} \\
= {} & \gamma_{(+)} m_{(+)} + \gamma_{(-)} m_{(-)} + E \left[\frac{4 \hat{u}}{r} \delta_{ab} \right] \nonumber \; ,
\end{align}
where $u = \hat u(\theta,\phi)/r  + {\cal O}(1/r^2)$.
The first two terms of Eq.\@~\ref{eq:adm_mass} are the ADM masses of
boosted Schwarzschild BHs and are thus equal to $\gamma_{(+)} m_{(+)}$ and
$\gamma_{(-)} m_{(-)}$, respectively. Finally,
$ E[4 \hat u/r \delta_{ab}] = \frac{1}{8 \pi} \oint\hat u \, \ud\Omega$.

The ADM momentum is given by
\begin{equation*}
  P_a[K_{a b}] = \frac{1}{8 \pi} \lim_{r \to \infty} \sum_{b = 1}^{3} \oint \left (K_{a b} - \delta_{a b} K \right ) \frac{x^b}{r} r^2 \, \ud \Omega \; ,
\end{equation*}
with $K = \sum_{a = 1}^{3} {K^a}_{a}$.  Using
Eqs.\@~\eqref{eq:Kij_decomp}, \eqref{eq:Aij_conformal},
\eqref{eq:A_superposition}, \eqref{eq:metric_superposition}, and~\eqref{eq:K_superposition} in the asymptotic region, the integrand becomes
\begin{equation*}
  K_{a b} - \delta_{a b} K = K_{a b}^{{(+)}} - \delta_{a b} K_{(+)} + K_{a b}^{{(-)}} - \delta_{a b} K_{(-)} + (\tilde{\mathbb{L}} b)_{a b} \; 
\end{equation*}
where $K_{a b}^{{(\pm)}}$ indicates the extrinsic curvature tensors for isolated BHs.  Given the momentum parameters $P_{a}^{(\pm)}$, the corresponding ADM momentum for an isolated BH is $P_{a}\left[K_{a b}^{(\pm)}\right] = P_{a}^{(\pm)}$.  Therefore, using the linearity of the integral, the ADM momentum for a BHB is
\begin{align*}
  P_{a}[K_{a b}] &= P_{a}\left[K_{a b}^{(+)}\right] + P_{a}\left[K_{a b}^{(-)}\right] + P_{a}\left[(\tilde{\mathbb{L}} b)_{a b}\right] \\
  &= P_{a}^{(+)} + P_{a}^{(-)} + P_{a}\left[(\tilde{\mathbb{L}} b)_{a b}\right] \; .
\end{align*}

The ADM angular momentum is given by the first moment of the ADM momentum:
\begin{equation*}
  J^{a}[K_{a b}] = \frac{\epsilon^{a b c}}{8 \pi} \lim_{r \to \infty} \sum_{b, c, d = 1}^{3} \oint x_{b} (K_{c d} - \delta_{c d} K) \frac{x^d}{r} r^2 \, \ud \Omega \; .
\end{equation*}
Using the same linearity and asymptotic properties, we can write this as
\begin{equation*}
  J^{a}[K_{a b}] = J_{(+)}^{a} + J_{(-)}^{a} + J^{a}\left[(\tilde{\mathbb{L}} b)_{a b}\right] \; .
\end{equation*}

\bibliographystyle{apsrev4-1}
\bibliography{../../../../Bibtex/references}

\begin{thebibliography}{97}%
\makeatletter
\providecommand \@ifxundefined [1]{%
 \@ifx{#1\undefined}
}%
\providecommand \@ifnum [1]{%
 \ifnum #1\expandafter \@firstoftwo
 \else \expandafter \@secondoftwo
 \fi
}%
\providecommand \@ifx [1]{%
 \ifx #1\expandafter \@firstoftwo
 \else \expandafter \@secondoftwo
 \fi
}%
\providecommand \natexlab [1]{#1}%
\providecommand \enquote  [1]{``#1''}%
\providecommand \bibnamefont  [1]{#1}%
\providecommand \bibfnamefont [1]{#1}%
\providecommand \citenamefont [1]{#1}%
\providecommand \href@noop [0]{\@secondoftwo}%
\providecommand \href [0]{\begingroup \@sanitize@url \@href}%
\providecommand \@href[1]{\@@startlink{#1}\@@href}%
\providecommand \@@href[1]{\endgroup#1\@@endlink}%
\providecommand \@sanitize@url [0]{\catcode `\\12\catcode `\$12\catcode
  `\&12\catcode `\#12\catcode `\^12\catcode `\_12\catcode `\%12\relax}%
\providecommand \@@startlink[1]{}%
\providecommand \@@endlink[0]{}%
\providecommand \url  [0]{\begingroup\@sanitize@url \@url }%
\providecommand \@url [1]{\endgroup\@href {#1}{\urlprefix }}%
\providecommand \urlprefix  [0]{URL }%
\providecommand \Eprint [0]{\href }%
\providecommand \doibase [0]{http://dx.doi.org/}%
\providecommand \selectlanguage [0]{\@gobble}%
\providecommand \bibinfo  [0]{\@secondoftwo}%
\providecommand \bibfield  [0]{\@secondoftwo}%
\providecommand \translation [1]{[#1]}%
\providecommand \BibitemOpen [0]{}%
\providecommand \bibitemStop [0]{}%
\providecommand \bibitemNoStop [0]{.\EOS\space}%
\providecommand \EOS [0]{\spacefactor3000\relax}%
\providecommand \BibitemShut  [1]{\csname bibitem#1\endcsname}%
\let\auto@bib@innerbib\@empty
\bibitem [{\citenamefont {Mroue}\ \emph {et~al.}(2013)\citenamefont {Mroue},
  \citenamefont {Scheel}, \citenamefont {Szilagyi}, \citenamefont {Pfeiffer},
  \citenamefont {Boyle} \emph {et~al.}}]{Mroue:2013xna}%
  \BibitemOpen
  \bibfield  {author} {\bibinfo {author} {\bibfnamefont {A.~H.}\ \bibnamefont
  {Mroue}}, \bibinfo {author} {\bibfnamefont {M.~A.}\ \bibnamefont {Scheel}},
  \bibinfo {author} {\bibfnamefont {B.}~\bibnamefont {Szilagyi}}, \bibinfo
  {author} {\bibfnamefont {H.~P.}\ \bibnamefont {Pfeiffer}}, \bibinfo {author}
  {\bibfnamefont {M.}~\bibnamefont {Boyle}},  \emph {et~al.},\ }\href {\doibase
  10.1103/PhysRevLett.111.241104} {\bibfield  {journal} {\bibinfo  {journal}
  {Phys. Rev. Lett.}\ }\textbf {\bibinfo {volume} {111}},\ \bibinfo {pages}
  {241104} (\bibinfo {year} {2013})},\ \Eprint {http://arxiv.org/abs/1304.6077}
  {arXiv:1304.6077 [gr-qc]} \BibitemShut {NoStop}%
\bibitem [{\citenamefont {Abbott}\ \emph
  {et~al.}(2016{\natexlab{a}})\citenamefont {Abbott} \emph
  {et~al.}}]{Abbott:2016blz}%
  \BibitemOpen
  \bibfield  {author} {\bibinfo {author} {\bibfnamefont {B.}~\bibnamefont
  {Abbott}} \emph {et~al.} (\bibinfo {collaboration} {Virgo, LIGO
  Scientific}),\ }\href {\doibase 10.1103/PhysRevLett.116.061102} {\bibfield
  {journal} {\bibinfo  {journal} {Phys. Rev. Lett.}\ }\textbf {\bibinfo
  {volume} {116}},\ \bibinfo {pages} {061102} (\bibinfo {year}
  {2016}{\natexlab{a}})},\ \Eprint {http://arxiv.org/abs/1602.03837}
  {arXiv:1602.03837 [gr-qc]} \BibitemShut {NoStop}%
\bibitem [{\citenamefont {Abbott}\ \emph
  {et~al.}(2016{\natexlab{b}})\citenamefont {Abbott} \emph
  {et~al.}}]{Abbott:2016nmj}%
  \BibitemOpen
  \bibfield  {author} {\bibinfo {author} {\bibfnamefont {B.~P.}\ \bibnamefont
  {Abbott}} \emph {et~al.} (\bibinfo {collaboration} {Virgo, LIGO
  Scientific}),\ }\href {\doibase 10.1103/PhysRevLett.116.241103} {\bibfield
  {journal} {\bibinfo  {journal} {Phys. Rev. Lett.}\ }\textbf {\bibinfo
  {volume} {116}},\ \bibinfo {pages} {241103} (\bibinfo {year}
  {2016}{\natexlab{b}})},\ \Eprint {http://arxiv.org/abs/1606.04855}
  {arXiv:1606.04855 [gr-qc]} \BibitemShut {NoStop}%
\bibitem [{\citenamefont {Pretorius}(2005)}]{Pretorius:2005gq}%
  \BibitemOpen
  \bibfield  {author} {\bibinfo {author} {\bibfnamefont {F.}~\bibnamefont
  {Pretorius}},\ }\href@noop {} {\bibfield  {journal} {\bibinfo  {journal}
  {Phys. Rev. Lett.}\ }\textbf {\bibinfo {volume} {95}},\ \bibinfo {pages}
  {121101} (\bibinfo {year} {2005})},\ \Eprint
  {http://arxiv.org/abs/gr-qc/0507014} {gr-qc/0507014} \BibitemShut {NoStop}%
\bibitem [{\citenamefont {Campanelli}\ \emph
  {et~al.}(2006{\natexlab{a}})\citenamefont {Campanelli}, \citenamefont
  {Lousto}, \citenamefont {Marronetti},\ and\ \citenamefont
  {Zlochower}}]{Campanelli:2005dd}%
  \BibitemOpen
  \bibfield  {author} {\bibinfo {author} {\bibfnamefont {M.}~\bibnamefont
  {Campanelli}}, \bibinfo {author} {\bibfnamefont {C.~O.}\ \bibnamefont
  {Lousto}}, \bibinfo {author} {\bibfnamefont {P.}~\bibnamefont {Marronetti}},
  \ and\ \bibinfo {author} {\bibfnamefont {Y.}~\bibnamefont {Zlochower}},\
  }\href@noop {} {\bibfield  {journal} {\bibinfo  {journal} {Phys. Rev. Lett.}\
  }\textbf {\bibinfo {volume} {96}},\ \bibinfo {pages} {111101} (\bibinfo
  {year} {2006}{\natexlab{a}})},\ \Eprint {http://arxiv.org/abs/gr-qc/0511048}
  {gr-qc/0511048} \BibitemShut {NoStop}%
\bibitem [{\citenamefont {Baker}\ \emph {et~al.}(2006)\citenamefont {Baker},
  \citenamefont {Centrella}, \citenamefont {Choi}, \citenamefont {Koppitz},\
  and\ \citenamefont {van Meter}}]{Baker:2005vv}%
  \BibitemOpen
  \bibfield  {author} {\bibinfo {author} {\bibfnamefont {J.~G.}\ \bibnamefont
  {Baker}}, \bibinfo {author} {\bibfnamefont {J.}~\bibnamefont {Centrella}},
  \bibinfo {author} {\bibfnamefont {D.-I.}\ \bibnamefont {Choi}}, \bibinfo
  {author} {\bibfnamefont {M.}~\bibnamefont {Koppitz}}, \ and\ \bibinfo
  {author} {\bibfnamefont {J.}~\bibnamefont {van Meter}},\ }\href@noop {}
  {\bibfield  {journal} {\bibinfo  {journal} {Phys. Rev. Lett.}\ }\textbf
  {\bibinfo {volume} {96}},\ \bibinfo {pages} {111102} (\bibinfo {year}
  {2006})},\ \Eprint {http://arxiv.org/abs/gr-qc/0511103} {gr-qc/0511103}
  \BibitemShut {NoStop}%
\bibitem [{\citenamefont {Abbott}\ \emph
  {et~al.}(2016{\natexlab{c}})\citenamefont {Abbott} \emph
  {et~al.}}]{TheLIGOScientific:2016src}%
  \BibitemOpen
  \bibfield  {author} {\bibinfo {author} {\bibfnamefont {B.~P.}\ \bibnamefont
  {Abbott}} \emph {et~al.} (\bibinfo {collaboration} {Virgo, LIGO
  Scientific}),\ }\href {\doibase 10.1103/PhysRevLett.116.221101} {\bibfield
  {journal} {\bibinfo  {journal} {Phys. Rev. Lett.}\ }\textbf {\bibinfo
  {volume} {116}},\ \bibinfo {pages} {221101} (\bibinfo {year}
  {2016}{\natexlab{c}})},\ \Eprint {http://arxiv.org/abs/1602.03841}
  {arXiv:1602.03841 [gr-qc]} \BibitemShut {NoStop}%
\bibitem [{\citenamefont {Abbott}\ \emph
  {et~al.}(2016{\natexlab{d}})\citenamefont {Abbott} \emph
  {et~al.}}]{TheLIGOScientific:2016pea}%
  \BibitemOpen
  \bibfield  {author} {\bibinfo {author} {\bibfnamefont {B.~P.}\ \bibnamefont
  {Abbott}} \emph {et~al.} (\bibinfo {collaboration} {Virgo, LIGO
  Scientific}),\ }\href@noop {} {\  (\bibinfo {year} {2016}{\natexlab{d}})},\
  \Eprint {http://arxiv.org/abs/1606.04856} {arXiv:1606.04856 [gr-qc]}
  \BibitemShut {NoStop}%
\bibitem [{\citenamefont {Lovelace}\ \emph {et~al.}(2016)\citenamefont
  {Lovelace} \emph {et~al.}}]{Lovelace:2016uwp}%
  \BibitemOpen
  \bibfield  {author} {\bibinfo {author} {\bibfnamefont {G.}~\bibnamefont
  {Lovelace}} \emph {et~al.},\ }\href@noop {} {\  (\bibinfo {year} {2016})},\
  \Eprint {http://arxiv.org/abs/1607.05377} {arXiv:1607.05377 [gr-qc]}
  \BibitemShut {NoStop}%
\bibitem [{\citenamefont {Abbott}\ \emph
  {et~al.}(2016{\natexlab{e}})\citenamefont {Abbott} \emph
  {et~al.}}]{Abbott:2016apu}%
  \BibitemOpen
  \bibfield  {author} {\bibinfo {author} {\bibfnamefont {B.~P.}\ \bibnamefont
  {Abbott}} \emph {et~al.} (\bibinfo {collaboration} {Virgo, LIGO
  Scientific}),\ }\href@noop {} {\  (\bibinfo {year} {2016}{\natexlab{e}})},\
  \Eprint {http://arxiv.org/abs/1606.01262} {arXiv:1606.01262 [gr-qc]}
  \BibitemShut {NoStop}%
\bibitem [{\citenamefont {Reynolds}(2013)}]{Reynolds:2013rva}%
  \BibitemOpen
  \bibfield  {author} {\bibinfo {author} {\bibfnamefont {C.~S.}\ \bibnamefont
  {Reynolds}},\ }\href {\doibase 10.1088/0264-9381/30/24/244004} {\bibfield
  {journal} {\bibinfo  {journal} {Class. Quant. Grav.}\ }\textbf {\bibinfo
  {volume} {30}},\ \bibinfo {pages} {244004} (\bibinfo {year} {2013})},\
  \Eprint {http://arxiv.org/abs/1307.3246} {arXiv:1307.3246 [astro-ph.HE]}
  \BibitemShut {NoStop}%
\bibitem [{\citenamefont {Campanelli}\ \emph
  {et~al.}(2006{\natexlab{b}})\citenamefont {Campanelli}, \citenamefont
  {Lousto},\ and\ \citenamefont {Zlochower}}]{Campanelli:2006uy}%
  \BibitemOpen
  \bibfield  {author} {\bibinfo {author} {\bibfnamefont {M.}~\bibnamefont
  {Campanelli}}, \bibinfo {author} {\bibfnamefont {C.~O.}\ \bibnamefont
  {Lousto}}, \ and\ \bibinfo {author} {\bibfnamefont {Y.}~\bibnamefont
  {Zlochower}},\ }\href@noop {} {\bibfield  {journal} {\bibinfo  {journal}
  {Phys. Rev.}\ }\textbf {\bibinfo {volume} {D74}},\ \bibinfo {pages}
  {041501(R)} (\bibinfo {year} {2006}{\natexlab{b}})},\ \Eprint
  {http://arxiv.org/abs/gr-qc/0604012} {gr-qc/0604012} \BibitemShut {NoStop}%
\bibitem [{\citenamefont {Lousto}\ and\ \citenamefont
  {Healy}(2015)}]{Lousto:2014ida}%
  \BibitemOpen
  \bibfield  {author} {\bibinfo {author} {\bibfnamefont {C.~O.}\ \bibnamefont
  {Lousto}}\ and\ \bibinfo {author} {\bibfnamefont {J.}~\bibnamefont {Healy}},\
  }\href {\doibase 10.1103/PhysRevLett.114.141101} {\bibfield  {journal}
  {\bibinfo  {journal} {Phys. Rev. Lett.}\ }\textbf {\bibinfo {volume} {114}},\
  \bibinfo {pages} {141101} (\bibinfo {year} {2015})},\ \Eprint
  {http://arxiv.org/abs/1410.3830} {arXiv:1410.3830 [gr-qc]} \BibitemShut
  {NoStop}%
\bibitem [{\citenamefont {Campanelli}\ \emph
  {et~al.}(2007{\natexlab{a}})\citenamefont {Campanelli}, \citenamefont
  {Lousto}, \citenamefont {Zlochower},\ and\ \citenamefont
  {Merritt}}]{Campanelli:2007ew}%
  \BibitemOpen
  \bibfield  {author} {\bibinfo {author} {\bibfnamefont {M.}~\bibnamefont
  {Campanelli}}, \bibinfo {author} {\bibfnamefont {C.~O.}\ \bibnamefont
  {Lousto}}, \bibinfo {author} {\bibfnamefont {Y.}~\bibnamefont {Zlochower}}, \
  and\ \bibinfo {author} {\bibfnamefont {D.}~\bibnamefont {Merritt}},\
  }\href@noop {} {\bibfield  {journal} {\bibinfo  {journal} {Astrophys. J.}\
  }\textbf {\bibinfo {volume} {659}},\ \bibinfo {pages} {L5} (\bibinfo {year}
  {2007}{\natexlab{a}})},\ \Eprint {http://arxiv.org/abs/gr-qc/0701164}
  {gr-qc/0701164} \BibitemShut {NoStop}%
\bibitem [{\citenamefont {Campanelli}\ \emph
  {et~al.}(2007{\natexlab{b}})\citenamefont {Campanelli}, \citenamefont
  {Lousto}, \citenamefont {Zlochower},\ and\ \citenamefont
  {Merritt}}]{Campanelli:2007cga}%
  \BibitemOpen
  \bibfield  {author} {\bibinfo {author} {\bibfnamefont {M.}~\bibnamefont
  {Campanelli}}, \bibinfo {author} {\bibfnamefont {C.~O.}\ \bibnamefont
  {Lousto}}, \bibinfo {author} {\bibfnamefont {Y.}~\bibnamefont {Zlochower}}, \
  and\ \bibinfo {author} {\bibfnamefont {D.}~\bibnamefont {Merritt}},\
  }\href@noop {} {\bibfield  {journal} {\bibinfo  {journal} {Phys. Rev. Lett.}\
  }\textbf {\bibinfo {volume} {98}},\ \bibinfo {pages} {231102} (\bibinfo
  {year} {2007}{\natexlab{b}})},\ \Eprint {http://arxiv.org/abs/gr-qc/0702133}
  {gr-qc/0702133} \BibitemShut {NoStop}%
\bibitem [{\citenamefont {Lousto}\ \emph
  {et~al.}(2012{\natexlab{a}})\citenamefont {Lousto}, \citenamefont
  {Zlochower}, \citenamefont {Dotti},\ and\ \citenamefont
  {Volonteri}}]{Lousto:2012su}%
  \BibitemOpen
  \bibfield  {author} {\bibinfo {author} {\bibfnamefont {C.~O.}\ \bibnamefont
  {Lousto}}, \bibinfo {author} {\bibfnamefont {Y.}~\bibnamefont {Zlochower}},
  \bibinfo {author} {\bibfnamefont {M.}~\bibnamefont {Dotti}}, \ and\ \bibinfo
  {author} {\bibfnamefont {M.}~\bibnamefont {Volonteri}},\ }\href@noop {}
  {\bibfield  {journal} {\bibinfo  {journal} {Phys. Rev.}\ }\textbf {\bibinfo
  {volume} {D85}},\ \bibinfo {pages} {084015} (\bibinfo {year}
  {2012}{\natexlab{a}})},\ \Eprint {http://arxiv.org/abs/1201.1923}
  {arXiv:1201.1923 [gr-qc]} \BibitemShut {NoStop}%
\bibitem [{\citenamefont {Lousto}\ and\ \citenamefont
  {Zlochower}(2013{\natexlab{a}})}]{Lousto:2012gt}%
  \BibitemOpen
  \bibfield  {author} {\bibinfo {author} {\bibfnamefont {C.~O.}\ \bibnamefont
  {Lousto}}\ and\ \bibinfo {author} {\bibfnamefont {Y.}~\bibnamefont
  {Zlochower}},\ }\href {\doibase 10.1103/PhysRevD.87.084027} {\bibfield
  {journal} {\bibinfo  {journal} {Phys. Rev.}\ }\textbf {\bibinfo {volume}
  {D87}},\ \bibinfo {pages} {084027} (\bibinfo {year} {2013}{\natexlab{a}})},\
  \Eprint {http://arxiv.org/abs/1211.7099} {arXiv:1211.7099 [gr-qc]}
  \BibitemShut {NoStop}%
\bibitem [{\citenamefont {Hemberger}\ \emph {et~al.}(2013)\citenamefont
  {Hemberger}, \citenamefont {Lovelace}, \citenamefont {Loredo}, \citenamefont
  {Kidder}, \citenamefont {Scheel}, \citenamefont {Szil{\'a}gyi}, \citenamefont
  {Taylor},\ and\ \citenamefont {Teukolsky}}]{Hemberger:2013hsa}%
  \BibitemOpen
  \bibfield  {author} {\bibinfo {author} {\bibfnamefont {D.~A.}\ \bibnamefont
  {Hemberger}}, \bibinfo {author} {\bibfnamefont {G.}~\bibnamefont {Lovelace}},
  \bibinfo {author} {\bibfnamefont {T.~J.}\ \bibnamefont {Loredo}}, \bibinfo
  {author} {\bibfnamefont {L.~E.}\ \bibnamefont {Kidder}}, \bibinfo {author}
  {\bibfnamefont {M.~A.}\ \bibnamefont {Scheel}}, \bibinfo {author}
  {\bibfnamefont {B.}~\bibnamefont {Szil{\'a}gyi}}, \bibinfo {author}
  {\bibfnamefont {N.~W.}\ \bibnamefont {Taylor}}, \ and\ \bibinfo {author}
  {\bibfnamefont {S.~A.}\ \bibnamefont {Teukolsky}},\ }\href {\doibase
  10.1103/PhysRevD.88.064014} {\bibfield  {journal} {\bibinfo  {journal} {Phys.
  Rev.}\ }\textbf {\bibinfo {volume} {D88}},\ \bibinfo {pages} {064014}
  (\bibinfo {year} {2013})},\ \Eprint {http://arxiv.org/abs/1305.5991}
  {arXiv:1305.5991 [gr-qc]} \BibitemShut {NoStop}%
\bibitem [{\citenamefont {Healy}\ \emph {et~al.}(2014)\citenamefont {Healy},
  \citenamefont {Lousto},\ and\ \citenamefont {Zlochower}}]{Healy:2014yta}%
  \BibitemOpen
  \bibfield  {author} {\bibinfo {author} {\bibfnamefont {J.}~\bibnamefont
  {Healy}}, \bibinfo {author} {\bibfnamefont {C.~O.}\ \bibnamefont {Lousto}}, \
  and\ \bibinfo {author} {\bibfnamefont {Y.}~\bibnamefont {Zlochower}},\ }\href
  {\doibase 10.1103/PhysRevD.90.104004} {\bibfield  {journal} {\bibinfo
  {journal} {Phys. Rev.}\ }\textbf {\bibinfo {volume} {D90}},\ \bibinfo {pages}
  {104004} (\bibinfo {year} {2014})},\ \Eprint {http://arxiv.org/abs/1406.7295}
  {arXiv:1406.7295 [gr-qc]} \BibitemShut {NoStop}%
\bibitem [{\citenamefont {Aasi}\ \emph {et~al.}(2014)\citenamefont {Aasi} \emph
  {et~al.}}]{Aasi:2014tra}%
  \BibitemOpen
  \bibfield  {author} {\bibinfo {author} {\bibfnamefont {J.}~\bibnamefont
  {Aasi}} \emph {et~al.} (\bibinfo {collaboration} {LIGO Scientific
  Collaboration, Virgo Collaboration, NINJA-2 Collaboration}),\ }\href
  {\doibase 10.1088/0264-9381/31/11/115004} {\bibfield  {journal} {\bibinfo
  {journal} {Class. Quant. Grav.}\ }\textbf {\bibinfo {volume} {31}},\ \bibinfo
  {pages} {115004} (\bibinfo {year} {2014})},\ \Eprint
  {http://arxiv.org/abs/1401.0939} {arXiv:1401.0939 [gr-qc]} \BibitemShut
  {NoStop}%
\bibitem [{\citenamefont {Aylott}\ \emph
  {et~al.}(2009{\natexlab{a}})\citenamefont {Aylott} \emph
  {et~al.}}]{Aylott:2009tn}%
  \BibitemOpen
  \bibfield  {author} {\bibinfo {author} {\bibfnamefont {B.}~\bibnamefont
  {Aylott}} \emph {et~al.},\ }\href {\doibase 10.1088/0264-9381/26/11/114008}
  {\bibfield  {journal} {\bibinfo  {journal} {Class. Quant. Grav.}\ }\textbf
  {\bibinfo {volume} {26}},\ \bibinfo {pages} {114008} (\bibinfo {year}
  {2009}{\natexlab{a}})},\ \Eprint {http://arxiv.org/abs/0905.4227}
  {arXiv:0905.4227 [gr-qc]} \BibitemShut {NoStop}%
\bibitem [{\citenamefont {Aylott}\ \emph
  {et~al.}(2009{\natexlab{b}})\citenamefont {Aylott} \emph
  {et~al.}}]{Aylott:2009ya}%
  \BibitemOpen
  \bibfield  {author} {\bibinfo {author} {\bibfnamefont {B.}~\bibnamefont
  {Aylott}} \emph {et~al.},\ }\href {\doibase 10.1088/0264-9381/26/16/165008}
  {\bibfield  {journal} {\bibinfo  {journal} {Class. Quant. Grav.}\ }\textbf
  {\bibinfo {volume} {26}},\ \bibinfo {pages} {165008} (\bibinfo {year}
  {2009}{\natexlab{b}})},\ \Eprint {http://arxiv.org/abs/0901.4399}
  {arXiv:0901.4399 [gr-qc]} \BibitemShut {NoStop}%
\bibitem [{\citenamefont {Ajith}\ \emph {et~al.}(2012)\citenamefont {Ajith}
  \emph {et~al.}}]{Ajith:2012az}%
  \BibitemOpen
  \bibfield  {author} {\bibinfo {author} {\bibfnamefont {P.}~\bibnamefont
  {Ajith}} \emph {et~al.},\ }\href {\doibase 10.1088/0264-9381/29/12/124001}
  {\bibfield  {journal} {\bibinfo  {journal} {Class. Quant. Grav.}\ }\textbf
  {\bibinfo {volume} {29}},\ \bibinfo {pages} {124001} (\bibinfo {year}
  {2012})},\ \Eprint {http://arxiv.org/abs/1201.5319} {arXiv:1201.5319 [gr-qc]}
  \BibitemShut {NoStop}%
\bibitem [{\citenamefont {Hinder}\ \emph {et~al.}(2014)\citenamefont {Hinder},
  \citenamefont {Buonanno}, \citenamefont {Boyle}, \citenamefont {Etienne},
  \citenamefont {Healy}, \citenamefont {Johnson-McDaniel}, \citenamefont
  {Nagar}, \citenamefont {Nakano}, \citenamefont {Pan}, \citenamefont
  {Pfeiffer}, \citenamefont {P{\"u}rrer}, \citenamefont {Reisswig},
  \citenamefont {Scheel}, \citenamefont {Schnetter}, \citenamefont {Sperhake},
  \citenamefont {Szil{\'a}gyi}, \citenamefont {Tichy}, \citenamefont {Wardell},
  \citenamefont {Zengino\u{g}lu}, \citenamefont {Alic}, \citenamefont
  {Bernuzzi}, \citenamefont {Bode}, \citenamefont {Br{\"u}gmann}, \citenamefont
  {Buchman}, \citenamefont {Campanelli}, \citenamefont {Chu}, \citenamefont
  {Damour}, \citenamefont {Grigsby}, \citenamefont {Hannam}, \citenamefont
  {Haas}, \citenamefont {Hemberger}, \citenamefont {Husa}, \citenamefont
  {Kidder}, \citenamefont {Laguna}, \citenamefont {London}, \citenamefont
  {Lovelace}, \citenamefont {Lousto}, \citenamefont {Marronetti}, \citenamefont
  {Matzner}, \citenamefont {M{\"o}sta}, \citenamefont {Mrou{\'e}},
  \citenamefont {M{\"u}ller}, \citenamefont {Mundim}, \citenamefont {Nerozzi},
  \citenamefont {Paschalidis}, \citenamefont {Pollney}, \citenamefont
  {Reifenberger}, \citenamefont {Rezzolla}, \citenamefont {Shapiro},
  \citenamefont {Shoemaker}, \citenamefont {Taracchini}, \citenamefont
  {Taylor}, \citenamefont {Teukolsky}, \citenamefont {Thierfelder},
  \citenamefont {Witek},\ and\ \citenamefont {Zlochower}}]{Hinder:2013oqa}%
  \BibitemOpen
  \bibfield  {author} {\bibinfo {author} {\bibfnamefont {I.}~\bibnamefont
  {Hinder}}, \bibinfo {author} {\bibfnamefont {A.}~\bibnamefont {Buonanno}},
  \bibinfo {author} {\bibfnamefont {M.}~\bibnamefont {Boyle}}, \bibinfo
  {author} {\bibfnamefont {Z.~B.}\ \bibnamefont {Etienne}}, \bibinfo {author}
  {\bibfnamefont {J.}~\bibnamefont {Healy}}, \bibinfo {author} {\bibfnamefont
  {N.~K.}\ \bibnamefont {Johnson-McDaniel}}, \bibinfo {author} {\bibfnamefont
  {A.}~\bibnamefont {Nagar}}, \bibinfo {author} {\bibfnamefont
  {H.}~\bibnamefont {Nakano}}, \bibinfo {author} {\bibfnamefont
  {Y.}~\bibnamefont {Pan}}, \bibinfo {author} {\bibfnamefont {H.~P.}\
  \bibnamefont {Pfeiffer}}, \bibinfo {author} {\bibfnamefont {M.}~\bibnamefont
  {P{\"u}rrer}}, \bibinfo {author} {\bibfnamefont {C.}~\bibnamefont
  {Reisswig}}, \bibinfo {author} {\bibfnamefont {M.~A.}\ \bibnamefont
  {Scheel}}, \bibinfo {author} {\bibfnamefont {E.}~\bibnamefont {Schnetter}},
  \bibinfo {author} {\bibfnamefont {U.}~\bibnamefont {Sperhake}}, \bibinfo
  {author} {\bibfnamefont {B.}~\bibnamefont {Szil{\'a}gyi}}, \bibinfo {author}
  {\bibfnamefont {W.}~\bibnamefont {Tichy}}, \bibinfo {author} {\bibfnamefont
  {B.}~\bibnamefont {Wardell}}, \bibinfo {author} {\bibfnamefont
  {A.}~\bibnamefont {Zengino\u{g}lu}}, \bibinfo {author} {\bibfnamefont
  {D.}~\bibnamefont {Alic}}, \bibinfo {author} {\bibfnamefont {S.}~\bibnamefont
  {Bernuzzi}}, \bibinfo {author} {\bibfnamefont {T.}~\bibnamefont {Bode}},
  \bibinfo {author} {\bibfnamefont {B.}~\bibnamefont {Br{\"u}gmann}}, \bibinfo
  {author} {\bibfnamefont {L.~T.}\ \bibnamefont {Buchman}}, \bibinfo {author}
  {\bibfnamefont {M.}~\bibnamefont {Campanelli}}, \bibinfo {author}
  {\bibfnamefont {T.}~\bibnamefont {Chu}}, \bibinfo {author} {\bibfnamefont
  {T.}~\bibnamefont {Damour}}, \bibinfo {author} {\bibfnamefont {J.~D.}\
  \bibnamefont {Grigsby}}, \bibinfo {author} {\bibfnamefont {M.}~\bibnamefont
  {Hannam}}, \bibinfo {author} {\bibfnamefont {R.}~\bibnamefont {Haas}},
  \bibinfo {author} {\bibfnamefont {D.~A.}\ \bibnamefont {Hemberger}}, \bibinfo
  {author} {\bibfnamefont {S.}~\bibnamefont {Husa}}, \bibinfo {author}
  {\bibfnamefont {L.~E.}\ \bibnamefont {Kidder}}, \bibinfo {author}
  {\bibfnamefont {P.}~\bibnamefont {Laguna}}, \bibinfo {author} {\bibfnamefont
  {L.}~\bibnamefont {London}}, \bibinfo {author} {\bibfnamefont
  {G.}~\bibnamefont {Lovelace}}, \bibinfo {author} {\bibfnamefont {C.~O.}\
  \bibnamefont {Lousto}}, \bibinfo {author} {\bibfnamefont {P.}~\bibnamefont
  {Marronetti}}, \bibinfo {author} {\bibfnamefont {R.~A.}\ \bibnamefont
  {Matzner}}, \bibinfo {author} {\bibfnamefont {P.}~\bibnamefont {M{\"o}sta}},
  \bibinfo {author} {\bibfnamefont {A.}~\bibnamefont {Mrou{\'e}}}, \bibinfo
  {author} {\bibfnamefont {D.}~\bibnamefont {M{\"u}ller}}, \bibinfo {author}
  {\bibfnamefont {B.~C.}\ \bibnamefont {Mundim}}, \bibinfo {author}
  {\bibfnamefont {A.}~\bibnamefont {Nerozzi}}, \bibinfo {author} {\bibfnamefont
  {V.}~\bibnamefont {Paschalidis}}, \bibinfo {author} {\bibfnamefont
  {D.}~\bibnamefont {Pollney}}, \bibinfo {author} {\bibfnamefont
  {G.}~\bibnamefont {Reifenberger}}, \bibinfo {author} {\bibfnamefont
  {L.}~\bibnamefont {Rezzolla}}, \bibinfo {author} {\bibfnamefont {S.~L.}\
  \bibnamefont {Shapiro}}, \bibinfo {author} {\bibfnamefont {D.}~\bibnamefont
  {Shoemaker}}, \bibinfo {author} {\bibfnamefont {A.}~\bibnamefont
  {Taracchini}}, \bibinfo {author} {\bibfnamefont {N.~W.}\ \bibnamefont
  {Taylor}}, \bibinfo {author} {\bibfnamefont {S.~A.}\ \bibnamefont
  {Teukolsky}}, \bibinfo {author} {\bibfnamefont {M.}~\bibnamefont
  {Thierfelder}}, \bibinfo {author} {\bibfnamefont {H.}~\bibnamefont {Witek}},
  \ and\ \bibinfo {author} {\bibfnamefont {Y.}~\bibnamefont {Zlochower}},\
  }\href {\doibase 10.1088/0264-9381/31/2/025012} {\bibfield  {journal}
  {\bibinfo  {journal} {Class. Quant. Grav.}\ }\textbf {\bibinfo {volume}
  {31}},\ \bibinfo {pages} {025012} (\bibinfo {year} {2014})},\ \Eprint
  {http://arxiv.org/abs/1307.5307} {arXiv:1307.5307 [gr-qc]} \BibitemShut
  {NoStop}%
\bibitem [{\citenamefont {Lousto}\ and\ \citenamefont
  {Zlochower}(2011{\natexlab{a}})}]{Lousto:2010ut}%
  \BibitemOpen
  \bibfield  {author} {\bibinfo {author} {\bibfnamefont {C.~O.}\ \bibnamefont
  {Lousto}}\ and\ \bibinfo {author} {\bibfnamefont {Y.}~\bibnamefont
  {Zlochower}},\ }\href {\doibase 10.1103/PhysRevLett.106.041101} {\bibfield
  {journal} {\bibinfo  {journal} {Phys. Rev. Lett.}\ }\textbf {\bibinfo
  {volume} {106}},\ \bibinfo {pages} {041101} (\bibinfo {year}
  {2011}{\natexlab{a}})},\ \Eprint {http://arxiv.org/abs/1009.0292}
  {arXiv:1009.0292 [gr-qc]} \BibitemShut {NoStop}%
\bibitem [{\citenamefont {Lousto}\ and\ \citenamefont
  {Zlochower}(2013{\natexlab{b}})}]{Lousto:2013oza}%
  \BibitemOpen
  \bibfield  {author} {\bibinfo {author} {\bibfnamefont {C.~O.}\ \bibnamefont
  {Lousto}}\ and\ \bibinfo {author} {\bibfnamefont {Y.}~\bibnamefont
  {Zlochower}},\ }\href {\doibase 10.1103/PhysRevD.88.024001} {\bibfield
  {journal} {\bibinfo  {journal} {Phys. Rev.}\ }\textbf {\bibinfo {volume}
  {D88}},\ \bibinfo {pages} {024001} (\bibinfo {year} {2013}{\natexlab{b}})},\
  \Eprint {http://arxiv.org/abs/1304.3937} {arXiv:1304.3937 [gr-qc]}
  \BibitemShut {NoStop}%
\bibitem [{\citenamefont {Lousto}\ and\ \citenamefont
  {Zlochower}(2008)}]{Lousto:2007rj}%
  \BibitemOpen
  \bibfield  {author} {\bibinfo {author} {\bibfnamefont {C.~O.}\ \bibnamefont
  {Lousto}}\ and\ \bibinfo {author} {\bibfnamefont {Y.}~\bibnamefont
  {Zlochower}},\ }\href {\doibase 10.1103/PhysRevD.77.024034} {\bibfield
  {journal} {\bibinfo  {journal} {Phys. Rev.}\ }\textbf {\bibinfo {volume}
  {D77}},\ \bibinfo {pages} {024034} (\bibinfo {year} {2008})},\ \Eprint
  {http://arxiv.org/abs/0711.1165} {arXiv:0711.1165 [gr-qc]} \BibitemShut
  {NoStop}%
\bibitem [{\citenamefont {Shibata}\ and\ \citenamefont
  {Uryu}(2006)}]{Shibata:2006ks}%
  \BibitemOpen
  \bibfield  {author} {\bibinfo {author} {\bibfnamefont {M.}~\bibnamefont
  {Shibata}}\ and\ \bibinfo {author} {\bibfnamefont {K.}~\bibnamefont {Uryu}},\
  }\href@noop {} {\bibfield  {journal} {\bibinfo  {journal} {Phys. Rev.}\
  }\textbf {\bibinfo {volume} {D74}},\ \bibinfo {pages} {121503} (\bibinfo
  {year} {2006})},\ \Eprint {http://arxiv.org/abs/gr-qc/0612142}
  {gr-qc/0612142} \BibitemShut {NoStop}%
\bibitem [{\citenamefont {Baiotti}\ and\ \citenamefont
  {Rezzolla}(2006)}]{Baiotti:2006wm}%
  \BibitemOpen
  \bibfield  {author} {\bibinfo {author} {\bibfnamefont {L.}~\bibnamefont
  {Baiotti}}\ and\ \bibinfo {author} {\bibfnamefont {L.}~\bibnamefont
  {Rezzolla}},\ }\href@noop {} {\bibfield  {journal} {\bibinfo  {journal}
  {Phys. Rev. Lett.}\ }\textbf {\bibinfo {volume} {97}},\ \bibinfo {pages}
  {141101} (\bibinfo {year} {2006})},\ \Eprint
  {http://arxiv.org/abs/gr-qc/0608113} {gr-qc/0608113} \BibitemShut {NoStop}%
\bibitem [{\citenamefont {Bowen}\ and\ \citenamefont
  {York}(1980)}]{Bowen:1980yu}%
  \BibitemOpen
  \bibfield  {author} {\bibinfo {author} {\bibfnamefont {J.~M.}\ \bibnamefont
  {Bowen}}\ and\ \bibinfo {author} {\bibfnamefont {J.~W.}\ \bibnamefont {York},
  \bibfnamefont {Jr.}},\ }\href {\doibase 10.1103/PhysRevD.21.2047} {\bibfield
  {journal} {\bibinfo  {journal} {Phys. Rev.}\ }\textbf {\bibinfo {volume}
  {D21}},\ \bibinfo {pages} {2047} (\bibinfo {year} {1980})}\BibitemShut
  {NoStop}%
\bibitem [{\citenamefont {Cook}\ and\ \citenamefont
  {York}(1990{\natexlab{a}})}]{Cook:1989fb}%
  \BibitemOpen
  \bibfield  {author} {\bibinfo {author} {\bibfnamefont {G.~B.}\ \bibnamefont
  {Cook}}\ and\ \bibinfo {author} {\bibfnamefont {J.}~\bibnamefont {York},
  \bibfnamefont {James~W.}},\ }\href@noop {} {\bibfield  {journal} {\bibinfo
  {journal} {Phys. Rev.}\ }\textbf {\bibinfo {volume} {D41}},\ \bibinfo {pages}
  {1077} (\bibinfo {year} {1990}{\natexlab{a}})}\BibitemShut {NoStop}%
\bibitem [{\citenamefont {Dain}\ \emph {et~al.}(2002)\citenamefont {Dain},
  \citenamefont {Lousto},\ and\ \citenamefont {Takahashi}}]{Dain:2002ee}%
  \BibitemOpen
  \bibfield  {author} {\bibinfo {author} {\bibfnamefont {S.}~\bibnamefont
  {Dain}}, \bibinfo {author} {\bibfnamefont {C.~O.}\ \bibnamefont {Lousto}}, \
  and\ \bibinfo {author} {\bibfnamefont {R.}~\bibnamefont {Takahashi}},\ }\href
  {\doibase 10.1103/PhysRevD.65.104038} {\bibfield  {journal} {\bibinfo
  {journal} {Phys. Rev.}\ }\textbf {\bibinfo {volume} {D65}},\ \bibinfo {pages}
  {104038} (\bibinfo {year} {2002})},\ \Eprint
  {http://arxiv.org/abs/gr-qc/0201062} {arXiv:gr-qc/0201062} \BibitemShut
  {NoStop}%
\bibitem [{\citenamefont {Lousto}\ \emph
  {et~al.}(2012{\natexlab{b}})\citenamefont {Lousto}, \citenamefont {Nakano},
  \citenamefont {Zlochower}, \citenamefont {Mundim},\ and\ \citenamefont
  {Campanelli}}]{Lousto:2012es}%
  \BibitemOpen
  \bibfield  {author} {\bibinfo {author} {\bibfnamefont {C.~O.}\ \bibnamefont
  {Lousto}}, \bibinfo {author} {\bibfnamefont {H.}~\bibnamefont {Nakano}},
  \bibinfo {author} {\bibfnamefont {Y.}~\bibnamefont {Zlochower}}, \bibinfo
  {author} {\bibfnamefont {B.~C.}\ \bibnamefont {Mundim}}, \ and\ \bibinfo
  {author} {\bibfnamefont {M.}~\bibnamefont {Campanelli}},\ }\href@noop {}
  {\bibfield  {journal} {\bibinfo  {journal} {Phys. Rev.}\ }\textbf {\bibinfo
  {volume} {D85}},\ \bibinfo {pages} {124013} (\bibinfo {year}
  {2012}{\natexlab{b}})},\ \Eprint {http://arxiv.org/abs/1203.3223}
  {arXiv:1203.3223 [gr-qc]} \BibitemShut {NoStop}%
\bibitem [{\citenamefont {Dain}(2001)}]{Dain:2000hk}%
  \BibitemOpen
  \bibfield  {author} {\bibinfo {author} {\bibfnamefont {S.}~\bibnamefont
  {Dain}},\ }\href@noop {} {\bibfield  {journal} {\bibinfo  {journal} {Phys.
  Rev. Lett.}\ }\textbf {\bibinfo {volume} {87}},\ \bibinfo {pages} {121102}
  (\bibinfo {year} {2001})},\ \Eprint {http://arxiv.org/abs/gr-qc/0012023}
  {gr-qc/0012023} \BibitemShut {NoStop}%
\bibitem [{\citenamefont {York}(1999)}]{York99}%
  \BibitemOpen
  \bibfield  {author} {\bibinfo {author} {\bibfnamefont {J.~W.}\ \bibnamefont
  {York}},\ }\href@noop {} {\bibfield  {journal} {\bibinfo  {journal} {Phys.
  Rev. Lett.}\ }\textbf {\bibinfo {volume} {82}},\ \bibinfo {pages} {1350}
  (\bibinfo {year} {1999})}\BibitemShut {NoStop}%
\bibitem [{\citenamefont {Cook}(2000)}]{Cook:2000vr}%
  \BibitemOpen
  \bibfield  {author} {\bibinfo {author} {\bibfnamefont {G.~B.}\ \bibnamefont
  {Cook}},\ }\href@noop {} {\bibfield  {journal} {\bibinfo  {journal} {Living
  Rev. Rel.}\ }\textbf {\bibinfo {volume} {3}},\ \bibinfo {pages} {5} (\bibinfo
  {year} {2000})},\ \Eprint {http://arxiv.org/abs/gr-qc/0007085}
  {arXiv:gr-qc/0007085 [gr-qc]} \BibitemShut {NoStop}%
\bibitem [{\citenamefont {Pfeiffer}\ and\ \citenamefont
  {York}(2003)}]{Pfeiffer:2002iy}%
  \BibitemOpen
  \bibfield  {author} {\bibinfo {author} {\bibfnamefont {H.~P.}\ \bibnamefont
  {Pfeiffer}}\ and\ \bibinfo {author} {\bibfnamefont {J.}~\bibnamefont {York},
  \bibfnamefont {James~W.}},\ }\href@noop {} {\bibfield  {journal} {\bibinfo
  {journal} {Phys. Rev. D}\ }\textbf {\bibinfo {volume} {67}},\ \bibinfo
  {pages} {044022} (\bibinfo {year} {2003})},\ \Eprint
  {http://arxiv.org/abs/gr-qc/0207095} {gr-qc/0207095} \BibitemShut {NoStop}%
\bibitem [{\citenamefont {Lovelace}\ \emph {et~al.}(2012)\citenamefont
  {Lovelace}, \citenamefont {Boyle}, \citenamefont {Scheel},\ and\
  \citenamefont {Szilagyi}}]{Lovelace:2011nu}%
  \BibitemOpen
  \bibfield  {author} {\bibinfo {author} {\bibfnamefont {G.}~\bibnamefont
  {Lovelace}}, \bibinfo {author} {\bibfnamefont {M.}~\bibnamefont {Boyle}},
  \bibinfo {author} {\bibfnamefont {M.~A.}\ \bibnamefont {Scheel}}, \ and\
  \bibinfo {author} {\bibfnamefont {B.}~\bibnamefont {Szilagyi}},\ }\href@noop
  {} {\bibfield  {journal} {\bibinfo  {journal} {Class. Quant. Grav.}\ }\textbf
  {\bibinfo {volume} {29}},\ \bibinfo {pages} {045003} (\bibinfo {year}
  {2012})},\ \Eprint {http://arxiv.org/abs/1110.2229} {arXiv:1110.2229 [gr-qc]}
  \BibitemShut {NoStop}%
\bibitem [{\citenamefont {Lovelace}\ \emph {et~al.}(2015)\citenamefont
  {Lovelace}, \citenamefont {Scheel}, \citenamefont {Owen}, \citenamefont
  {Giesler}, \citenamefont {Katebi}, \citenamefont {Szil{\'a}gyi},
  \citenamefont {Chu}, \citenamefont {Demos}, \citenamefont {Hemberger},
  \citenamefont {Kidder}, \citenamefont {Pfeiffer},\ and\ \citenamefont
  {Afshari}}]{Lovelace:2014twa}%
  \BibitemOpen
  \bibfield  {author} {\bibinfo {author} {\bibfnamefont {G.}~\bibnamefont
  {Lovelace}}, \bibinfo {author} {\bibfnamefont {M.~A.}\ \bibnamefont
  {Scheel}}, \bibinfo {author} {\bibfnamefont {R.}~\bibnamefont {Owen}},
  \bibinfo {author} {\bibfnamefont {M.}~\bibnamefont {Giesler}}, \bibinfo
  {author} {\bibfnamefont {R.}~\bibnamefont {Katebi}}, \bibinfo {author}
  {\bibfnamefont {B.}~\bibnamefont {Szil{\'a}gyi}}, \bibinfo {author}
  {\bibfnamefont {T.}~\bibnamefont {Chu}}, \bibinfo {author} {\bibfnamefont
  {N.}~\bibnamefont {Demos}}, \bibinfo {author} {\bibfnamefont {D.~A.}\
  \bibnamefont {Hemberger}}, \bibinfo {author} {\bibfnamefont {L.~E.}\
  \bibnamefont {Kidder}}, \bibinfo {author} {\bibfnamefont {H.~P.}\
  \bibnamefont {Pfeiffer}}, \ and\ \bibinfo {author} {\bibfnamefont
  {N.}~\bibnamefont {Afshari}},\ }\href {\doibase
  10.1088/0264-9381/32/6/065007} {\bibfield  {journal} {\bibinfo  {journal}
  {Class. Quant. Grav.}\ }\textbf {\bibinfo {volume} {32}},\ \bibinfo {pages}
  {065007} (\bibinfo {year} {2015})},\ \Eprint {http://arxiv.org/abs/1411.7297}
  {arXiv:1411.7297 [gr-qc]} \BibitemShut {NoStop}%
\bibitem [{\citenamefont {Scheel}\ \emph {et~al.}(2015)\citenamefont {Scheel},
  \citenamefont {Giesler}, \citenamefont {Hemberger}, \citenamefont {Lovelace},
  \citenamefont {Kuper}, \citenamefont {Boyle}, \citenamefont {Szil{\'a}gyi},\
  and\ \citenamefont {Kidder}}]{Scheel:2014ina}%
  \BibitemOpen
  \bibfield  {author} {\bibinfo {author} {\bibfnamefont {M.~A.}\ \bibnamefont
  {Scheel}}, \bibinfo {author} {\bibfnamefont {M.}~\bibnamefont {Giesler}},
  \bibinfo {author} {\bibfnamefont {D.~A.}\ \bibnamefont {Hemberger}}, \bibinfo
  {author} {\bibfnamefont {G.}~\bibnamefont {Lovelace}}, \bibinfo {author}
  {\bibfnamefont {K.}~\bibnamefont {Kuper}}, \bibinfo {author} {\bibfnamefont
  {M.}~\bibnamefont {Boyle}}, \bibinfo {author} {\bibfnamefont
  {B.}~\bibnamefont {Szil{\'a}gyi}}, \ and\ \bibinfo {author} {\bibfnamefont
  {L.~E.}\ \bibnamefont {Kidder}},\ }\href {\doibase
  10.1088/0264-9381/32/10/105009} {\bibfield  {journal} {\bibinfo  {journal}
  {Class. Quant. Grav.}\ }\textbf {\bibinfo {volume} {32}},\ \bibinfo {pages}
  {105009} (\bibinfo {year} {2015})},\ \Eprint {http://arxiv.org/abs/1412.1803}
  {arXiv:1412.1803 [gr-qc]} \BibitemShut {NoStop}%
\bibitem [{\citenamefont {Hannam}\ \emph {et~al.}(2007)\citenamefont {Hannam},
  \citenamefont {Husa}, \citenamefont {Bruegmann}, \citenamefont {Gonzalez},\
  and\ \citenamefont {Sperhake}}]{Hannam:2006zt}%
  \BibitemOpen
  \bibfield  {author} {\bibinfo {author} {\bibfnamefont {M.}~\bibnamefont
  {Hannam}}, \bibinfo {author} {\bibfnamefont {S.}~\bibnamefont {Husa}},
  \bibinfo {author} {\bibfnamefont {B.}~\bibnamefont {Bruegmann}}, \bibinfo
  {author} {\bibfnamefont {J.~A.}\ \bibnamefont {Gonzalez}}, \ and\ \bibinfo
  {author} {\bibfnamefont {U.}~\bibnamefont {Sperhake}},\ }\href {\doibase
  10.1088/0264-9381/24/12/S02} {\bibfield  {journal} {\bibinfo  {journal}
  {Class. Quant. Grav.}\ }\textbf {\bibinfo {volume} {24}},\ \bibinfo {pages}
  {S15} (\bibinfo {year} {2007})},\ \Eprint
  {http://arxiv.org/abs/gr-qc/0612001} {arXiv:gr-qc/0612001 [gr-qc]}
  \BibitemShut {NoStop}%
\bibitem [{\citenamefont {L{\"o}ffler}\ \emph {et~al.}(2012)\citenamefont
  {L{\"o}ffler}, \citenamefont {Faber}, \citenamefont {Bentivegna},
  \citenamefont {Bode}, \citenamefont {Diener}, \citenamefont {Haas},
  \citenamefont {Hinder}, \citenamefont {Mundim}, \citenamefont {Ott},
  \citenamefont {Schnetter}, \citenamefont {Allen}, \citenamefont
  {Campanelli},\ and\ \citenamefont {Laguna}}]{Loffler:2011ay}%
  \BibitemOpen
  \bibfield  {author} {\bibinfo {author} {\bibfnamefont {F.}~\bibnamefont
  {L{\"o}ffler}}, \bibinfo {author} {\bibfnamefont {J.}~\bibnamefont {Faber}},
  \bibinfo {author} {\bibfnamefont {E.}~\bibnamefont {Bentivegna}}, \bibinfo
  {author} {\bibfnamefont {T.}~\bibnamefont {Bode}}, \bibinfo {author}
  {\bibfnamefont {P.}~\bibnamefont {Diener}}, \bibinfo {author} {\bibfnamefont
  {R.}~\bibnamefont {Haas}}, \bibinfo {author} {\bibfnamefont {I.}~\bibnamefont
  {Hinder}}, \bibinfo {author} {\bibfnamefont {B.~C.}\ \bibnamefont {Mundim}},
  \bibinfo {author} {\bibfnamefont {C.~D.}\ \bibnamefont {Ott}}, \bibinfo
  {author} {\bibfnamefont {E.}~\bibnamefont {Schnetter}}, \bibinfo {author}
  {\bibfnamefont {G.}~\bibnamefont {Allen}}, \bibinfo {author} {\bibfnamefont
  {M.}~\bibnamefont {Campanelli}}, \ and\ \bibinfo {author} {\bibfnamefont
  {P.}~\bibnamefont {Laguna}},\ }\href@noop {} {\bibfield  {journal} {\bibinfo
  {journal} {Class. Quant. Grav.}\ }\textbf {\bibinfo {volume} {29}},\ \bibinfo
  {pages} {115001} (\bibinfo {year} {2012})},\ \Eprint
  {http://arxiv.org/abs/1111.3344} {arXiv:1111.3344 [gr-qc]} \BibitemShut
  {NoStop}%
\bibitem [{ein()}]{einsteintoolkit}%
  \BibitemOpen
  \href@noop {} {}\bibinfo {note} {Einstein Toolkit home page: {\tt
  http://einsteintoolkit.org}}\BibitemShut {NoStop}%
\bibitem [{cac()}]{cactus_web}%
  \BibitemOpen
  \href@noop {} {}\bibinfo {note} {Cactus Computational Toolkit home page: {\tt
  http://cactuscode.org}}\BibitemShut {NoStop}%
\bibitem [{\citenamefont {Schnetter}\ \emph {et~al.}(2004)\citenamefont
  {Schnetter}, \citenamefont {Hawley},\ and\ \citenamefont
  {Hawke}}]{Schnetter-etal-03b}%
  \BibitemOpen
  \bibfield  {author} {\bibinfo {author} {\bibfnamefont {E.}~\bibnamefont
  {Schnetter}}, \bibinfo {author} {\bibfnamefont {S.~H.}\ \bibnamefont
  {Hawley}}, \ and\ \bibinfo {author} {\bibfnamefont {I.}~\bibnamefont
  {Hawke}},\ }\href@noop {} {\bibfield  {journal} {\bibinfo  {journal} {Class.
  Quant. Grav.}\ }\textbf {\bibinfo {volume} {21}},\ \bibinfo {pages} {1465}
  (\bibinfo {year} {2004})},\ \Eprint {http://arxiv.org/abs/gr-qc/0310042}
  {gr-qc/0310042} \BibitemShut {NoStop}%
\bibitem [{\citenamefont {Marronetti}\ \emph {et~al.}(2000)\citenamefont
  {Marronetti}, \citenamefont {Huq}, \citenamefont {Laguna}, \citenamefont
  {Lehner}, \citenamefont {Matzner} \emph {et~al.}}]{Marronetti:2000rw}%
  \BibitemOpen
  \bibfield  {author} {\bibinfo {author} {\bibfnamefont {P.}~\bibnamefont
  {Marronetti}}, \bibinfo {author} {\bibfnamefont {M.}~\bibnamefont {Huq}},
  \bibinfo {author} {\bibfnamefont {P.}~\bibnamefont {Laguna}}, \bibinfo
  {author} {\bibfnamefont {L.}~\bibnamefont {Lehner}}, \bibinfo {author}
  {\bibfnamefont {R.~A.}\ \bibnamefont {Matzner}},  \emph {et~al.},\ }\href
  {\doibase 10.1103/PhysRevD.62.024017} {\bibfield  {journal} {\bibinfo
  {journal} {Phys. Rev.}\ }\textbf {\bibinfo {volume} {D62}},\ \bibinfo {pages}
  {024017} (\bibinfo {year} {2000})},\ \Eprint
  {http://arxiv.org/abs/gr-qc/0001077} {arXiv:gr-qc/0001077 [gr-qc]}
  \BibitemShut {NoStop}%
\bibitem [{\citenamefont {Bonning}\ \emph {et~al.}(2003)\citenamefont
  {Bonning}, \citenamefont {Marronetti}, \citenamefont {Neilsen},\ and\
  \citenamefont {Matzner}}]{Bonning:2003im}%
  \BibitemOpen
  \bibfield  {author} {\bibinfo {author} {\bibfnamefont {E.}~\bibnamefont
  {Bonning}}, \bibinfo {author} {\bibfnamefont {P.}~\bibnamefont {Marronetti}},
  \bibinfo {author} {\bibfnamefont {D.}~\bibnamefont {Neilsen}}, \ and\
  \bibinfo {author} {\bibfnamefont {R.}~\bibnamefont {Matzner}},\ }\href
  {\doibase 10.1103/PhysRevD.68.044019} {\bibfield  {journal} {\bibinfo
  {journal} {Phys. Rev.}\ }\textbf {\bibinfo {volume} {D68}},\ \bibinfo {pages}
  {044019} (\bibinfo {year} {2003})},\ \Eprint
  {http://arxiv.org/abs/gr-qc/0305071} {arXiv:gr-qc/0305071 [gr-qc]}
  \BibitemShut {NoStop}%
\bibitem [{\citenamefont {Ansorg}\ \emph {et~al.}(2004)\citenamefont {Ansorg},
  \citenamefont {Br\"ugmann},\ and\ \citenamefont {Tichy}}]{Ansorg:2004ds}%
  \BibitemOpen
  \bibfield  {author} {\bibinfo {author} {\bibfnamefont {M.}~\bibnamefont
  {Ansorg}}, \bibinfo {author} {\bibfnamefont {B.}~\bibnamefont {Br\"ugmann}},
  \ and\ \bibinfo {author} {\bibfnamefont {W.}~\bibnamefont {Tichy}},\
  }\href@noop {} {\bibfield  {journal} {\bibinfo  {journal} {Phys. Rev.}\
  }\textbf {\bibinfo {volume} {D70}},\ \bibinfo {pages} {064011} (\bibinfo
  {year} {2004})},\ \Eprint {http://arxiv.org/abs/gr-qc/0404056}
  {gr-qc/0404056} \BibitemShut {NoStop}%
\bibitem [{\citenamefont {Tichy}\ \emph {et~al.}(2003)\citenamefont {Tichy},
  \citenamefont {Br{\"u}gmann}, \citenamefont {Campanelli},\ and\ \citenamefont
  {Diener}}]{Tichy:2002ec}%
  \BibitemOpen
  \bibfield  {author} {\bibinfo {author} {\bibfnamefont {W.}~\bibnamefont
  {Tichy}}, \bibinfo {author} {\bibfnamefont {B.}~\bibnamefont {Br{\"u}gmann}},
  \bibinfo {author} {\bibfnamefont {M.}~\bibnamefont {Campanelli}}, \ and\
  \bibinfo {author} {\bibfnamefont {P.}~\bibnamefont {Diener}},\ }\href@noop {}
  {\bibfield  {journal} {\bibinfo  {journal} {Phys. Rev.}\ }\textbf {\bibinfo
  {volume} {D67}},\ \bibinfo {pages} {064008} (\bibinfo {year} {2003})},\
  \Eprint {http://arxiv.org/abs/gr-qc/0207011} {gr-qc/0207011} \BibitemShut
  {NoStop}%
\bibitem [{\citenamefont {Kelly}\ \emph {et~al.}(2007)\citenamefont {Kelly},
  \citenamefont {Tichy}, \citenamefont {Campanelli},\ and\ \citenamefont
  {Whiting}}]{Kelly:2007uc}%
  \BibitemOpen
  \bibfield  {author} {\bibinfo {author} {\bibfnamefont {B.~J.}\ \bibnamefont
  {Kelly}}, \bibinfo {author} {\bibfnamefont {W.}~\bibnamefont {Tichy}},
  \bibinfo {author} {\bibfnamefont {M.}~\bibnamefont {Campanelli}}, \ and\
  \bibinfo {author} {\bibfnamefont {B.~F.}\ \bibnamefont {Whiting}},\ }\href
  {\doibase 10.1103/PhysRevD.76.024008} {\bibfield  {journal} {\bibinfo
  {journal} {Phys. Rev.}\ }\textbf {\bibinfo {volume} {D76}},\ \bibinfo {pages}
  {024008} (\bibinfo {year} {2007})},\ \Eprint {http://arxiv.org/abs/0704.0628}
  {arXiv:0704.0628 [gr-qc]} \BibitemShut {NoStop}%
\bibitem [{\citenamefont {Kelly}\ \emph {et~al.}(2010)\citenamefont {Kelly},
  \citenamefont {Tichy}, \citenamefont {Zlochower}, \citenamefont
  {Campanelli},\ and\ \citenamefont {Whiting}}]{Kelly:2009js}%
  \BibitemOpen
  \bibfield  {author} {\bibinfo {author} {\bibfnamefont {B.~J.}\ \bibnamefont
  {Kelly}}, \bibinfo {author} {\bibfnamefont {W.}~\bibnamefont {Tichy}},
  \bibinfo {author} {\bibfnamefont {Y.}~\bibnamefont {Zlochower}}, \bibinfo
  {author} {\bibfnamefont {M.}~\bibnamefont {Campanelli}}, \ and\ \bibinfo
  {author} {\bibfnamefont {B.~F.}\ \bibnamefont {Whiting}},\ }\href {\doibase
  10.1088/0264-9381/27/11/114005} {\bibfield  {journal} {\bibinfo  {journal}
  {Class. Quant. Grav.}\ }\textbf {\bibinfo {volume} {27}},\ \bibinfo {pages}
  {114005} (\bibinfo {year} {2010})},\ \Eprint {http://arxiv.org/abs/0912.5311}
  {arXiv:0912.5311 [gr-qc]} \BibitemShut {NoStop}%
\bibitem [{\citenamefont {Mundim}\ \emph {et~al.}(2011)\citenamefont {Mundim},
  \citenamefont {Kelly}, \citenamefont {Zlochower}, \citenamefont {Nakano},\
  and\ \citenamefont {Campanelli}}]{Mundim:2010hu}%
  \BibitemOpen
  \bibfield  {author} {\bibinfo {author} {\bibfnamefont {B.~C.}\ \bibnamefont
  {Mundim}}, \bibinfo {author} {\bibfnamefont {B.~J.}\ \bibnamefont {Kelly}},
  \bibinfo {author} {\bibfnamefont {Y.}~\bibnamefont {Zlochower}}, \bibinfo
  {author} {\bibfnamefont {H.}~\bibnamefont {Nakano}}, \ and\ \bibinfo {author}
  {\bibfnamefont {M.}~\bibnamefont {Campanelli}},\ }\href {\doibase
  10.1088/0264-9381/28/13/134003} {\bibfield  {journal} {\bibinfo  {journal}
  {Class. Quant. Grav.}\ }\textbf {\bibinfo {volume} {28}},\ \bibinfo {pages}
  {134003} (\bibinfo {year} {2011})},\ \Eprint {http://arxiv.org/abs/1012.0886}
  {arXiv:1012.0886 [gr-qc]} \BibitemShut {NoStop}%
\bibitem [{\citenamefont {Cook}\ and\ \citenamefont
  {York}(1990{\natexlab{b}})}]{Cook90a}%
  \BibitemOpen
  \bibfield  {author} {\bibinfo {author} {\bibfnamefont {G.}~\bibnamefont
  {Cook}}\ and\ \bibinfo {author} {\bibfnamefont {J.~W.}\ \bibnamefont
  {York}},\ }\href@noop {} {\bibfield  {journal} {\bibinfo  {journal} {Phys.
  Rev. D}\ }\textbf {\bibinfo {volume} {41}},\ \bibinfo {pages} {1077}
  (\bibinfo {year} {1990}{\natexlab{b}})}\BibitemShut {NoStop}%
\bibitem [{\citenamefont {Healy}\ \emph {et~al.}(2015)\citenamefont {Healy},
  \citenamefont {Ruchlin},\ and\ \citenamefont {Lousto}}]{Healy:2015mla}%
  \BibitemOpen
  \bibfield  {author} {\bibinfo {author} {\bibfnamefont {J.}~\bibnamefont
  {Healy}}, \bibinfo {author} {\bibfnamefont {I.}~\bibnamefont {Ruchlin}}, \
  and\ \bibinfo {author} {\bibfnamefont {C.~O.}\ \bibnamefont {Lousto}},\
  }\href@noop {} {\  (\bibinfo {year} {2015})},\ \Eprint
  {http://arxiv.org/abs/1506.06153} {arXiv:1506.06153 [gr-qc]} \BibitemShut
  {NoStop}%
\bibitem [{SXS()}]{SXS:catalog}%
  \BibitemOpen
  \href@noop {} {}\bibinfo {howpublished}
  {\url{http://www.black-holes.org/waveforms}}\BibitemShut {NoStop}%
\bibitem [{\citenamefont {Alcubierre}(2003)}]{Alcubierre02b}%
  \BibitemOpen
  \bibfield  {author} {\bibinfo {author} {\bibfnamefont {M.}~\bibnamefont
  {Alcubierre}},\ }\href@noop {} {\bibfield  {journal} {\bibinfo  {journal}
  {Class. Quant. Grav.}\ }\textbf {\bibinfo {volume} {20}},\ \bibinfo {pages}
  {607} (\bibinfo {year} {2003})},\ \Eprint
  {http://arxiv.org/abs/gr-qc/0210050} {gr-qc/0210050} \BibitemShut {NoStop}%
\bibitem [{\citenamefont {Arnowitt}\ \emph {et~al.}(1962)\citenamefont
  {Arnowitt}, \citenamefont {Deser},\ and\ \citenamefont
  {Misner}}]{Arnowitt62}%
  \BibitemOpen
  \bibfield  {author} {\bibinfo {author} {\bibfnamefont {R.}~\bibnamefont
  {Arnowitt}}, \bibinfo {author} {\bibfnamefont {S.}~\bibnamefont {Deser}}, \
  and\ \bibinfo {author} {\bibfnamefont {C.~W.}\ \bibnamefont {Misner}},\ }in\
  \href@noop {} {\emph {\bibinfo {booktitle} {Gravitation: An Introduction to
  Current Research}}},\ \bibinfo {editor} {edited by\ \bibinfo {editor}
  {\bibfnamefont {L.}~\bibnamefont {Witten}}}\ (\bibinfo  {publisher} {John
  Wiley},\ \bibinfo {address} {New York},\ \bibinfo {year} {1962})\ pp.\
  \bibinfo {pages} {227--265},\ \Eprint {http://arxiv.org/abs/gr-qc/0405109}
  {gr-qc/0405109} \BibitemShut {NoStop}%
\bibitem [{\citenamefont {{Alcubierre}}(2008)}]{AlcubierreBook2008}%
  \BibitemOpen
  \bibfield  {author} {\bibinfo {author} {\bibfnamefont {M.}~\bibnamefont
  {{Alcubierre}}},\ }\href@noop {} {\emph {\bibinfo {title} {Introduction to
  3+1 Numerical Relativity, by Miguel Alcubierre.~ISBN 978-0-19-920567-7
  (HB).~Published by Oxford University Press, Oxford, UK, 2008.}}}\ (\bibinfo
  {publisher} {Oxford University Press},\ \bibinfo {year} {2008})\BibitemShut
  {NoStop}%
\bibitem [{\citenamefont {Brandt}\ and\ \citenamefont
  {Br{\"u}gmann}(1997)}]{Brandt97b}%
  \BibitemOpen
  \bibfield  {author} {\bibinfo {author} {\bibfnamefont {S.}~\bibnamefont
  {Brandt}}\ and\ \bibinfo {author} {\bibfnamefont {B.}~\bibnamefont
  {Br{\"u}gmann}},\ }\href@noop {} {\bibfield  {journal} {\bibinfo  {journal}
  {Phys. Rev. Lett.}\ }\textbf {\bibinfo {volume} {78}},\ \bibinfo {pages}
  {3606} (\bibinfo {year} {1997})},\ \Eprint
  {http://arxiv.org/abs/gr-qc/9703066} {gr-qc/9703066} \BibitemShut {NoStop}%
\bibitem [{\citenamefont {Brandt}\ and\ \citenamefont
  {Seidel}(1996)}]{Brandt:1996si}%
  \BibitemOpen
  \bibfield  {author} {\bibinfo {author} {\bibfnamefont {S.~R.}\ \bibnamefont
  {Brandt}}\ and\ \bibinfo {author} {\bibfnamefont {E.}~\bibnamefont
  {Seidel}},\ }\href {\doibase 10.1103/PhysRevD.54.1403} {\bibfield  {journal}
  {\bibinfo  {journal} {Phys. Rev.}\ }\textbf {\bibinfo {volume} {D54}},\
  \bibinfo {pages} {1403} (\bibinfo {year} {1996})},\ \Eprint
  {http://arxiv.org/abs/gr-qc/9601010} {arXiv:gr-qc/9601010} \BibitemShut
  {NoStop}%
\bibitem [{\citenamefont {Krivan}\ and\ \citenamefont
  {Price}(1998)}]{Krivan:1998td}%
  \BibitemOpen
  \bibfield  {author} {\bibinfo {author} {\bibfnamefont {W.}~\bibnamefont
  {Krivan}}\ and\ \bibinfo {author} {\bibfnamefont {R.~H.}\ \bibnamefont
  {Price}},\ }\href@noop {} {\bibfield  {journal} {\bibinfo  {journal} {Phys.
  Rev. D}\ }\textbf {\bibinfo {volume} {58}},\ \bibinfo {pages} {104003}
  (\bibinfo {year} {1998})}\BibitemShut {NoStop}%
\bibitem [{\citenamefont {Baker}\ \emph {et~al.}(2002)\citenamefont {Baker},
  \citenamefont {Campanelli},\ and\ \citenamefont {Lousto}}]{Baker:2001sf}%
  \BibitemOpen
  \bibfield  {author} {\bibinfo {author} {\bibfnamefont {J.~G.}\ \bibnamefont
  {Baker}}, \bibinfo {author} {\bibfnamefont {M.}~\bibnamefont {Campanelli}}, \
  and\ \bibinfo {author} {\bibfnamefont {C.~O.}\ \bibnamefont {Lousto}},\
  }\href {\doibase 10.1103/PhysRevD.65.044001} {\bibfield  {journal} {\bibinfo
  {journal} {Phys. Rev.}\ }\textbf {\bibinfo {volume} {D65}},\ \bibinfo {pages}
  {044001} (\bibinfo {year} {2002})},\ \Eprint
  {http://arxiv.org/abs/gr-qc/0104063} {arXiv:gr-qc/0104063 [gr-qc]}
  \BibitemShut {NoStop}%
\bibitem [{\citenamefont {Zlochower}\ \emph {et~al.}(2005)\citenamefont
  {Zlochower}, \citenamefont {Baker}, \citenamefont {Campanelli},\ and\
  \citenamefont {Lousto}}]{Zlochower:2005bj}%
  \BibitemOpen
  \bibfield  {author} {\bibinfo {author} {\bibfnamefont {Y.}~\bibnamefont
  {Zlochower}}, \bibinfo {author} {\bibfnamefont {J.~G.}\ \bibnamefont
  {Baker}}, \bibinfo {author} {\bibfnamefont {M.}~\bibnamefont {Campanelli}}, \
  and\ \bibinfo {author} {\bibfnamefont {C.~O.}\ \bibnamefont {Lousto}},\
  }\href {\doibase 10.1103/PhysRevD.72.024021} {\bibfield  {journal} {\bibinfo
  {journal} {Phys. Rev.}\ }\textbf {\bibinfo {volume} {D72}},\ \bibinfo {pages}
  {024021} (\bibinfo {year} {2005})},\ \Eprint
  {http://arxiv.org/abs/gr-qc/0505055} {arXiv:gr-qc/0505055} \BibitemShut
  {NoStop}%
\bibitem [{\citenamefont {Zilhao}\ and\ \citenamefont
  {Noble}(2014)}]{Zilhao:2013dta}%
  \BibitemOpen
  \bibfield  {author} {\bibinfo {author} {\bibfnamefont {M.}~\bibnamefont
  {Zilhao}}\ and\ \bibinfo {author} {\bibfnamefont {S.~C.}\ \bibnamefont
  {Noble}},\ }\href {\doibase 10.1088/0264-9381/31/6/065013} {\bibfield
  {journal} {\bibinfo  {journal} {Class. Quant. Grav.}\ }\textbf {\bibinfo
  {volume} {31}},\ \bibinfo {pages} {065013} (\bibinfo {year} {2014})},\
  \Eprint {http://arxiv.org/abs/1309.2960} {arXiv:1309.2960 [gr-qc]}
  \BibitemShut {NoStop}%
\bibitem [{\citenamefont {Liu}\ \emph {et~al.}(2009)\citenamefont {Liu},
  \citenamefont {Etienne},\ and\ \citenamefont {Shapiro}}]{Liu:2009al}%
  \BibitemOpen
  \bibfield  {author} {\bibinfo {author} {\bibfnamefont {Y.~T.}\ \bibnamefont
  {Liu}}, \bibinfo {author} {\bibfnamefont {Z.~B.}\ \bibnamefont {Etienne}}, \
  and\ \bibinfo {author} {\bibfnamefont {S.~L.}\ \bibnamefont {Shapiro}},\
  }\href {\doibase 10.1103/PhysRevD.80.121503} {\bibfield  {journal} {\bibinfo
  {journal} {Phys. Rev.}\ }\textbf {\bibinfo {volume} {D80}},\ \bibinfo {pages}
  {121503} (\bibinfo {year} {2009})},\ \Eprint {http://arxiv.org/abs/1001.4077}
  {arXiv:1001.4077 [gr-qc]} \BibitemShut {NoStop}%
\bibitem [{Note1()}]{Note1}%
  \BibitemOpen
  \bibinfo {note} {Http://www.holoborodko.com/pavel/mpfr/}\BibitemShut
  {NoStop}%
\bibitem [{Note2()}]{Note2}%
  \BibitemOpen
  \bibinfo {note} {Http://www.mpfr.org/}\BibitemShut {NoStop}%
\bibitem [{\citenamefont {Pfeiffer}\ \emph {et~al.}(2002)\citenamefont
  {Pfeiffer}, \citenamefont {Cook},\ and\ \citenamefont
  {Teukolsky}}]{Pfeiffer:2002xz}%
  \BibitemOpen
  \bibfield  {author} {\bibinfo {author} {\bibfnamefont {H.~P.}\ \bibnamefont
  {Pfeiffer}}, \bibinfo {author} {\bibfnamefont {G.~B.}\ \bibnamefont {Cook}},
  \ and\ \bibinfo {author} {\bibfnamefont {S.~A.}\ \bibnamefont {Teukolsky}},\
  }\href@noop {} {\bibfield  {journal} {\bibinfo  {journal} {Phys. Rev. D}\
  }\textbf {\bibinfo {volume} {66}},\ \bibinfo {pages} {024047} (\bibinfo
  {year} {2002})},\ \Eprint {http://arxiv.org/abs/gr-qc/0203085}
  {gr-qc/0203085} \BibitemShut {NoStop}%
\bibitem [{\citenamefont {Brown}\ \emph {et~al.}(2007)\citenamefont {Brown},
  \citenamefont {Sarbach}, \citenamefont {Schnetter}, \citenamefont {Tiglio},
  \citenamefont {Diener} \emph {et~al.}}]{Brown:2007pg}%
  \BibitemOpen
  \bibfield  {author} {\bibinfo {author} {\bibfnamefont {D.}~\bibnamefont
  {Brown}}, \bibinfo {author} {\bibfnamefont {O.}~\bibnamefont {Sarbach}},
  \bibinfo {author} {\bibfnamefont {E.}~\bibnamefont {Schnetter}}, \bibinfo
  {author} {\bibfnamefont {M.}~\bibnamefont {Tiglio}}, \bibinfo {author}
  {\bibfnamefont {P.}~\bibnamefont {Diener}},  \emph {et~al.},\ }\href
  {\doibase 10.1103/PhysRevD.76.081503} {\bibfield  {journal} {\bibinfo
  {journal} {Phys. Rev.}\ }\textbf {\bibinfo {volume} {D76}},\ \bibinfo {pages}
  {081503} (\bibinfo {year} {2007})},\ \Eprint {http://arxiv.org/abs/0707.3101}
  {arXiv:0707.3101 [gr-qc]} \BibitemShut {NoStop}%
\bibitem [{\citenamefont {Brown}\ \emph {et~al.}(2009)\citenamefont {Brown},
  \citenamefont {Diener}, \citenamefont {Sarbach}, \citenamefont {Schnetter},\
  and\ \citenamefont {Tiglio}}]{Brown:2008sb}%
  \BibitemOpen
  \bibfield  {author} {\bibinfo {author} {\bibfnamefont {D.}~\bibnamefont
  {Brown}}, \bibinfo {author} {\bibfnamefont {P.}~\bibnamefont {Diener}},
  \bibinfo {author} {\bibfnamefont {O.}~\bibnamefont {Sarbach}}, \bibinfo
  {author} {\bibfnamefont {E.}~\bibnamefont {Schnetter}}, \ and\ \bibinfo
  {author} {\bibfnamefont {M.}~\bibnamefont {Tiglio}},\ }\href {\doibase
  10.1103/PhysRevD.79.044023} {\bibfield  {journal} {\bibinfo  {journal} {Phys.
  Rev.}\ }\textbf {\bibinfo {volume} {D79}},\ \bibinfo {pages} {044023}
  (\bibinfo {year} {2009})},\ \Eprint {http://arxiv.org/abs/0809.3533}
  {arXiv:0809.3533 [gr-qc]} \BibitemShut {NoStop}%
\bibitem [{\citenamefont {Lovelace}(2007)}]{Lovelace:2007zz}%
  \BibitemOpen
  \bibfield  {author} {\bibinfo {author} {\bibfnamefont {G.}~\bibnamefont
  {Lovelace}},\ }\emph {\bibinfo {title} {{Topics in Gravitational-Wave
  Physics}}},\ \href@noop {} {Ph.D. thesis},\ \bibinfo  {school} {Caltech}
  (\bibinfo {year} {2007})\BibitemShut {NoStop}%
\bibitem [{\citenamefont {Grandclement}\ \emph {et~al.}(2001)\citenamefont
  {Grandclement}, \citenamefont {Bonazzola}, \citenamefont {Gourgoulhon},\ and\
  \citenamefont {Marck}}]{Grandclement:2000xv}%
  \BibitemOpen
  \bibfield  {author} {\bibinfo {author} {\bibfnamefont {P.}~\bibnamefont
  {Grandclement}}, \bibinfo {author} {\bibfnamefont {S.}~\bibnamefont
  {Bonazzola}}, \bibinfo {author} {\bibfnamefont {E.}~\bibnamefont
  {Gourgoulhon}}, \ and\ \bibinfo {author} {\bibfnamefont {J.~A.}\ \bibnamefont
  {Marck}},\ }\href {\doibase 10.1006/jcph.2001.6734} {\bibfield  {journal}
  {\bibinfo  {journal} {J. Comput. Phys.}\ }\textbf {\bibinfo {volume} {170}},\
  \bibinfo {pages} {231} (\bibinfo {year} {2001})},\ \Eprint
  {http://arxiv.org/abs/gr-qc/0003072} {arXiv:gr-qc/0003072 [gr-qc]}
  \BibitemShut {NoStop}%
\bibitem [{\citenamefont {Marronetti}\ \emph {et~al.}(2008)\citenamefont
  {Marronetti}, \citenamefont {Tichy}, \citenamefont {Br{\"u}gmann},
  \citenamefont {Gonzalez},\ and\ \citenamefont
  {Sperhake}}]{Marronetti:2007wz}%
  \BibitemOpen
  \bibfield  {author} {\bibinfo {author} {\bibfnamefont {P.}~\bibnamefont
  {Marronetti}}, \bibinfo {author} {\bibfnamefont {W.}~\bibnamefont {Tichy}},
  \bibinfo {author} {\bibfnamefont {B.}~\bibnamefont {Br{\"u}gmann}}, \bibinfo
  {author} {\bibfnamefont {J.}~\bibnamefont {Gonzalez}}, \ and\ \bibinfo
  {author} {\bibfnamefont {U.}~\bibnamefont {Sperhake}},\ }\href {\doibase
  10.1103/PhysRevD.77.064010} {\bibfield  {journal} {\bibinfo  {journal} {Phys.
  Rev.}\ }\textbf {\bibinfo {volume} {D77}},\ \bibinfo {pages} {064010}
  (\bibinfo {year} {2008})},\ \Eprint {http://arxiv.org/abs/0709.2160}
  {arXiv:0709.2160 [gr-qc]} \BibitemShut {NoStop}%
\bibitem [{\citenamefont {Thornburg}(2004)}]{Thornburg2003:AH-finding}%
  \BibitemOpen
  \bibfield  {author} {\bibinfo {author} {\bibfnamefont {J.}~\bibnamefont
  {Thornburg}},\ }\href {\doibase 10.1088/0264-9381/21/2/026} {\bibfield
  {journal} {\bibinfo  {journal} {Class. Quant. Grav.}\ }\textbf {\bibinfo
  {volume} {21}},\ \bibinfo {pages} {743} (\bibinfo {year} {2004})},\ \Eprint
  {http://arxiv.org/abs/gr-qc/0306056} {gr-qc/0306056} \BibitemShut {NoStop}%
\bibitem [{\citenamefont {Dreyer}\ \emph {et~al.}(2003)\citenamefont {Dreyer},
  \citenamefont {Krishnan}, \citenamefont {Shoemaker},\ and\ \citenamefont
  {Schnetter}}]{Dreyer02a}%
  \BibitemOpen
  \bibfield  {author} {\bibinfo {author} {\bibfnamefont {O.}~\bibnamefont
  {Dreyer}}, \bibinfo {author} {\bibfnamefont {B.}~\bibnamefont {Krishnan}},
  \bibinfo {author} {\bibfnamefont {D.}~\bibnamefont {Shoemaker}}, \ and\
  \bibinfo {author} {\bibfnamefont {E.}~\bibnamefont {Schnetter}},\ }\href@noop
  {} {\bibfield  {journal} {\bibinfo  {journal} {Phys. Rev.}\ }\textbf
  {\bibinfo {volume} {D67}},\ \bibinfo {pages} {024018} (\bibinfo {year}
  {2003})},\ \Eprint {http://arxiv.org/abs/gr-qc/0206008} {gr-qc/0206008}
  \BibitemShut {NoStop}%
\bibitem [{\citenamefont {Winicour}(1980)}]{Winicour_AMGR}%
  \BibitemOpen
  \bibfield  {author} {\bibinfo {author} {\bibfnamefont {J.}~\bibnamefont
  {Winicour}},\ }in\ \href@noop {} {\emph {\bibinfo {booktitle} {General
  Relativity and Gravitation Vol 2}}},\ \bibinfo {editor} {edited by\ \bibinfo
  {editor} {\bibfnamefont {A.}~\bibnamefont {Held}}}\ (\bibinfo  {publisher}
  {Plenum},\ \bibinfo {address} {New York},\ \bibinfo {year} {1980})\ pp.\
  \bibinfo {pages} {71--96}\BibitemShut {NoStop}%
\bibitem [{\citenamefont {Campanelli}\ and\ \citenamefont
  {Lousto}(1999)}]{Campanelli:1998jv}%
  \BibitemOpen
  \bibfield  {author} {\bibinfo {author} {\bibfnamefont {M.}~\bibnamefont
  {Campanelli}}\ and\ \bibinfo {author} {\bibfnamefont {C.~O.}\ \bibnamefont
  {Lousto}},\ }\href {\doibase 10.1103/PhysRevD.59.124022} {\bibfield
  {journal} {\bibinfo  {journal} {Phys. Rev.}\ }\textbf {\bibinfo {volume}
  {D59}},\ \bibinfo {pages} {124022} (\bibinfo {year} {1999})},\ \Eprint
  {http://arxiv.org/abs/gr-qc/9811019} {arXiv:gr-qc/9811019 [gr-qc]}
  \BibitemShut {NoStop}%
\bibitem [{\citenamefont {Lousto}\ and\ \citenamefont
  {Zlochower}(2007)}]{Lousto:2007mh}%
  \BibitemOpen
  \bibfield  {author} {\bibinfo {author} {\bibfnamefont {C.~O.}\ \bibnamefont
  {Lousto}}\ and\ \bibinfo {author} {\bibfnamefont {Y.}~\bibnamefont
  {Zlochower}},\ }\href@noop {} {\bibfield  {journal} {\bibinfo  {journal}
  {Phys. Rev.}\ }\textbf {\bibinfo {volume} {D76}},\ \bibinfo {pages}
  {041502(R)} (\bibinfo {year} {2007})},\ \Eprint
  {http://arxiv.org/abs/gr-qc/0703061} {gr-qc/0703061} \BibitemShut {NoStop}%
\bibitem [{\citenamefont {Alcubierre}\ \emph {et~al.}(2003)\citenamefont
  {Alcubierre}, \citenamefont {Br\"ugmann}, \citenamefont {Diener},
  \citenamefont {Koppitz}, \citenamefont {Pollney}, \citenamefont {Seidel},\
  and\ \citenamefont {Takahashi}}]{Alcubierre02a}%
  \BibitemOpen
  \bibfield  {author} {\bibinfo {author} {\bibfnamefont {M.}~\bibnamefont
  {Alcubierre}}, \bibinfo {author} {\bibfnamefont {B.}~\bibnamefont
  {Br\"ugmann}}, \bibinfo {author} {\bibfnamefont {P.}~\bibnamefont {Diener}},
  \bibinfo {author} {\bibfnamefont {M.}~\bibnamefont {Koppitz}}, \bibinfo
  {author} {\bibfnamefont {D.}~\bibnamefont {Pollney}}, \bibinfo {author}
  {\bibfnamefont {E.}~\bibnamefont {Seidel}}, \ and\ \bibinfo {author}
  {\bibfnamefont {R.}~\bibnamefont {Takahashi}},\ }\href@noop {} {\bibfield
  {journal} {\bibinfo  {journal} {Phys. Rev.}\ }\textbf {\bibinfo {volume}
  {D67}},\ \bibinfo {pages} {084023} (\bibinfo {year} {2003})},\ \Eprint
  {http://arxiv.org/abs/gr-qc/0206072} {gr-qc/0206072} \BibitemShut {NoStop}%
\bibitem [{\citenamefont {van Meter}\ \emph {et~al.}(2006)\citenamefont {van
  Meter}, \citenamefont {Baker}, \citenamefont {Koppitz},\ and\ \citenamefont
  {Choi}}]{vanMeter:2006vi}%
  \BibitemOpen
  \bibfield  {author} {\bibinfo {author} {\bibfnamefont {J.~R.}\ \bibnamefont
  {van Meter}}, \bibinfo {author} {\bibfnamefont {J.~G.}\ \bibnamefont
  {Baker}}, \bibinfo {author} {\bibfnamefont {M.}~\bibnamefont {Koppitz}}, \
  and\ \bibinfo {author} {\bibfnamefont {D.-I.}\ \bibnamefont {Choi}},\
  }\href@noop {} {\bibfield  {journal} {\bibinfo  {journal} {Phys. Rev.}\
  }\textbf {\bibinfo {volume} {D73}},\ \bibinfo {pages} {124011} (\bibinfo
  {year} {2006})},\ \Eprint {http://arxiv.org/abs/gr-qc/0605030}
  {gr-qc/0605030} \BibitemShut {NoStop}%
\bibitem [{\citenamefont {Zlochower}\ \emph {et~al.}(2012)\citenamefont
  {Zlochower}, \citenamefont {Ponce},\ and\ \citenamefont
  {Lousto}}]{Zlochower:2012fk}%
  \BibitemOpen
  \bibfield  {author} {\bibinfo {author} {\bibfnamefont {Y.}~\bibnamefont
  {Zlochower}}, \bibinfo {author} {\bibfnamefont {M.}~\bibnamefont {Ponce}}, \
  and\ \bibinfo {author} {\bibfnamefont {C.~O.}\ \bibnamefont {Lousto}},\
  }\href {\doibase 10.1103/PhysRevD.86.104056} {\bibfield  {journal} {\bibinfo
  {journal} {Phys. Rev.}\ }\textbf {\bibinfo {volume} {D86}},\ \bibinfo {pages}
  {104056} (\bibinfo {year} {2012})},\ \Eprint {http://arxiv.org/abs/1208.5494}
  {arXiv:1208.5494 [gr-qc]} \BibitemShut {NoStop}%
\bibitem [{\citenamefont {Etienne}\ \emph {et~al.}(2014)\citenamefont
  {Etienne}, \citenamefont {Baker}, \citenamefont {Paschalidis}, \citenamefont
  {Kelly},\ and\ \citenamefont {Shapiro}}]{Etienne:2014tia}%
  \BibitemOpen
  \bibfield  {author} {\bibinfo {author} {\bibfnamefont {Z.~B.}\ \bibnamefont
  {Etienne}}, \bibinfo {author} {\bibfnamefont {J.~G.}\ \bibnamefont {Baker}},
  \bibinfo {author} {\bibfnamefont {V.}~\bibnamefont {Paschalidis}}, \bibinfo
  {author} {\bibfnamefont {B.~J.}\ \bibnamefont {Kelly}}, \ and\ \bibinfo
  {author} {\bibfnamefont {S.~L.}\ \bibnamefont {Shapiro}},\ }\href {\doibase
  10.1103/PhysRevD.90.064032} {\bibfield  {journal} {\bibinfo  {journal} {Phys.
  Rev.}\ }\textbf {\bibinfo {volume} {D90}},\ \bibinfo {pages} {064032}
  (\bibinfo {year} {2014})},\ \Eprint {http://arxiv.org/abs/1404.6523}
  {arXiv:1404.6523 [astro-ph.HE]} \BibitemShut {NoStop}%
\bibitem [{\citenamefont {Campanelli}\ \emph
  {et~al.}(2006{\natexlab{c}})\citenamefont {Campanelli}, \citenamefont
  {Lousto},\ and\ \citenamefont {Zlochower}}]{Campanelli:2006fg}%
  \BibitemOpen
  \bibfield  {author} {\bibinfo {author} {\bibfnamefont {M.}~\bibnamefont
  {Campanelli}}, \bibinfo {author} {\bibfnamefont {C.~O.}\ \bibnamefont
  {Lousto}}, \ and\ \bibinfo {author} {\bibfnamefont {Y.}~\bibnamefont
  {Zlochower}},\ }\href@noop {} {\bibfield  {journal} {\bibinfo  {journal}
  {Phys. Rev.}\ }\textbf {\bibinfo {volume} {D74}},\ \bibinfo {pages} {084023}
  (\bibinfo {year} {2006}{\natexlab{c}})},\ \Eprint
  {http://arxiv.org/abs/astro-ph/0608275} {astro-ph/0608275} \BibitemShut
  {NoStop}%
\bibitem [{\citenamefont {Alic}\ \emph {et~al.}(2012)\citenamefont {Alic},
  \citenamefont {Bona-Casas}, \citenamefont {Bona}, \citenamefont {Rezzolla},\
  and\ \citenamefont {Palenzuela}}]{Alic:2011gg}%
  \BibitemOpen
  \bibfield  {author} {\bibinfo {author} {\bibfnamefont {D.}~\bibnamefont
  {Alic}}, \bibinfo {author} {\bibfnamefont {C.}~\bibnamefont {Bona-Casas}},
  \bibinfo {author} {\bibfnamefont {C.}~\bibnamefont {Bona}}, \bibinfo {author}
  {\bibfnamefont {L.}~\bibnamefont {Rezzolla}}, \ and\ \bibinfo {author}
  {\bibfnamefont {C.}~\bibnamefont {Palenzuela}},\ }\href {\doibase
  10.1103/PhysRevD.85.064040} {\bibfield  {journal} {\bibinfo  {journal} {Phys.
  Rev.}\ }\textbf {\bibinfo {volume} {D85}},\ \bibinfo {pages} {064040}
  (\bibinfo {year} {2012})},\ \Eprint {http://arxiv.org/abs/1106.2254}
  {arXiv:1106.2254 [gr-qc]} \BibitemShut {NoStop}%
\bibitem [{\citenamefont {Lousto}\ \emph {et~al.}(2016)\citenamefont {Lousto},
  \citenamefont {Healy},\ and\ \citenamefont {Nakano}}]{Lousto:2015uwa}%
  \BibitemOpen
  \bibfield  {author} {\bibinfo {author} {\bibfnamefont {C.~O.}\ \bibnamefont
  {Lousto}}, \bibinfo {author} {\bibfnamefont {J.}~\bibnamefont {Healy}}, \
  and\ \bibinfo {author} {\bibfnamefont {H.}~\bibnamefont {Nakano}},\ }\href
  {\doibase 10.1103/PhysRevD.93.044031} {\bibfield  {journal} {\bibinfo
  {journal} {Phys. Rev.}\ }\textbf {\bibinfo {volume} {D93}},\ \bibinfo {pages}
  {044031} (\bibinfo {year} {2016})},\ \Eprint
  {http://arxiv.org/abs/1506.04768} {arXiv:1506.04768 [gr-qc]} \BibitemShut
  {NoStop}%
\bibitem [{\citenamefont {Lousto}\ and\ \citenamefont
  {Healy}(2016)}]{Lousto:2016nlp}%
  \BibitemOpen
  \bibfield  {author} {\bibinfo {author} {\bibfnamefont {C.~O.}\ \bibnamefont
  {Lousto}}\ and\ \bibinfo {author} {\bibfnamefont {J.}~\bibnamefont {Healy}},\
  }\href@noop {} {\  (\bibinfo {year} {2016})},\ \Eprint
  {http://arxiv.org/abs/1601.05086} {arXiv:1601.05086 [gr-qc]} \BibitemShut
  {NoStop}%
\bibitem [{\citenamefont {Lousto}\ and\ \citenamefont
  {Zlochower}(2011{\natexlab{b}})}]{Lousto:2011kp}%
  \BibitemOpen
  \bibfield  {author} {\bibinfo {author} {\bibfnamefont {C.~O.}\ \bibnamefont
  {Lousto}}\ and\ \bibinfo {author} {\bibfnamefont {Y.}~\bibnamefont
  {Zlochower}},\ }\href {\doibase 10.1103/PhysRevLett.107.231102} {\bibfield
  {journal} {\bibinfo  {journal} {Phys. Rev. Lett.}\ }\textbf {\bibinfo
  {volume} {107}},\ \bibinfo {pages} {231102} (\bibinfo {year}
  {2011}{\natexlab{b}})},\ \Eprint {http://arxiv.org/abs/1108.2009}
  {arXiv:1108.2009 [gr-qc]} \BibitemShut {NoStop}%
\bibitem [{\citenamefont {Anderson}\ \emph {et~al.}(2008)\citenamefont
  {Anderson} \emph {et~al.}}]{Anderson:2007kz}%
  \BibitemOpen
  \bibfield  {author} {\bibinfo {author} {\bibfnamefont {M.}~\bibnamefont
  {Anderson}} \emph {et~al.},\ }\href@noop {} {\bibfield  {journal} {\bibinfo
  {journal} {Phys. Rev.}\ }\textbf {\bibinfo {volume} {D77}},\ \bibinfo {pages}
  {024006} (\bibinfo {year} {2008})},\ \Eprint
  {http://arxiv.org/abs/arXiv:0708.2720 [gr-qc]} {arXiv:0708.2720 [gr-qc]}
  \BibitemShut {NoStop}%
\bibitem [{\citenamefont {Baiotti}\ \emph {et~al.}(2008)\citenamefont
  {Baiotti}, \citenamefont {Giacomazzo},\ and\ \citenamefont
  {Rezzolla}}]{Baiotti:2008ra}%
  \BibitemOpen
  \bibfield  {author} {\bibinfo {author} {\bibfnamefont {L.}~\bibnamefont
  {Baiotti}}, \bibinfo {author} {\bibfnamefont {B.}~\bibnamefont {Giacomazzo}},
  \ and\ \bibinfo {author} {\bibfnamefont {L.}~\bibnamefont {Rezzolla}},\
  }\href {\doibase 10.1103/PhysRevD.78.084033} {\bibfield  {journal} {\bibinfo
  {journal} {Phys. Rev.}\ }\textbf {\bibinfo {volume} {D78}},\ \bibinfo {pages}
  {084033} (\bibinfo {year} {2008})},\ \Eprint {http://arxiv.org/abs/0804.0594}
  {arXiv:0804.0594 [gr-qc]} \BibitemShut {NoStop}%
\bibitem [{\citenamefont {Kiuchi}\ \emph {et~al.}(2009)\citenamefont {Kiuchi},
  \citenamefont {Sekiguchi}, \citenamefont {Shibata},\ and\ \citenamefont
  {Taniguchi}}]{Kiuchi:2009jt}%
  \BibitemOpen
  \bibfield  {author} {\bibinfo {author} {\bibfnamefont {K.}~\bibnamefont
  {Kiuchi}}, \bibinfo {author} {\bibfnamefont {Y.}~\bibnamefont {Sekiguchi}},
  \bibinfo {author} {\bibfnamefont {M.}~\bibnamefont {Shibata}}, \ and\
  \bibinfo {author} {\bibfnamefont {K.}~\bibnamefont {Taniguchi}},\ }\href
  {\doibase 10.1103/PhysRevD.80.064037} {\bibfield  {journal} {\bibinfo
  {journal} {Phys. Rev.}\ }\textbf {\bibinfo {volume} {D80}},\ \bibinfo {pages}
  {064037} (\bibinfo {year} {2009})},\ \Eprint {http://arxiv.org/abs/0904.4551}
  {arXiv:0904.4551 [gr-qc]} \BibitemShut {NoStop}%
\bibitem [{\citenamefont {Bona}\ \emph {et~al.}(1995)\citenamefont {Bona},
  \citenamefont {Mass{\'o}}, \citenamefont {Seidel},\ and\ \citenamefont
  {Stela}}]{Bona94b}%
  \BibitemOpen
  \bibfield  {author} {\bibinfo {author} {\bibfnamefont {C.}~\bibnamefont
  {Bona}}, \bibinfo {author} {\bibfnamefont {J.}~\bibnamefont {Mass{\'o}}},
  \bibinfo {author} {\bibfnamefont {E.}~\bibnamefont {Seidel}}, \ and\ \bibinfo
  {author} {\bibfnamefont {J.}~\bibnamefont {Stela}},\ }\href@noop {}
  {\bibfield  {journal} {\bibinfo  {journal} {Phys. Rev. Lett.}\ }\textbf
  {\bibinfo {volume} {75}},\ \bibinfo {pages} {600} (\bibinfo {year} {1995})},\
  \Eprint {http://arxiv.org/abs/gr-qc/9412071} {gr-qc/9412071} \BibitemShut
  {NoStop}%
\bibitem [{\citenamefont {Br{\"u}gmann}(2009)}]{Brugmann:2009gc}%
  \BibitemOpen
  \bibfield  {author} {\bibinfo {author} {\bibfnamefont {B.}~\bibnamefont
  {Br{\"u}gmann}},\ }\href {\doibase 10.1007/s10714-009-0818-6} {\bibfield
  {journal} {\bibinfo  {journal} {Gen. Rel. Grav.}\ }\textbf {\bibinfo {volume}
  {41}},\ \bibinfo {pages} {2131} (\bibinfo {year} {2009})},\ \Eprint
  {http://arxiv.org/abs/0904.4418} {arXiv:0904.4418 [gr-qc]} \BibitemShut
  {NoStop}%
\bibitem [{\citenamefont {Sperhake}\ \emph {et~al.}(2008)\citenamefont
  {Sperhake}, \citenamefont {Cardoso}, \citenamefont {Pretorius}, \citenamefont
  {Berti},\ and\ \citenamefont {Gonzalez}}]{Sperhake:2008ga}%
  \BibitemOpen
  \bibfield  {author} {\bibinfo {author} {\bibfnamefont {U.}~\bibnamefont
  {Sperhake}}, \bibinfo {author} {\bibfnamefont {V.}~\bibnamefont {Cardoso}},
  \bibinfo {author} {\bibfnamefont {F.}~\bibnamefont {Pretorius}}, \bibinfo
  {author} {\bibfnamefont {E.}~\bibnamefont {Berti}}, \ and\ \bibinfo {author}
  {\bibfnamefont {J.~A.}\ \bibnamefont {Gonzalez}},\ }\href {\doibase
  10.1103/PhysRevLett.101.161101} {\bibfield  {journal} {\bibinfo  {journal}
  {Phys. Rev. Lett.}\ }\textbf {\bibinfo {volume} {101}},\ \bibinfo {pages}
  {161101} (\bibinfo {year} {2008})},\ \Eprint {http://arxiv.org/abs/0806.1738}
  {arXiv:0806.1738 [gr-qc]} \BibitemShut {NoStop}%
\bibitem [{\citenamefont {Shibata}\ \emph {et~al.}(2008)\citenamefont
  {Shibata}, \citenamefont {Okawa},\ and\ \citenamefont
  {Yamamoto}}]{Shibata:2008rq}%
  \BibitemOpen
  \bibfield  {author} {\bibinfo {author} {\bibfnamefont {M.}~\bibnamefont
  {Shibata}}, \bibinfo {author} {\bibfnamefont {H.}~\bibnamefont {Okawa}}, \
  and\ \bibinfo {author} {\bibfnamefont {T.}~\bibnamefont {Yamamoto}},\ }\href
  {\doibase 10.1103/PhysRevD.78.101501} {\bibfield  {journal} {\bibinfo
  {journal} {Phys. Rev.}\ }\textbf {\bibinfo {volume} {D78}},\ \bibinfo {pages}
  {101501} (\bibinfo {year} {2008})},\ \Eprint {http://arxiv.org/abs/0810.4735}
  {arXiv:0810.4735 [gr-qc]} \BibitemShut {NoStop}%
\bibitem [{\citenamefont {Sperhake}\ \emph {et~al.}(2009)\citenamefont
  {Sperhake}, \citenamefont {Cardoso}, \citenamefont {Pretorius}, \citenamefont
  {Berti}, \citenamefont {Hinderer},\ and\ \citenamefont
  {Yunes}}]{Sperhake:2009jz}%
  \BibitemOpen
  \bibfield  {author} {\bibinfo {author} {\bibfnamefont {U.}~\bibnamefont
  {Sperhake}}, \bibinfo {author} {\bibfnamefont {V.}~\bibnamefont {Cardoso}},
  \bibinfo {author} {\bibfnamefont {F.}~\bibnamefont {Pretorius}}, \bibinfo
  {author} {\bibfnamefont {E.}~\bibnamefont {Berti}}, \bibinfo {author}
  {\bibfnamefont {T.}~\bibnamefont {Hinderer}}, \ and\ \bibinfo {author}
  {\bibfnamefont {N.}~\bibnamefont {Yunes}},\ }\href {\doibase
  10.1103/PhysRevLett.103.131102} {\bibfield  {journal} {\bibinfo  {journal}
  {Phys. Rev. Lett.}\ }\textbf {\bibinfo {volume} {103}},\ \bibinfo {pages}
  {131102} (\bibinfo {year} {2009})},\ \Eprint {http://arxiv.org/abs/0907.1252}
  {arXiv:0907.1252 [gr-qc]} \BibitemShut {NoStop}%
\bibitem [{\citenamefont {Sperhake}\ \emph {et~al.}(2013)\citenamefont
  {Sperhake}, \citenamefont {Berti}, \citenamefont {Cardoso},\ and\
  \citenamefont {Pretorius}}]{Sperhake:2012me}%
  \BibitemOpen
  \bibfield  {author} {\bibinfo {author} {\bibfnamefont {U.}~\bibnamefont
  {Sperhake}}, \bibinfo {author} {\bibfnamefont {E.}~\bibnamefont {Berti}},
  \bibinfo {author} {\bibfnamefont {V.}~\bibnamefont {Cardoso}}, \ and\
  \bibinfo {author} {\bibfnamefont {F.}~\bibnamefont {Pretorius}},\ }\href
  {\doibase 10.1103/PhysRevLett.111.041101} {\bibfield  {journal} {\bibinfo
  {journal} {Phys. Rev. Lett.}\ }\textbf {\bibinfo {volume} {111}},\ \bibinfo
  {pages} {041101} (\bibinfo {year} {2013})},\ \Eprint
  {http://arxiv.org/abs/1211.6114} {arXiv:1211.6114 [gr-qc]} \BibitemShut
  {NoStop}%
\bibitem [{\citenamefont {Wald}(1984)}]{Wald84}%
  \BibitemOpen
  \bibfield  {author} {\bibinfo {author} {\bibfnamefont {R.~M.}\ \bibnamefont
  {Wald}},\ }\href@noop {} {\emph {\bibinfo {title} {General Relativity}}},\
  Wald84\ (\bibinfo  {publisher} {The University of Chicago Press},\ \bibinfo
  {address} {Chicago},\ \bibinfo {year} {1984})\BibitemShut {NoStop}%
\end{thebibliography}%

\end{document}